%% file: main.tex
\documentclass[notitlepage,onecolumn,pra,superscriptaddress,showkey,aps,nofootinbib,nobibnotes,10pt]{revtex4-2}
\usepackage[table,xcdraw]{xcolor}
\usepackage{titlesec}
\usepackage{comment}
\usepackage{tabularray}
\usepackage{physics}
\usepackage{array}
\UseTblrLibrary{booktabs}

\usepackage{float} 
\renewcommand\thesection{\arabic{section}}

\titleformat{\section}
  [hang]
  {\normalfont\bfseries}
  {\thesection.}
  {0.5em}
  {}

\renewcommand{\thesubsection}{\!.\arabic{subsection}}
\titleformat{\subsection}[hang]
  {\normalfont\bfseries}
  {\thesection.\arabic{subsection}} 
  {0.5em}{}

\input{preamble.tex}
\usetikzlibrary{decorations.pathreplacing}

\makeatother

\begin{document}

\title{Local Operations and Field Mediated Entanglement\\
without a Local Tensor Product Structure}

\author{Alberto Spalvieri}
\affiliation{University of Vienna, Faculty of Physics, Vienna Doctoral School in Physics, and Vienna Center for Quantum Science and Technology (VCQ), Boltzmanngasse 5, A-1090 Vienna, Austria}
\affiliation{Institute for Quantum Optics and Quantum Information (IQOQI),
Austrian Academy of Sciences, Boltzmanngasse 3, A-1090 Vienna, Austria}
\affiliation{Institute for Theoretical Physics, %
ETH Zurich, Wolfgang-Pauli-Strasse 27, 8093 Zurich, Switzerland}

\author{Sébastien C. Garmier}
\affiliation{Institute for Theoretical Physics, %
ETH Zurich, Wolfgang-Pauli-Strasse 27, 8093 Zurich, Switzerland}

\author{Flaminia Giacomini}
\affiliation{Institute for Theoretical Physics, %
ETH Zurich, Wolfgang-Pauli-Strasse 27, 8093 Zurich, Switzerland}
\affiliation{Dipartimento di Fisica, Universit{\`a} di Roma Tor Vergata and Sezione INFN Roma2, Via della Ricerca Scientifica 1, 00133, Roma, Italy}

\begin{abstract}
{\noindent Quantum information has become a powerful tool for probing the structure of quantum field theories, yet its application to gauge theories remains subtle. On the one hand, quantum information theory assumes subsystem locality, i.e.~the factorization of the total Hilbert space into subsystems. On the other hand, gauge constraints prevent the total Hilbert space to decompose into a spacetime-local tensor product structure. Because the Hilbert space structure of gauge theories does not accommodate the subsystem decomposition used in quantum information theory, standard information-theoretic results, such as the Local Operations and Classical Communication (LOCC) theorem, cannot be used straightforwardly in the context of gauge theories.
In this work, we bridge this gap in the case of a two-dimensional lattice gauge model that captures key features of electromagnetism. In particular, we construct gauge-invariant local algebras and derive a physically meaningful decomposition of the Hilbert space, providing an operationally consistent notion of locality in the absence of a local tensor-product structure.
We apply this framework to field-mediated entanglement protocols relevant to proposed tests of the quantum nature of gravity. We show that the discretized version of electromagnetism satisfies an analogue of the LOCC theorem: entanglement cannot be generated without genuine quantum field interactions, even in the absence of a spacetime-local tensor product factorization of the Hilbert space. This may point towards an operational way to define a subsystem structure for gauge theories.}
\end{abstract}
\maketitle
\section{Introduction}

Quantum information theory (QIT) has provided powerful operational tools for analysing composite quantum systems and the correlations shared between their subsystems 
\cite{PhysicsPhysiqueFizika.1.195,INGARDEN197643,PhysRevLett.69.2881,bennett1993teleporting,peres1996separability,preskill1998lecture,nielsen_chuang_2000,horodecki2009quantum}. These tools rely on the assumption of \emph{subsystem locality}, namely that 
the total Hilbert space can be factorized into smaller subsystems with the standard tensor product, and that the subsystems can be fully characterized by performing a full set of quantum operations in a laboratory. Although this characterization is fully abstract, and QIT does not prescribe a preferred physical realization, in laboratory situations this structure has to be embedded in space and time. Hence, subsystem locality typically coincides operationally with the spatial separation between laboratories and the spatial localization of physical systems therein.

Gauge theories challenge this picture. Because physical states must satisfy gauge constraints, localization in bounded regions of space(time) cannot in general be achieved. Mathematically, this is equivalent to the impossibility of factorizing the gauge-invariant Hilbert space into a \emph{spacetime local} tensor product. This hinders the application of standard information-theoretic protocols for preparing and measuring quantum states and, ultimately, the operational definition of a subsystem structure. As a consequence, one cannot straightforwardly define entanglement, a central resource for QIT. These difficulties have motivated several approaches to defining entanglement in gauge theories \cite{dec_ent_donnelly_2012,Casini_2014,casini2023lectures,Ghosh_2015,Soni_2016,Van_Acoleyen_2016,Panizza_2022,Donnelly_2015,Zanardi_2004,Bianchi_2024,PhysRevD.90.105013,BUIVIDOVICH2008141,radicevic2014notesentanglementabeliangauge,Donnelly_2014,Beem2015,Aoki2015,Donnelly_2016,Zanardi2001,delcamp2016entanglement,hung2015revisiting,dong2024holographic}. However, most of these developments only concern finite-dimensional models, and do not provide a complete operational description of local operations or of entanglement generation in infinite-dimensional gauge theories such as quantum electrodynamics (QED). 

These questions have gained renewed relevance due to proposals to test the quantum nature of gravity via field-mediated entanglement (FME) \cite{PhysRevLett.119.240401,PhysRevLett.119.240402,Galley2022nogotheoremnatureof,Hall_2018,anastopoulos_hu_2015,PhysRevD.98.126009,doi:10.1142/S0218271819430016,CHRISTODOULOU201964,PRXQuantum.2.010325,Marshman_2020,Krisnanda_2020,Marletto_2020,Pal_2021,Kent_2021,Danielson_2022,Zhou_2022,Yant2023}. There, one typically employs arguments based on the Local Operation and Classical Communication (LOCC) theorem \cite{Bennett_1996,RevModPhys.81.865}, stating that Local Operations (LO) and Classical Communication (CC) cannot generate entanglement, to conclude that gravity cannot be classical~\cite{PhysRevLett.119.240401, PhysRevLett.119.240402}. These arguments were originally formulated in finite-dimensional Hilbert spaces and rely on the subsystem structure used in QIT. However, since gravity is a gauge theory, these arguments cannot be straightforwardly applied to the FME scenario~\cite{Hall_2018,Christodoulou_2023,10.1116/5.0101334,Mart_n_Mart_nez_2023}. Recent work has clarified the structure of the quantum state of gauge fields sourced by quantum matter \cite{Giacomini2023quantumstatesof}, and even more recent studies \cite{ludescher2025gravitymediatedentanglementinfinitedimensionalsystems,yant2025operationalquantumfieldtheoretic,howl2025classicalgravityinducesentanglement,Marios2023,Bengyat2024} have made progress towards casting these proposals in a field-theoretic language. However, a fully operational account of locality and entanglement generation is still missing, and it is unknown whether the LOCC theorem can be extended to a scenario in which the mediating field has gauge redundancy.

In this work we address these issues by introducing a two-dimensional lattice toy model that captures key structural features of QED\footnote{Throughout this work, QED refers specifically to the theory of the quantized electromagnetic field coupled to a quantum source, not including the electron and positron spinor fields.}. Thanks to the discrete nature of the model we avoid other known difficulties in defining subsystem locality in quantum field theory \cite{haag1996local,doi:10.1142/S0129055X1450010X}. Building on Refs.\,\cite{Van_Acoleyen_2016,Bianchi_2024}, we construct gauge-invariant local operator algebras and the corresponding operational notion of locality, define an entanglement structure compatible with the gauge constraints, and recover a notion of local subsystems within superselection sectors which permits the formulation of a generalized LOCC theorem valid for non-factorizable Hilbert spaces. Such a toy model allows us to significantly simplify the derivations and provides us with a useful visualization of concepts such as locally accessible operations. Furthermore, through this explicit and simple approach, we aim to take a further step toward bridging the gap between the more abstract mathematical formalism, extensively explored in the literature \cite{dec_ent_donnelly_2012,Casini_2014,casini2023lectures,Ghosh_2015,Soni_2016,Van_Acoleyen_2016,Panizza_2022,Donnelly_2015,Zanardi_2004,Bianchi_2024,PhysRevD.90.105013,BUIVIDOVICH2008141,radicevic2014notesentanglementabeliangauge,Donnelly_2014,Beem2015,Aoki2015,Donnelly_2016,Zanardi2001,delcamp2016entanglement,hung2015revisiting,dong2024holographic}, and an intuitive operational description in the spirit of quantum information theory. For example, we derive the gauge-theoretic description of the operations involved in creating a superposition of sources in the presence of constraints, showing precisely in what sense they can still be regarded as local operations in such a non-trivial setting. We also establish an extended version of the protocol of FME in Ref.\,\cite{PhysRevLett.119.240401} valid for our gauge toy model.

\section{\label{sec:toymodel}Classical toy model}
 The simplest choice of a model that reproduces the non-factorizability of the Hilbert space is to have a two-dimensional lattice reducing to two-dimensional QED in the appropriate continuum limit. Such a model is compatible with known realizations of QED on a lattice \cite{CHANDRASEKHARAN1997455,QCD_brower99,Wiese2013}.

\begin{figure}[ht]
    \centering
    \includegraphics[scale=0.8]{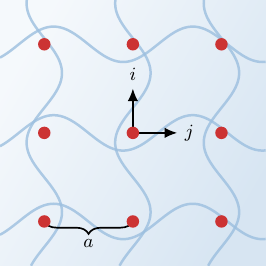}
    \caption{Graphical representation of a portion of the lattice, with waves representing a non-trivial interaction between the sites.
    }
    \label{fig:lattice}
\end{figure}

In particular, we consider a square lattice of size $N \times N$, labeled by two indices $i$ and $j$ (see Fig.\,\ref{fig:lattice}), and lattice spacing $a$ with periodic boundary conditions. In practice, we focus on the regime $N\gg1$, so that all regions of interest lie far from the boundary.
Each site of the lattice contains a field degree of freedom. For the sake of intuition, we may think of it as a particle; however, this physical interpretation is not essential to our analysis. Each particle, located on a lattice site $(i, j)$, is associated with a position vector in the lattice plane, denoted by $\mathbf{q}^{\,i,j} = (q_x^{i,j}, q_y^{i,j})$, where
$x, y$ denote the horizontal and vertical directions respectively, and an additional abstract component $q_0^{i,j}$ with physical dimensions of time. The presence of $q_0^{i,j}$ is related to the emergence of a Gauss-like constraint, analogous to the role played by $A_0$ in electromagnetic theory. Furthermore, we add sources in our toy model by associating a matter density function $\rho^{i,j}$ and current vector $\mathbf{J}^{i,j}$ at each site of the lattice. 

Here, we provide the essential elements to map our model to the electromagnetic theory. A complete derivation is given in Appendices~\ref{app:discrete_deriv} and \ref{app:class_model}. First, we derive a suitable Lagrangian in terms of discrete variables, followed by a Legendre transform to obtain the corresponding Hamiltonian. This procedure establishes a direct correspondence between electromagnetic quantities and those of the discrete model, as summarized in Tab.\,\ref{tab:e-m}.

\begin{table}[ht]
    \centering
    \begin{tblr}{colspec  = {|c|c|},
    colsep   = 2pt,
    rowsep   = 2pt,
    row{odd} = {bg=chillred!20},
    row{1}   = {bg=chillred!30},
}
        \hline
        \textbf{Electromagnetism} & \textbf{Toy model} \\ 
        \hline
        \hline
        $\mathbf{A}(\mathbf{x}),\quad A_0(\mathbf{x})$ & $\mathbf{q}^{i,j},\quad q_0^{i,j}$ \\
        
        $\mathbf{E}(\mathbf{x})$ & $-\mathbf{p}^{i,j}$ \\
    
        $\mathbf{B}(\mathbf{x}) = \nabla \times \mathbf{A}(\mathbf{x})$ & $b^{i,j} = \bar{\partial}_x q_y^{i,j} - \bar{\partial}_y q_x^{i,j}$ \\
        $\rho(\mathbf{x}),\quad \mathbf{J}(\mathbf{x})$&$\rho^{i,j},\quad \mathbf{J}^{i,j}$
        \\
        $ H = \frac{1}{2} \int d\mathbf{x} \left[ |\mathbf{E}|^2 + |\mathbf{B}|^2+(\rho-\nabla \cdot \mathbf{E}) A_0-\mathbf{J}\cdot\mathbf{A}\right]$ 
        &    
        $ H = \frac{1}{2} \sum_{i,j} \left[ p_s^{i,j}p_s^{i,j} + (b^{i,j})^2 + (\rho^{i,j}+ \bar{\partial}_s p_s^{i,j}) q_0^{i,j} - J_s^{i,j} q_s^{i,j} \right]$\\
        \hline
    \end{tblr}
    \caption{Correspondence between key quantities in electromagnetic theory and their discrete analogues in the toy model. In the last line, the repeated index $s$ denotes an implicit summation over the components $s \in \{x, y\}$.}
    \label{tab:e-m}
\end{table}

As expected, the discrete position field corresponds to the vector potential. The conjugate momentum $\mathbf{p}^{i,j}$ of $\mathbf{q}^{i,j}$ represents the electric field, up to a minus sign. The discrete analogue of the magnetic field is obtained by taking the $z$ component of the discrete curl of the position vector (this is the only component in two dimensions), and it is defined as
\begin{equation}\label{b_def}
b^{i,j} := \bar{\partial}_x q_y^{i,j} - \bar{\partial}_y q_x^{i,j} ,
\end{equation}
where we have introduced the discrete derivatives 
\begin{equation}\label{disc_der_def}
        \bar{\partial}_x f^{i,j} = \frac{1}{2a} \left(f^{i,j+1} - f^{i,j-1} \right), \qquad
        \bar{\partial}_y f^{i,j} = \frac{1}{2a} \left(f^{i+1,j} - f^{i-1,j} \right),
\end{equation}
on an arbitrary function $f^{i,j}$ defined at the site $(i,j)$ of the lattice. This definition of partial derivatives preserves key properties of their continuum counterparts, such as Schwarz's theorem, linearity, and the product rule. For details, see Appendix~\ref{app:discrete_deriv}. 

The inclusion of source-related terms completes the parallelism between electromagnetic quantities and those of the discrete model (see Table \ref{tab:e-m}). This yields a precise correspondence between the two Hamiltonians, as shown in the table, still expressed prior to any gauge fixing.
Moreover, we find that the momentum conjugate to the abstract temporal component of the position field vanishes identically:
\begin{equation}\label{prim_constr}
p_0^{i,j} = \frac{\delta L}{\delta \dot{q}_0^{i,j}} = 0 \qquad \forall, i,j.
\end{equation}
This condition represents the \textit{primary constraint} of the model, in the sense of the standard formulation of gauge theories (see, for example, Ref.\,\cite{quant_gauge_sys}).
Having established the Hamiltonian, we now analyze the corresponding Hamilton equations. The main results are summarized in Tab.\,\ref{tab:eom}. 
\begin{table}[ht]
    \centering
    \begin{tblr}{colspec  = {|c|c|},
    colsep   = 2pt,
    rowsep   = 2pt,
    row{odd} = {bg=chillred!20},
    row{1}   = {bg=chillred!30},
}
    \hline
    \textbf{Electromagnetism} & \textbf{Toy model} \\ 
    \hline
    \hline
    $\partial_t \mathbf{E} = \nabla \times \mathbf{B} - \mathbf{J}$ &
    $\begin{cases}
        \dot{p}_x^{i,j} = -\bar{\partial}_y b^{i,j} + J_x^{i,j} \\
        \dot{p}_y^{i,j} = \bar{\partial}_x b^{i,j} + J_y^{i,j}
    \end{cases}$\\
    
    $\partial_t\mathbf{B}=-\mathbf{\nabla}\times\mathbf{E}$ & $\dot{b}^{i,j} = \bar{\partial}_x p_y^{i,j} - \bar{\partial}_y p_x^{i,j}$ \\

    $\mathcal{G}_{\rho}:=\mathbf{\nabla}\cdot\mathbf{E}(\mathbf{x})-\rho(\mathbf{x})=0$ & $\mathcal{C}_{\rho}^{i,j} := \bar{\partial}_x p_x^{i,j} + \bar{\partial}_y p_y^{i,j} + \rho^{i,j} = 0$ \\
    $\mathbf{\nabla}\cdot\mathbf{J}(\mathbf{x})+\dot{\rho}(\mathbf{x})=0$&$\bar{\partial}_x J_x^{i,j} + \bar{\partial}_y J_y^{i,j} + \dot{\rho}^{i,j} = 0$
    \\
    \hline
    \end{tblr}
    \caption{Table comparing the equations of motion and constraints of the toy model with their corresponding analogues in electromagnetic theory.}
    \label{tab:eom}
\end{table}
Notably, the equations of motion can be recast in a form directly analogous to Maxwell’s equations in $2+1$ dimensions.

Following the standard treatment of gauge theories (see again Ref.\,\cite{quant_gauge_sys}), we first require the conservation of the primary constraint defined in Eq.\,\eqref{prim_constr} along dynamical trajectories. This leads to an additional condition, known as the secondary constraint, which takes the form of a discrete Gauss law:
\begin{equation}\label{constr_source}
\dot{p}_0^{i,j} = -\frac{\delta H}{\delta q_0^{i,j}} = 0
\quad \Longrightarrow \quad
\mathcal{C}_{\rho}^{i,j} := \bar{\partial}_x p_x^{i,j} + \bar{\partial}_y p_y^{i,j} + \rho^{i,j} = 0.
\end{equation}
Requiring conservation of this secondary constraint yields a further condition, equivalent to the continuity equation. Since this final constraint involves only the matter degrees of freedom, no additional constraints need to be imposed on the field variables. As a result, in the vacuum case ($\rho = 0$, $\mathbf{J} = 0$), the Gauss-like constraint is automatically preserved and represents the sole secondary constraint of the theory.

Up to this point, we have implicitly treated $p$ and $b$ as the physical observables, by analogy with electromagnetism. We now briefly justify this choice by discussing gauge transformations and gauge-invariant quantities. A full derivation is provided in Appendix~\ref{app:gauge_inv}.

According to Dirac’s formalism \cite{quant_gauge_sys}, the primary and secondary constraints, being first-class (mutually commuting), act as generators of infinitesimal gauge transformations. The variation of a generic function $f$ under a transformation generated by a constraint $\phi$ is given by
\begin{equation}\label{classical_gauge_transf}
\delta f = \epsilon \{f, \phi\},
\end{equation}
where $\epsilon$ is the infinitesimal transformation parameter and $\{\,\cdot\,,\,\cdot\,\}$ denotes the Poisson bracket. These transformations leave physical states invariant; hence, only gauge-invariant quantities correspond to physical observables.

To identify such observables, we compute how the canonical variables transform under the constraints of the model. The results are summarized in Tab.\,\ref{tab:g_trans}.
\begin{table}[ht]
    \centering
    \begin{tblr}{colspec  = {|c|c|c|},
    colsep   = 2pt,
    rowsep   = 2pt,
    row{odd} = {bg=chillred!20},
    row{1}   = {bg=chillred!30},
} 
    \hline
    \textbf{Quantity} & \textbf{Variation ($\delta$) under $p^{i,j}_0$}&\textbf{Variation ($\delta$) under $\mathcal{C}^{i,j}$} \\ 
    \hline
    \hline
    $q_0$ & $\epsilon^{i,j}_1$ & 0\\
    
    $q_s^{i,j}$ & 0&$-\bar{\partial}_s\epsilon_2^{i,j}$ \\

    $p_s^{i,j},b^{i,j}$ & 0 & 0 \\
    \hline
    \end{tblr}
    \caption{Variation of key quantities in the toy model under gauge transformations generated by the primary and secondary constraints, in the absence of sources. Here $\epsilon^{i,j}_{\bullet}$ represents the transformation parameter and 0 indicates the quantity is invariant.}
    \label{tab:g_trans}
\end{table}
As shown in the table, the primary constraints generate gauge transformations that affect only the temporal component of the position variable, where the transformation parameter represents just a linear shift. This allows us to partially fix the gauge by selecting a specific value for this parameter without altering the physical content of the theory. Throughout the rest of this work, we adopt the choice $q^{i,j}_0 = 0$ for every $i,j$, which corresponds to the temporal gauge in electromagnetism ($A_0 = 0$)\cite{QFT-fradkin}. Under this gauge choice, the full Hamiltonian simplifies to\footnote{In this gauge, the Gauss-like secondary constraint no longer appears explicitly in the Hamiltonian. It must therefore be enforced alongside the equations of motion to obtain the correct physical solutions.}
\begin{equation}\label{ham_temp_gauge}
H = \frac{1}{2} \sum_{i,j} \left[ p_s^{i,j} p_s^{i,j} + (b^{i,j})^2 - J_s^{i,j} q_s^{i,j} \right].
\end{equation}

We now consider the gauge transformations generated by the secondary constraints. We focus here on the sourceless case, as the presence of sources does not affect the gauge symmetry of the field. In this setting, the only secondary constraints are the Gauss-like conditions, which from Eq.\,\eqref{constr_source} reduce to
\begin{equation}\label{sourceless_constr}
\mathcal{C}^{i,j} := \bar{\partial}_x p_x^{i,j} + \bar{\partial}_y p_y^{i,j},
\end{equation}
valid by setting $\rho^{i,j} = 0$ and $\mathbf{J}^{i,j} = 0$ throughout the lattice.
In this case, while the momenta $p$ are gauge-invariant and thus directly represent physical observables, the spatial components of $q$ undergo nontrivial transformations   (see second line in Table \ref{tab:g_trans}). This indicates that the $q_s^{i,j}$ variables contain gauge redundancy. To extract genuine physical information, one must therefore construct combinations of $q$'s that are invariant under gauge transformations.

The relevant quantities are the $b^{i,j}$ fields, which are not only gauge-invariant, but also represent the most localized gauge-invariant combinations of $q$—involving the minimal number of variables over the smallest region around each lattice site (see Appendix~\ref{app:b_inv} for details).

\section{\label{sec:quant}Quantizing the toy model}
Quantization is carried out via the canonical procedure in the temporal gauge ($q_0 = 0$), following the approach adopted for electromagnetism in Ref.\,\cite{QFT-fradkin}. The classical fields $q_s^{i,j}$ and $p_s^{i,j}$ are promoted to operators satisfying the canonical commutation relations
\begin{equation}
    [\hat{q}_s^{i,j},\hat{p}_r^{n,m}]=\text{i}\hbar \cdot\delta_{s,r}\delta^{i,n}\delta^{j,m}\,\,\,\, \forall s,r=\{x,y\}\,\,\,\,\,\forall i,j.
\end{equation}
Here, the imaginary unit ``i'' is to be distinguished from the site index $i$.
In this section, we focus on quantizing only the field degrees of freedom and consider the sourceless case. We introduce a quantum model of matter, crucial for analyzing the FME setup, in Section~\ref{sec:q_matter}.

The equations of motion are obtained by replacing classical Poisson brackets with quantum commutators, resulting in operator equations. In that case, the Hamiltonian in the temporal gauge becomes
\begin{equation}\label{quantum_ham}
\hat{H} = \frac{1}{2} \sum_{i,j} \left[ (\hat{p}_x^{i,j})^2 + (\hat{p}_y^{i,j})^2 + (\hat{b}^{i,j})^2 \right].
\end{equation}
Similarly, the Gauss-like constraint in the sourceless case becomes an operator condition that must be satisfied by physical states:
\begin{equation}\label{constr_cond_quant}
\hat{\mathcal{C}}^{i,j} \ket{\psi}_{\text{phys}} = 0 \qquad \forall i,j.
\end{equation}
Before imposing this constraint, the kinematical Hilbert space of the system is defined as the tensor product of local Hilbert spaces at each lattice site:
\begin{equation}\label{kin}
    \mathcal{H}_{\text{kin}}=\bigotimes_{i,j}\mathcal{H}^\text{kin}_{i,j}.
\end{equation}
This space contains all possible states $\ket{\psi}$ with independent wave functions at each point in space, which in the $q$ or $p$ field representations respectively have the forms $\psi(q^{i,j}_s)$ and $\psi(p^{i,j}_s)$.
Furthermore, when acting on these states, written in the $q$ field basis, the momentum operator can be defined as
\begin{equation}\label{p_q_rep}
    \hat{p}^{i,j}_s:=-\text{i}\hbar\frac{\partial}{\partial q_s^{i,j}}.
\end{equation}

To obtain the physical Hilbert space, we project each kinematical state onto the subspace satisfying the constraint in Eq.\,\eqref{constr_cond_quant}. This is done via the (improper) projector
\begin{equation}\label{proj}
    \hat{\Pi}=\prod_{i,j}\delta(\hat{\mathcal{C}}^{i,j})=\prod_{i,j}\int_{-\infty}^{+\infty}{dz^{i,j} \,\,\text{e}^{\text{i}z^{i,j}\cdot\,\hat{\mathcal{C}}^{i,j}}}
    ,
\end{equation}
where the previous expression is defined up to an overall, irrelevant normalization factor.
Applying this projector to the kinematical state yields the physical state
\begin{equation}
\ket{\psi}_{\text{phys}} = \hat{\Pi} \ket{\psi}_{\text{kin}}.
\end{equation}

Technically, since the constraint is a linear combination of momentum operators, whose spectra are continuous, it has a continuous spectrum around the eigenvalue zero. Therefore we need to carefully normalize the physical states by appropriately redefining their inner product. This ensures that such a projection is always possible\footnote{This is a standard procedure in gauge theories, see e.g. Ref.\,\cite{quant_gauge_sys}. The new inner product for physical states can be defined as $\left\langle\psi|\phi\right\rangle_{\text{ph}} = \bra{\psi} \hat{\Pi}\ket{\phi}_{\text{kin}}$, where $\hat{\Pi}$ is the full improper projector as defined in Eq.\,\eqref{proj} and $\langle \cdot | \cdot \rangle$ is the original inner product of $\mathcal{H}_{\text{kin}}$\cite{Giacomini2023quantumstatesof,Vanrietvelde_2020}.}.

Crucially, as in electromagnetic theory, the constraints act non-locally, linking nearby points on the lattice. As a result, the physical Hilbert space does not admit a local tensor product structure like in Eq.\,\eqref{kin}. The loss of locality implies that standard notions from quantum information theory become ill-defined. Addressing this structural limitation is the central goal of the present work.

\subsection{\label{sec:q_matter}Quantum matter model}
Since the main application of this work is field-mediated entanglement between quantum sources (Section \ref{sec:FME}), we now introduce a model for quantum matter within the toy model framework. We restrict to static sources by setting $\mathbf{J}^{i,j}=0$ at every lattice site and assign to each site a matter degree of freedom represented by a single qubit. In the computational basis $\{\ket{0}^{i,j}_M, \ket{1}^{i,j}_M\}$, the state $\ket{1}^{i,j}_M$ denotes the presence of a unit charge at site $(i,j)$, while $\ket{0}^{i,j}_M$ corresponds to its absence.  

This minimal model captures only semiclassical sources, namely superpositions of approximately localized states. A description of delocalized quantum sources would require a more elaborate framework, which lies beyond the scope of this work.  
Within this setting, the matter densities $\rho^{i,j}$ are promoted to operators on the matter Hilbert space $\mathcal{H}^M$:
\begin{equation}\label{matter_dens}  
    \hat{\rho}^{i,j} = \ket{1}\!\bra{1}^{i,j}_M.  
\end{equation}  
The quantized Gauss-like constraint in Eq.\,\eqref{constr_source} then acquires a matter contribution, namely
\begin{equation}\label{physical_projection_matt}  
    \hat{\mathcal{C}}^{i,j}_{\rho}\ket{\psi}_{\text{phy}} = \left(\hat{\mathcal{C}}^{i,j} + \hat{\rho}^{i,j}\right)\ket{\psi}_{\text{phy}} = 0 \qquad \forall i,j,  
\end{equation}  
while the matter Hilbert space decomposes as  
\begin{equation}\label{matt_hilb}  
    \mathcal{H}^M = \bigoplus_{n=0}^{N^2} \underset{s_n}{\text{span}} \left\{  \ket{s_n}_M \right\}= \bigoplus_{n=0}^{N^2} \bigoplus_{s_n}\ket{s_n}_M\,  
\end{equation}  
where $\ket{s_n}_M$ denotes a matter state with exactly $n$ particles, and $s_n$ specifies one of the possible configurations of $n$ occupied sites on the $N \times N$ lattice, written in the computational basis.  
A more detailed discussion of this model is provided in Appendix \ref{app:qmatter_model}.

\subsection{\label{ground_state}Ground state}
We now derive the ground state of the quantum model, first in the source-less case, and then in the presence of a static classical source in the field basis. Technically, our result for the ground state is the discrete analogue of what was obtained in Ref.\,\cite{Giacomini2023quantumstatesof} for the case of electromagnetism. These results will later prove essential in the context of the field mediated entanglement (FME) analysis. As before, we summarize the key findings here and refer to Appendix~\ref{app:g_state} for a detailed derivation.

To solve the sourceless case, we adopt a procedure analogous to the one used in Ref.\,\cite{QFT-fradkin} for quantum electromagnetism. Specifically, we enforce the Gauss-like constraint and solve the Schr{\"o}dinger equation in Fourier space by employing the discrete Fourier transform (DFT), as defined in detail in Appendix~\ref{app:sec:DFT}. In the field representation (obtained by DFT), this leads to a dynamical equation that is equivalent to a system of independent quantum harmonic oscillators at each site of the Fourier space lattice (note that these are \emph{not} the particles of the toy model). The ground state energy is then given by
\begin{equation}\label{g_energy}
    \mathcal{E}_0= \frac{1}{2}\hbar\sum_{\alpha,\beta}|\mathbf{\bar{k}}(\alpha, \beta)|,
\end{equation}
where $\mathbf{\bar{k}}$ denotes the discrete lattice wave vector defined by the DFT momentum modes $\alpha$ and $\beta$, spanning the same discrete range as $i$ and $j$. 
The total ground state is simply given by the product of all Gaussians solutions of such DFT oscillators centered at the classical energy minimum. 
Transforming back to position space, in the $p$-field representation we get the wave function
\begin{equation}\label{g_state_p_basis}
    \Psi_0[p]=A\,\delta(\mathcal{C})\,\exp\bigg[-\frac{1}{2\hbar }\sum_{i,j}\sum_{n,m}G(i-n,j-m)\,p^{i,j}_s\,p^{n,m}_s\bigg],    
\end{equation}
where $A$ is normalization constatn, $G$ is the Green function of Schr{\"o}dinger equation, defined as
\begin{equation}\label{G_func}
  G(i-n,j-m)=\frac{1}{N^2}\sum_{\alpha,\beta}\frac{1}{|\mathbf{\bar{k}}|} \,\text{e}^{-\text{i}\frac{2\pi}{N}[(i-n)\alpha+(j-m)\beta]},
\end{equation}
and $\delta(\mathcal{C})$ enforces the Gauss-like constraint:
\begin{equation}
    \delta(\mathcal{C})=\prod_{i,j}\delta(\bar{\partial}_x p^{i,j}_x+\bar{\partial}_y p^{i,j}_y).
\end{equation}
This structure closely mirrors the known vacuum solution in electromagnetism \cite{QFT-fradkin, Giacomini2023quantumstatesof}, with the main distinction lying in the lattice-specific form of the Green function $G$. 

The full vacuum ground state in the $p$-field basis is then
\begin{equation}\label{g_state_vacuum}
    \ket{\psi^0}_F =\, A \int \mathcal{D}p\, \Psi_{0}[p] \prod_{I,J} \ket{p_x^{I,J}, p_y^{I,J}}_F 
    =A \int \mathcal{D}p\, \delta(\mathcal{C})\, \exp\left[-\frac{1}{2\hbar} \sum_{i,j} \sum_{n,m} G(i-n,j-m)\, p^{i,j}_s p^{n,m}_s \right]  \prod_{I,J} \ket{p_x^{I,J}, p_y^{I,J}}_F \,,
\end{equation}
where $A > 0$ is a constant.
This state is manifestly highly entangled, further reinforcing the analogy with the electromagnetic vacuum, which is also known to exhibit strong entanglement \cite{haag1996local, shi2004entanglement}.

We now introduce a static classical source $\rho^{i,j}$, with $J_x^{i,j} = J_y^{i,j} = 0$ at all sites. This is equivalent of restricting ourselves to a specific classical matter configuration sector from the decomposition in Eq.\,\eqref{matt_hilb}. 

By incorporating the modified Gauss-like constraint and repeating the above procedure, we obtain a correction to the ground state energy given by
\begin{equation}\label{energy_differe}
   \mathcal{E}_\rho - \mathcal{E}_0 = \frac{1}{2}\sum_{i,j}\sum_{n,m} D(i-n,j-m)\, \rho^{i,j}\, \rho^{n,m}, 
\end{equation}
where $D$ is defined analogously to $G$:
\begin{equation}
    D(i-n,j-m) = \frac{1}{N^2}\sum_{\alpha,\beta}\frac{1}{|\mathbf{\bar{k}}|^2} \,\text{e}^{-\text{i}\frac{2\pi}{N}[(i-n)\alpha+(j-m)\beta]}.
\end{equation}
Note that this differs from Eq.\,\eqref{G_func} by a $1/|k|$ factor. Again this can be related to the electromagnetic analogue by performing a continuum limit (see again Appendix~\ref{app:cont_limit}).

This energy difference corresponds precisely to the Coulomb potential of electromagnetism in $2+1$ dimensions, yielding an energy that scales with the logarithm of the distance.

The wave function of the ground state in the $p$-field representation is
\begin{equation}\label{source_g_state_p}
\Psi_{0,\rho}[p]  = A\,\delta(\mathcal{C}_\rho)\, \exp\bigg[ 
- \frac{1}{2\hbar} \sum_{i,j}\sum_{n,m} G(i-n,j-m)  (p^{i,j}_s - p^{i,j}_{s,\rho})(p^{n,m}_s - p^{n,m}_{s,\rho}) 
\bigg],
\end{equation}
with the background shift due to the source given by the static electric field with charge $\rho$:
\begin{equation} \label{p_rho} 
p^{i,j}_{s,\rho} = \sum_{n,m} \bar{\partial}_s^{\{i,j\}} D(i-n,j-m)\, \rho^{n,m},
\end{equation}
and the modified projector is
\begin{equation}
    \delta(\mathcal{C}_\rho) = \prod_{i,j} \delta(\bar{\partial}_x p^{i,j}_x + \bar{\partial}_y p^{i,j}_y + \rho^{i,j}).
\end{equation}
As in Ref.\,\cite{Giacomini2023quantumstatesof}, we observe that the presence of a static source shifts the center of the Gaussian in $p$-space, effectively translating the field configuration by the classical solution. The final expression for the ground state with a source is
\begin{equation}\label{g_state_source}
\begin{split}
\ket{\psi^0_{\rho}}_F 
=&\, A \int \mathcal{D}p\, \Psi_{0,\rho}[p] 
    \prod_{i,j} \ket{p_x^{i,j}, p_y^{i,j}}_F \\
=&\, A \int \mathcal{D}p\, \delta(\mathcal{C}_\rho)\,
    \exp\bigg[ -\frac{1}{2\hbar} \sum_{i,j} \sum_{n,m} G(i-n,j-m)\, 
    \Delta p^{i,j}_{s,\rho}\, \Delta p^{n,m}_{s,\rho} \bigg]  \prod_{i,j} \ket{p_x^{i,j}, p_y^{i,j}}_F\,,
\end{split}
\end{equation}
where $\Delta p^{i,j}_{s,\rho} := p^{i,j}_s - p^{i,j}_{s,\rho}$. As expected, the ground state remains highly entangled also in the presence of external sources.

\section{\label{sec:la_and_Hst}Local algebras and Hilbert space
structure}
We now turn to the main focus of this work, namely to understand the role of local operations in a model without a local tensor product structure. As previously noted, the physical Hilbert space lacks a local tensor product structure, preventing direct adoption of standard definitions for local operations and entanglement. In this section, we employ the simplified gauge model to first derive an algebra-induced Hilbert space structure, which then enables a well-defined notion of local operations and others useful notions of quantum information theory. For clarity of exposition, we present here the main features and results of this construction, while a more detailed and pedagogical discussion is provided in Appendices \ref{app:la} and \ref{app:Hss}. 
 
\subsection{\label{sec:alg_def}Lattice portions and local algebras}
We need to identify the set of gauge-invariant operations, one can perform locally within a specific region of the lattice in our toy model. This means to identify the \textit{local algebra} of observables associated to a certain region. 

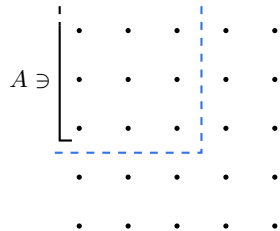
\begin{figure}[ht]
    \centering
    \begin{tikzpicture}[scale=0.65]

        \definecolor{b}{rgb}{0.9,0.55,0}

        \def\rows{5}
        \def\cols{5}

        \foreach \x in {1,...,\cols} {
            \foreach \y in {1,...,\rows} {
                \fill[black] (\x,\y) circle (1.5pt);
            }
        }

        \draw[rect,dashed, thick] (3.5, 5.5) -- (3.5, 2.5) -- (0.5, 2.5);
        \draw[black, thick] (0.6,5.1)--(0.6,2.75)--(0.85, 2.75);
        \draw[black, dashed, thick] (0.6,5.5) -- (0.6,5.1);

        \node at (0,4) {$A \ni$};

    \end{tikzpicture}
    \caption{Representation of a portion of the region $A$.}
    \label{fig:sigle_A}
\end{figure}

We consider a square region of linear size $M$, which we denote by $A$.
We think of $A$ as the set of all lattice sites contained within the boundaries of the region (see Fig.\,\ref{fig:sigle_A}).
This enclosed portion of the lattice represents the ``laboratory'' where a hypothetical observer or experimenter can perform operations. Due to the periodic boundary conditions, the absolute location of this region within the lattice is physically irrelevant. In scenarios involving multiple regions, such as in the context of FME, only the relative positions of regions carry physical significance.

We now define the local algebra of field and matter operators accessible to a region $A$, denoted by $\mathcal{A}_A$. The main points are summarized here, while a detailed analysis is provided in Appendix \ref{app:sec:alg_def}.
As in lattice gauge theories (see, e.g., Ref.\,\cite{Casini_2014,casini2023lectures}), we begin with the full operator algebra $\mathcal{O}$, in general not gauge-invariant, defined as the union of matter and field algebras:
\begin{equation}
    \mathcal{O} = \mathcal{O}^M \vee \mathcal{O}^F .
\end{equation}
The matter algebra $\mathcal{O}^M$ is generated by all operators acting on the qubits assigned to each lattice site, while the field algebra $\mathcal{O}^F$ is generated by position and momentum operators, i.e. by all linear operators on the kinematical Hilbert space $\mathcal{L}(\mathcal{H}_{\text{kin}})$. We then restrict to the subalgebra $\mathcal{O}_A$, consisting of operators supported on sites $(i,j) \in A$. The local algebra $\mathcal{A}_A$ is defined as the gauge-invariant subalgebra of $\mathcal{O}_A$, namely the set of operators commuting with the full constraint $\hat{\mathcal{C}}_{\rho}^{i,j}$ at every lattice site. These constitute the physical observables accessible to an experimenter in region $A$\footnote{This choice of local algebra is not unique; however, for regions that are well separated, different choices are equivalent (i.e. up to boundary effects).}.

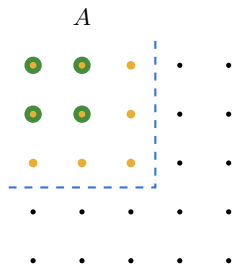
\begin{figure}[ht]
    \centering
   \begin{tikzpicture}[scale=0.65]
        \definecolor{rect}{rgb}{0.2,0.45,0.9}

        \def\rows{5}
        \def\cols{5}

        \foreach \x in {1,...,\cols} {
            \foreach \y in {1,...,\rows} {
                \fill[black] (\x,\y) circle (1.5pt);
            }
        }

        \draw[rect,dashed, thick] (3.5, 5.5) -- (3.5, 2.5) -- (0.5, 2.5);

        \node at (2,6) { $A$};

        \foreach \x in {1,2,3} {
            \foreach \y in {3,4,5} {
                \fill[myyellow] (\x,\y) circle (2.5pt);
            }
        }

        \foreach \x in {1,2} {
            \foreach \y in {4,5} {
                \draw[lightgreen, line width=2pt] (\x,\y) circle (3.5pt);
            }
        }
    \end{tikzpicture}
    \caption{Graphical representation of the generators of $\mathcal{A}_A$ based on the \emph{Graphs} notations introduced in Appendix \ref{app:graph_def}. The yellow circles and green rings label respectively the $\hat{p}$ and $\hat{b}$ operators at each site.}
    \label{graph_algrebra_def}
\end{figure}

In the sourceless case, $\mathcal{A}_A$ admits a simple graphical representation in terms of the Graphs notation introduced in Appendix \ref{app:graph_def}. As discussed in Section \ref{app:gauge_inv} and Appendix \ref{app:b_inv}, the momentum operators $\hat{p}$ automatically commute with the constraints, while the operators $\hat{b}$ provide the simplest gauge-invariant combinations of $\hat{q}$'s. Hence, $\mathcal{A}_A$ is generated by products of $\hat{p}^{i,j}$ and $\hat{b}^{i,j}$ fully contained in $A$, as illustrated in Fig.\,\ref{graph_algrebra_def}.  

In the presence of quantum sources, the intuitive graphical picture is no longer directly applicable. This is because matter operators enter the local algebra, modifying the commutation conditions, which now depend on the density operators, to identify the true gauge-invariant physical algebra (see Appendix \ref{app:sec:alg_def}). Nonetheless, the graphical representation remains a useful heuristic guide.

\subsection{\label{sec:struct_hilb}Local Hilbert space structure}
Having defined the local algebra of observables accessible to region $A$, we now proceed to study its associated Hilbert space structure.

It is well known that, under suitable conditions, a Hilbert space structure can be induced from the operator algebra acting on it (see Ref.\,\cite{Zanardi_2004} and Appendix \ref{app:sec:simple_Hss}). In standard quantum mechanics, this typically corresponds to the emergence of a local tensor product decomposition, where an algebra and its commutant act independently on complementary subsystems. This is the conventional quantum-information-theoretic notion of operational locality, not to be identified with relativistic locality. In gauge theories, however, the presence of constraints makes the center of the algebra nontrivial, and, as a result, such a decomposition is no longer available. As mentioned earlier, this complicates the definition of standard notions such as entanglement and locality. 

To overcome this issue, we adopt a method inspired by previous work on gauge theories \cite{dec_ent_donnelly_2012,Casini_2014,casini2023lectures,Van_Acoleyen_2016,kijowski1998structure, kijowski1997algebra}, which enables us to define a meaningful local decomposition. This, in turn, provides an operational notion of both entanglement and local operations in our toy model. The procedure begins by diagonalizing the nontrivial center of the local algebra $\mathcal{A}_A$. This induces a block-diagonal form of $\mathcal{A}_A$ together with its commutant, thereby giving rise to a corresponding structure of \textit{superselection sectors} in the Hilbert space. To obtain the final physical result, we must then project onto the physical Hilbert space by enforcing the constraint over the whole lattice. This is carried out in two steps: first, by diagonalizing the complete set of commuting constraint operators on the lattice, and second, by restricting to the joint eigenspaces associated with the eigenvalues that satisfy the constraint conditions.

This method is usually rigorously implemented in the framework of Type I von Neumann algebras of bounded operators but customarily also used in finite sets of harmonic oscillators. In our case, indeed, the relevant algebras are generated by unbounded position and momentum operators. Technically this step could be justified by passing to the corresponding Weyl operators \cite{Dereziński_Gérard_2013}, which are bounded unitaries satisfying an exponential version of canonical commutation relations and thus Type I. By Stone’s theorem, the original self-adjoint operators are then recovered from their exponentials, and for a finite number of harmonic oscillators the Stone-von Neumann theorem ensures that the associated von Neumann algebra is also of Type I \cite{Hall2013, KadisonRingroseII,grundling2000localquantumconstraints, KadisonRingroseII, BratteliRobinson, Takesaki2003}. In the rest of the discussion, we do not focus on this specific construction and we choose to keep the mathematical details to the minimum, as our goal is to construct the simplest physical setting in which the Hilbert space structure of a gauge theory on a lattice can be explored, with the purpose of exploring the notion of locality and the generation of entanglement.

In the presence of quantum sources, the procedure previously discussed leads to two distinct decompositions of the Hilbert space: one obtained by treating matter and field degrees of freedom jointly, and another by considering them separately. Importantly, both constructions are based on the same gauge-invariant local algebra introduced in the previous section. The crucial difference, as we shall show, is that only the former naturally induces a subsystem decomposition, thereby providing a generalized notion of locality.
Nevertheless, both decompositions play a crucial role in the analysis of the FME experiment in Section \ref{sec:FME}. In what follows, we discuss these two results and their main consequences, leaving the detailed derivation to Appendix \ref{app:sec:source_struct}.

\paragraph*{Operational Decomposition.}
We begin to treat matter and field degrees of freedom jointly, by considering the local algebra of all gauge-invariant operators in region $A$, denoted $\mathcal{A}_A$, together with its commutant $\mathcal{A}'_A$ (the set of operators commuting with all elements of $\mathcal{A}_A$). As discussed previously, the first step is to diagonalize the center of $\mathcal{A}_A$ (abelian by definition). The center contains all full constraint operators $\hat{\mathcal{C}}^{i,j}_{\rho}$ fully supported within $A$, as well as additional \emph{edge terms} corresponding to truncated constraints crossing the boundary of the region, which can be interpreted as the components of the electric field normal to the boundary. Diagonalizing these operators yields a block-diagonal decomposition of both $\mathcal{A}_A$ and its commutant, which in turn induces a corresponding structure on the full kinematical Hilbert space.  

Proceeding with this construction, we diagonalize the complete set of constraint operators $\hat{\mathcal{C}}^{i,j}_{\rho}$ over the entire lattice. This decomposition makes explicit the sectors of the Hilbert space on which the constraints take definite eigenvalues. In this way, the algebras and Hilbert space decompose as\footnote{More precisely, since the spectra of the operators in our model are continuous, ``direct sums'' $\bigoplus$ must be seen as a direct integrals (see~\cite{von_neumann}). However, for the purposes of this work it is not necessary to go further into such mathematical technicalities. Note also that in the usual language of representation theory the two factors in the first decomposition are referred respectively as the irreducible representation and the multiplicity (vice-versa for the second expression).  }\medskip
\begin{equation}\label{first_dec}
    \mathcal{A}_A = \bigoplus_{K,R} \mathcal{O}^{K}_{A} \otimes \mathbb{I}^{R,K}_{\bar{A}}, \qquad
    \bar{\mathcal{A}}_A = \bigoplus_{K,R} \mathbb{I}^K_A \otimes \mathcal{O}^{R,K}_{\bar{A}}, \qquad
    \mathcal{H} = \bigoplus_{K,R} \mathcal{H}_{A}^K \otimes \mathcal{H}_{\bar{A}}^{R,K},
\end{equation}
where $\bar{\mathcal{A}}_A$ denotes the gauge-invariant subalgebra of $\mathcal{A}'_A$, corresponding to operations localized outside $A$. Here, $K$ and $R$ label, respectively, the simultaneous eigenspaces of the center of $\mathcal{A}_A$ and of the full set of lattice constraints. In this decomposition, $\mathcal{O}^{K}_{A}$ and $\mathcal{O}^{R,K}_{\bar{A}}$ are the full operator algebras acting on $\mathcal{H}_{A}^K$ and $\mathcal{H}_{\bar{A}}^{R,K}$, which include both matter and field degrees of freedom. 

We now turn to the final step of the construction, namely the projection from the kinematical Hilbert space onto the physical subspace. This is achieved by enforcing the constraint,\smallskip
\begin{equation}\label{physical_projection}
    \hat{\mathcal{C}}^{i,j}_{\rho}\ket{\psi}_{\text{phy}} 
    = \left(\hat{\mathcal{C}}^{i,j} + \hat{\rho}^{i,j}\right) \ket{\psi}_{\text{phy}} = 0 
    \qquad \forall\, i,j,
\end{equation}
which amounts to restricting the Hilbert space to the subspace where all constraints have vanishing eigenvalue over the whole lattice. Within the decomposition introduced above, this corresponds to projecting onto a specific value of $R$,
namely the one corresponding to the simultaneous eigenspace where all constraints vanish.

Technically, since the spectrum of the constraint is continuous around zero, only improper eigenvectors exist at eigenvalue zero. This raises potential complications with the normalization of physical states. The issue can, however, be resolved by redefining the inner product on the physical Hilbert space so that the projection is well defined, \text{mycyan}{as it is explained in detail for the analogous electromagnetic case in \cite{Giacomini2023quantumstatesof}}.\footnote{Specifically this is achieved by introducing a new inner product for physical states,
\[
    \langle \psi | \phi \rangle_{\text{ph}} = \bra{\psi}\hat{\Pi}\ket{\phi}_{\text{kin}},
\]
where $\hat{\Pi}$ is the improper projector of Eq.\,\eqref{proj}, and $\langle \cdot | \cdot \rangle$ denotes the original inner product on $\mathcal{H}_{\text{kin}}$ \cite{Giacomini2023quantumstatesof,Vanrietvelde_2020}.}
We denote this choice of eigenspace by $R=0$.

While this projection uniquely determines $R$, residual freedom remains in the choice of $K$. Indeed, $K$ labels not only the eigenspaces of constraints fully contained in $A$ (which are fixed once $R=0$ is imposed), but also the edge terms. Since edge terms involve only truncated constraints crossing the boundary, they may take nontrivial eigenvalues even if the full constraint vanishes.

In summary, the physical Hilbert space is obtained by projecting onto the $R=0$ sector, with $K$ encoding the remaining edge terms. This yields
\begin{equation}\label{oper_dec}
    \mathcal{H}_{\text{phy}} 
    = \bigoplus_{K} \mathcal{H}_{A}^K \otimes \mathcal{H}_{\bar{A}}^K .
\end{equation}
We refer to this as the \emph{Operational Decomposition}, since, as discussed in more detail below, it provides a built-in notion of the complete set of operationally local operations on both matter and field degrees of freedom accessible within region $A$.
The corresponding structure of such local algebras acting on the physical Hilbert space is, in fact, given by 
\begin{equation}\label{oper_algebras}
    \mathcal{A}_A = \bigoplus_{K} \mathcal{O}^{K}_{A} \otimes \mathbb{I}^{K}_{\bar{A}}, \qquad
    \bar{\mathcal{A}}_A = \bigoplus_{K} \mathbb{I}^K_A \otimes \mathcal{O}^{K}_{\bar{A}}.
\end{equation}

\paragraph*{Split Decomposition.}
We now turn to treating matter and field degrees of freedom separately, starting with the local algebra of region $A$ restricted to field operators (i.e. operators acting acting only on the field degrees of freedom and as the identity matter), denoted by $\mathcal{A}_A^F$. This case is essentially analogous to the sourceless setting (see Appendix \ref{app:sec:sourceless_struct}): the algebra coincides with the one represented in Fig.\,\ref{graph_algrebra_def}.  
We therefore repeat a procedure similar to that of the previous paragraph. First, we diagonalize the center of $\mathcal{A}_A^F$, which now contains only the pure field constraints $\hat{\mathcal{C}}^{i,j}$ fully contained in $A$ together with the corresponding pure field edge terms. We then perform a second decomposition by diagonalizing the full set of such constraints over the whole lattice. This yields exactly the same decomposition as in Eq.\,\eqref{first_dec}, but now restricted to only the field Hilbert space.  

Since we have not yet performed any physical projection, the matter and field Hilbert spaces can be combined using the standard tensor product. Recalling the decomposition of the matter Hilbert space introduced in Eq.\,\eqref{matt_hilb}, we obtain the following structure for the total kinematical Hilbert space,  
\begin{equation}\label{source_dec_kin}
    \mathcal{H} = \bigoplus_{n=0}^{N^2} \bigoplus_{s_n}  \ket{s_n}_M \otimes \left( \bigoplus_{k,r} \mathcal{H}_A^k \otimes \mathcal{H}_{\bar{A}}^{r,k} \right).
\end{equation}
Here, $r$ and $k$ label, respectively, the joint eigenspaces of the pure field constraints and of the center of $\mathcal{A}_A^F$. The factors $\mathcal{H}_A^k$ and $\mathcal{H}_{\bar{A}}^{r,k}$ now denote the Hilbert spaces on which only the pure field local algebras act.

We now need to conclude by projecting onto the physical Hilbert space, which is again achieved by enforcing the constraint in Eq.\,\eqref{physical_projection}. In this case, however, since matter and field degrees of freedom are treated separately, the projection explicitly induces a coupling between them (matter-field entanglement). Unlike the previous setting, where $R$ and $K$ jointly labeled both matter and field eigenspaces, here the matter configurations and the field superselection sectors give rise to two distinct decompositions.  

Since the configuration states $\ket{s_n}_M$ in Eq.\,\eqref{source_dec_kin} are eigenstates of the density operators, each $\hat{\rho}^{i,j}$ in Eq.\,\eqref{physical_projection} acts as a number, taking values $0$ or $1$ on such states. Consequently, the constraint equation reduces to a condition solely on the pure field constraints $\hat{\mathcal{C}}^{i,j}$, with the matter configuration entering only as a classical parameter. This parameter uniquely identifies the eigenspace of the pure field constraint operators onto which the projection must be performed. In terms of the decomposition in Eq.\,\eqref{source_dec_kin}, this amounts to fixing the parameter $r$ according to the matter configuration, i.e., $r = r(s_n)$.  
Projecting accordingly, we obtain the physical Hilbert space as
\begin{equation}\label{split_dec}
    \mathcal{H}_{\text{phy}} 
    = \bigoplus_{n=0}^{N^2} \bigoplus_{s_n}  
    \ket{s_n}_M \otimes 
    \left( \bigoplus_{k(s_n)} 
    \mathcal{H}_A^{k(s_n)} \otimes 
    \mathcal{H}_{\bar{A}}^{s_n,k(s_n)} \right),
\end{equation}
which we refer to as the \textit{Split Decomposition}, because it makes the split between field and matter degrees of freedom explicit.
Here, for simplicity of notation, we wrote directly $s_n$ instead of $r(s_n)$ to label the Hilbert space. Furthermore, $k$ again labels the edge terms, which in general depend parametrically on the matter configuration. One could, for instance, restrict attention to cases where matter is localized well within the interior of region $A$, so that it does not directly influence the edge terms. Nevertheless, even in such cases, which of the remaining terms in the direct sum over $k$ survive after projection still depend on the matter configuration. 

\paragraph*{Comparison.}
Let us now briefly compare the two decompositions obtained for the source case, highlighting their respective strengths and limitations.

The \textit{Operational Decomposition}, as the name suggests, provides a natural operational notion of all admissible operations inside and outside region $A$, represented by the algebras in Eq.\,\eqref{oper_algebras}. In this setting, the local algebras act block-diagonally and cannot mix different $K$ sectors, which thus play the role of superselection sectors. As we discuss shortly, this feature is central to our goal of formulating an extended operational definition of entanglement and locality. However, by treating matter and field jointly, this decomposition incorporates their interaction only implicitly.  

The \textit{Split Decomposition}, in contrast, lacks a clear operational characterization of physical operations, but it offers a direct and explicit description of how the field Hilbert space and its structure depend on a given semiclassical superposition of matter configurations. This proves particularly useful in the study of FME, as it allows us to explicitly track the mechanism by which two quantum sources become entangled through their interaction with the field. For instance, given the decomposition in Eq.\,\eqref{split_dec}, a general physical state in $\mathcal{H}_{\text{phy}}$ can be written as
\begin{equation}\label{phy_state_model}
    \ket{\Upsilon} = \eta \sum_{n=0}^{N^2} \sum_{s_n} \alpha_{s_n} \, \ket{s_n}_M \ket{\Psi_{s_n}}_F,
\end{equation}
with $\eta$ being a constant, which closely resembles the corresponding structure in the QED case studied in Ref.\,\cite{Giacomini2023quantumstatesof}.

\subsection{\label{sec:ent_loc}Operational entanglement and local operations}
Having compared the two decompositions and their respective strengths, we now turn to the 
Operational Decomposition to extract concrete operational consequences. 

In particular, we conclude this section by providing our first main finding: an operational notion of entanglement and locality valid for our toy model, together with an extended formulation of the LOCC theorem. This construction is based on the Operational Decomposition introduced in the previous section (Eq.\,\eqref{oper_dec}).  

By definition, the local algebra $\mathcal{A}_A$ (resp.\ $\bar{\mathcal{A}}_A$) contains all operations accessible to a \emph{local observer} within region $A$ (or its complement, respectively). As shown in Eq.\,\eqref{oper_algebras}, these algebras act block-diagonally and cannot induce transitions between different $K$-sectors. Following \cite{Van_Acoleyen_2016}, this implies that a local observer cannot operationally distinguish a coherent superposition of sectors from the corresponding incoherent mixture.  

A direct consequence is that the entanglement accessible to the observer is restricted to that within each $K$-sector. This sector-wise entanglement between $A$ and $\bar{A}$ is well defined, since each block exhibits a tensor product structure of the form $\mathcal{H}_A^K \otimes \mathcal{H}_{\bar{A}}^K$. This provides an operational notion of entanglement consistent with the standard formalism of quantum information theory.
The price we pay for this, of course, is that entanglement must be considered in each $K$-sector separately.

Moreover, by measuring the edge terms, the observer may project the state onto a definite superselection sector. Once $K$ is fixed, the Hilbert space reduces to a standard local tensor product structure, and the familiar notions of subsystems, entanglement, local operations, and LOCC apply directly. In this sense, the operators in $\mathcal{A}_A$ and $\bar{\mathcal{A}}_A$ are best regarded as \emph{generalized local operations}, appropriate for Hilbert spaces without a global tensor product structure.  

We can thus state our generalized notions of entanglement and locality to our toy model:  
\begin{itemize}
    \item \textit{Generalized local operations} belong to the accessible algebras in Eq.\,\eqref{oper_algebras}, which act block-diagonally and, within each sector, coincide with the action of standard local operations in quantum mechanics;  
    \item The \textit{operationally accessible entanglement} is precisely the entanglement usually defined in standard quantum mechanics between elements of a tensor product, but instead applied independently in each $K$ block ($\mathcal{H}_A^K \otimes \mathcal{H}_{\bar{A}}^K$).  
\end{itemize}  
These notions can be extended also to other gauge theories where no standard local tensor product structure is available, but a decomposition as the one in Eqs.\,\eqref{oper_dec} and \eqref{oper_algebras} is still valid. Similar sector-wise approaches to entanglement can be also found in \cite{Van_Acoleyen_2016,dec_ent_donnelly_2012,Casini_2014,casini2023lectures,Bianchi_2024}.

Together, these results allow us to retain a central theorem of quantum information theory, the LOCC theorem, which plays a key role in FME-type arguments for field-mediated entanglement experiments (see Section \ref{sec:FME}). Specifically, the standard LOCC theorem now applies independently to the spatially local subsystems within each $K$-sector of the decomposition in Eq.\,\eqref{oper_dec}. We therefore state the generalized theorem as follows:  
\begin{highlightbox}[Generalized LOCC Theorem]\label{box:general_LOCC}
No \emph{operationally accessible entanglement} can be generated through the exclusive use of classical communication and \emph{generalized local operations}, belonging to the local algebra accessible to a given spatial region.
\end{highlightbox}
\vspace{2.5mm}
The previous theorem can be proved by noticing that the generalized local operations and accessible entanglement introduced above coincide, within each sector, with the corresponding standard quantum-information-theoretic notions. Hence, the original LOCC theorem applies independently in every sector, from which the statement follows immediately \cite{Bennett_1996,nielsen_chuang_2000}.
\section{\label{sec:FME}Field mediated entanglement experiment}

In the previous section we established an algebra-induced Hilbert space structure that provides an operational meaning to local operations and entanglement, even in non-factorizable gauge theories such as our toy model. We are now ready to apply these results in the context of a concrete example: \emph{field-mediated entanglement}. Consistently with the rest of the paper, we report here only the main results and discussion, referring to Appendix~\ref{app:FME} for detailed derivations.

\subsection{The context}\label{sec:FME_recap}
\paragraph*{Field-mediated entanglement and the quantum nature of gravity.}
Since the 1957 Chapel Hill conference, the question of whether gravity can exhibit quantum features has remained central~\cite{dewitt_rickles_2011}. There, Feynman proposed a thought experiment in which, by preparing a source of gravity in a quantum superposition of positions and by letting it interact with a test mass, one could distinguish, via an interferometric experiment, a coherent superposition from a classical mixture~\cite{zeh_2011}. This idea has recently gained renewed attention through proposals for table-top experiments in which two sources of gravity become entangled via their gravitational interaction~\cite{PhysRevLett.119.240401,PhysRevLett.119.240402}.

\begin{figure}[ht]
    \centering
\begin{tikzpicture}[scale=0.8, transform shape]

\definecolor{myred}{rgb}{1.00,0.2,0.2}
\definecolor{myblue}{rgb}{0,0.4,0.9}

\begin{scope}[rotate=90]

    \foreach \y in {1.7,-2.4}{
    \begin{scope}[shift={(0,\y)}]
        \draw[thick,myred, <-] (-0.6,0) -- (-1,0) -- (-0.5,0);
        \draw[thick,myred] (-0.5,-1) rectangle (3.5,1);
        \draw[thick,myred,->] (3.5,0) -- (4,0);
    \end{scope}
    };

    \node[rotate=-90] at (0.7,3.1) {$\ket{L}^B$};
    \node[rotate=-90] at (0.7,0.2) {$\ket{R}^B$};

    \node[rotate=-90] at (0.7,-1) {$\ket{L}^A$};
    \node[rotate=-90] at (0.7,-3.9) {$\ket{R}^A$};

    \draw[myblue,<->] (2.8,2.7) -- node[below,rotate=-90] {$d$} (2.8,0.7);
    \draw[myblue,<->] (2.8,-3.4) -- node[below,rotate=-90] {$d$} (2.8,-1.4);
    \draw[myblue,<->] (2.2,-3.4) -- (2.2,0.7);
    \node[myblue, rotate=-90] at (1.9, -0.7) {$L$};
    \node[rotate=-90] at (-1.3,-2.4) {$A$};
    \node[rotate=-90] at (-1.3,1.7) {$B$};

\end{scope}

\end{tikzpicture}
    \caption{Schematic representation of the FME protocol. Two massive particles, each in a spatial superposition, interact solely via the gravitational field while isolated from any other influence. The difference in distance between the interferometric paths is responsible for the generation of entanglement.}
    \label{fig:BMV}
\end{figure}
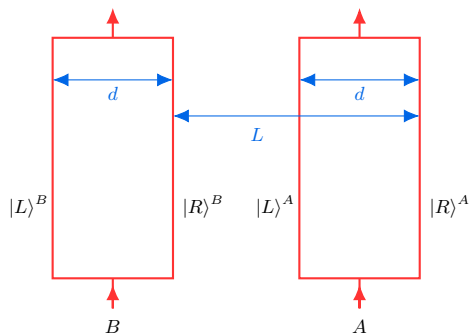

The setup proposed in~\cite{PhysRevLett.119.240401} consists of two macroscopic particles, each placed in a spatial superposition within an interferometer, allowed to interact gravitationally, and which are finally recombined (Fig.~\ref{fig:BMV}). The argument rests on two assumptions: 
\begin{itemize}
    \item the masses interact solely through the gravitational field, with no other mediation; 
    \item the LOCC theorem of quantum information theory, that entanglement cannot be generated by local operations and classical communication~\cite{Bennett_1996}, can be applied.
\end{itemize}
Under these assumptions, one would conclude that any observation of entanglement implies that the mediating field does itself possess degrees of freedom allowing more than classical communication \cite{PhysRevLett.119.240401,PhysRevLett.119.240402}.

In the Newtonian limit, the interaction is modeled by the Hamiltonian
\begin{equation}\label{newt_ham_short}
    \hat{H}_I = -G\frac{M_A m_B}{\hat{r}_{AB}},
\end{equation}
where $\hat{r}_{AB}$ denotes the distance operator between the two masses. Starting from an initial product state, time evolution under Eq.\,\eqref{newt_ham_short} produces relative phases ($\Delta\phi_{+}$ and $\Delta\phi_{-}$) depending on the distance between the particles that, in general, yields an entangled state provided that $\Delta \phi_+ \neq - \Delta \phi_-$, i.e.
\begin{equation}
   \frac{1}{2}(\ket{L}^A+\ket{R}^A)(\ket{L}^B+\ket{R}^B)\quad\rightarrow\quad\frac{\text{e}^{\text{i}\phi}}{2}[\ket{L}^A(\ket{L}^B+\text{e}^{\text{i}\Delta\phi_{-}}\ket{R}^B)+\ket{R}^A(\text{e}^{\text{i}\Delta\phi_{+}}\ket{L}^B+\ket{R}^B)].
\end{equation}

\paragraph*{LOCC and gauge theories.}
The application of the LOCC theorem, on which the argument of field-mediated entanglement (FME) relies, to gauge theories such as QED or linearized quantum gravity raises conceptual challenges. 
FME can be formally described within quantized gauge field theories. As shown in Ref.\,\cite{Giacomini2023quantumstatesof}, entanglement generation in interferometric setups like the FME protocol occurs in both QED and linearized quantum gravity. As we also found for our toy model in Section\,\ref{ground_state}, Gauss-like constraints restrict the physical Hilbert space so that the field ground state depends on the matter configuration. In the FME setup, the sources themselves are prepared in a spatial superposition, thus each configuration becomes correlated with a distinct field state, producing an overall superposition of correlated amplitudes in which matter and field are entangled due to the gauge constraint. The resulting energy differences between amplitudes, analogous to Eq.\,\eqref{energy_differe} in our toy model, generate phase shifts between the two spatially separated source states, equivalent to those found in Refs.\,\cite{PhysRevLett.119.240401, PhysRevLett.119.240402}, but now derived directly from the dynamics of the quantized field.

Furthermore, Gauss-type constraints couple matter and field degrees of freedom \emph{non-locally}, preventing the physical Hilbert space from factorizing both into independent matter and field sectors, and into a standard local tensor-product decomposition across spatial regions~\cite{haag1996local,Giddings_2015}. Instead, it exhibits a more intricate structure, as discussed in Section~\ref{sec:la_and_Hst}. Consequently, operations on the sources, most notably the creation of superposed configurations, become non-trivial and must consistently account for the coupling to the field. Moreover, the usual notions of \emph{local operations} and \emph{entanglement} cease to be well defined.

Thus, these structural features give rise to two main issues: 
\begin{itemize}
    \item the lack of an explicit mechanism describing how superposed quantum matter configurations can be prepared via local operations in the laboratory, consistently with the induced matter–field coupling,
    \item  the questionable validity of LOCC in the absence of a clear subsystem decomposition.
\end{itemize}

In the remainder of this section, we show that both issues can be resolved in the case of our toy model. Specifically the Hilbert space decompositions we previously developed enables the application of our operationally accessible entanglement, generalized local operations, and generalized LOCC theorem.

\subsection{The spatial setup}\label{sec:spatial_setup}
\begin{figure}[ht]
    \centering
    \begin{tikzpicture}[scale=0.7]

        \foreach \x in {-1,-0.8,...,8.2} {
            \foreach \y in {-0.4,-0.2,...,2.6} {
                \fill[black] (\x,\y) circle (0.5pt);
            }
        }
         \draw[rect] (0.1,0.1) rectangle (2.1,2.1);
        \draw[rect] (5.1,0.1) rectangle (7.1,2.1);
        \node at (1.1,1.1) {\Large $A$};
        \node at (6.1,1.1) {\Large $B$};

    \end{tikzpicture}
    \caption{Two square regions of dimension $M$, representing the laboratories in which the two parties $A$ and $B$ are located during a field-mediated entanglement (FME) experiment.}
    \label{fig:bmv_regions}
\end{figure}
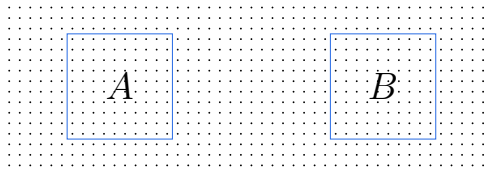

We now study the FME protocol within our quantized toy model, starting by identifying its spatial configuration in the context of the lattice gauge theory. The setup involves two parties located in spatially separated laboratories, represented in our lattice by two distant regions labeled $A$ and $B$. A convenient choice is to take two square regions of dimension $M$, as illustrated in Fig.~\ref{fig:bmv_regions}. 

The regions are chosen sufficiently far apart so that their local algebras have a trivial intersection, which is always achievable with the algebra definitions introduced in the previous chapter. This spatial separation should not be confused with the notion of spacelike separation.

We consider two static, point-like sources, each localized well within either of the two regions: one in $A$ and one in $B$. This configuration allows us to directly apply the results of Section~\ref{sec:struct_hilb} to the combined region $AB = A \cup B$, making use of the decompositions given in Eqs.\,\eqref{oper_dec} and~\eqref{split_dec}.

\paragraph*{Operational Decomposition.}
Starting from Eqs.\,\eqref{oper_dec} and~\eqref{oper_algebras} for the Hilbert space and local algebras in the operational decomposition, and imposing the condition $\mathcal{A}_A \cap \mathcal{A}_B = \mathbb{I}$ (spatially separated regions), we obtain
\begin{equation} \label{oper_dec_FME_nospin}
\mathcal{A}_{AB} = \bigoplus_{K} \mathcal{O}^{K}_{A} \otimes \mathcal{O}^{K}_{B} \otimes \mathbb{I}^{K}_{\overline{AB}}, \qquad \mathcal{H}_{\text{phy}} = \bigoplus_{K} \mathcal{H}_{A}^{K} \otimes \mathcal{H}_{B}^{K} \otimes \mathcal{H}_{\overline{AB}}^{K}, 
\end{equation}
where $\overline{AB}$ denotes the region complementary to $A \cup B$, i.e., the remainder of the lattice.

\paragraph*{Split Decomposition.}
Similarly, restricting Eq.\,\eqref{split_dec} to the sector with $n=2$, corresponding to the two-particle subspace, and analogously enforcing $\mathcal{A}^F_A \cap \mathcal{A}^F_B = \mathbb{I}$, we find
\begin{equation} \label{split_dec_FME_nospin}
\mathcal{H}_{\text{phy}} = \bigoplus_{s,k(s)} \ket{s}_M \otimes \mathcal{H}_A^{k(s)} \otimes \mathcal{H}_B^{k(s)} \otimes \mathcal{H}_{\overline{AB}}^{s,k(s)},
\end{equation}
where $s$ labels matter configurations with one particle in region $A$ and the other in region $B$ (see Appendix~\ref{app:sec:spatial_setup} for the detailed derivation).

These decompositions, obtained under the assumption $\mathcal{A}_A \cap \mathcal{A}_B = \mathbb{I}$, provide the structural basis for the analysis of the FME process: the first enables an extended FME-like argument within the generalized LOCC framework introduced in Box~\ref{box:general_LOCC}, while the second offers an explicit representation suitable for investigating entanglement generation.

\paragraph*{Introducing the spins}
Before discussing the protocol, we introduce one final element to our state-space structure. As outlined at the end of Section~\ref{sec:struct_hilb}, a generic physical state generally exhibits coupling between field and matter degrees of freedom (see Eq.\,\eqref{phy_state_model}). Consequently, a proper detection of entanglement between spatially separated systems would in principle demand simultaneous measurements on both degrees of freedom. A strategy to avoid direct field measurements was proposed in Ref.~\cite{PhysRevLett.119.240401}. This relies on the introduction of an additional, dynamically independent spin-$\frac{1}{2}$ degree of freedom $\sigma$. This spin acts as a controllable probe and allows for the detection of entanglement through spin measurements alone. Due to its independence from matter and field, the spin Hilbert space factors trivially as a tensor product, yielding the following Split Decomposition,
\begin{equation}\label{split_dec_FME}
      \mathcal{H}_{\text{phy}} = \mathcal{H}^{\sigma}\otimes\left[ \bigoplus_{s} \ket{s}_M \otimes \left( \bigoplus_{k(s)} \mathcal{H}_A^{k(s)} \otimes \mathcal{H}_B^{k(s)} \otimes \mathcal{H}_{\overline{AB}}^{s,k(s)} \right)\right],  
\end{equation}

The same is true for the Operational Decomposition, but in this case it is useful to include the spin degrees of freedom together with the other degrees of freedom. Since we consider two independent spins, each located with the corresponding matter source in its region, their joint Hilbert space trivially factorizes. By applying the distributive property of the direct sum, the total physical Hilbert space can be rewritten as
\begin{equation}
  \mathcal{H}_{\text{phy}}  =\bigoplus_{K} \underbrace{\left(\mathcal{H}_{A}^K\otimes\mathcal{H}_{\sigma}^A\right)}_{\displaystyle\mathcal{H}_{A}^{K,\text{tot}}} \otimes \underbrace{\left(\mathcal{H}_{B}^K\otimes\mathcal{H}_{\sigma}^B\right)}_{\displaystyle\mathcal{H}_{B}^{K,\text{tot}}} \otimes \mathcal{H}_{\overline{AB}}^K,
\end{equation}
where $\mathcal{H}_{\sigma}^A$ and $\mathcal{H}_{\sigma}^B$ are both spin-$\frac{1}{2}$ spaces.
Redefining the total local Hilbert spaces, we obtain the compact form
\begin{equation}\label{oper_dec_FME}
      \mathcal{H}_{\text{phy}}  =\bigoplus_{K} \mathcal{H}_{A}^{K,\text{tot}} \otimes \mathcal{H}_{B}^{K,\text{tot}} \otimes \mathcal{H}_{\overline{AB}}^K.
\end{equation}
Finally, the corresponding local algebras of gauge-invariant operations acting on the total Hilbert spaces of regions $A$ and $B$ are
\begin{equation}\label{local_algebras_FME}
        \mathcal{A}_{A}^{\text{tot}} = \bigoplus_{K} \mathcal{O}^{K,{\text{tot}}}_{A}\otimes \mathbb{I}^{K}_B \otimes \mathbb{I}^{K}_{\overline{AB}}, \qquad
        \mathcal{A}^{\text{tot}}_{B} = \bigoplus_{K} \mathbb{I}^K_A \otimes \mathcal{O}^{K,{\text{tot}}}_{B}\otimes \mathbb{I}^{K}_{\overline{AB}}.
\end{equation}
\subsection{The FME protocol and entanglement generation}\label{sec:FME_protocol}
\begin{figure}[ht]
    \centering
        \begin{tikzpicture}[scale=0.8, every node/.style={scale=0.8}]
        

        \draw[dashed] (-1,0.7) -- (3,0.7);
        \node[left] at (-1,0.7) {(0)};

        \draw[dashed] (-1,1) -- (3,1);
        \node[left] at (-1,1) {(1)};

        \draw[dashed] (-1,1.6) -- (3,1.6);
        \node[left] at (-1,1.6) {(2)};
        
        \draw[dashed] (-1,3.5) -- (3,3.5);
        \node[left] at (-1,3.5) {(3)};

        \draw[dashed] (-1,3.8) -- (3,3.8);
        \node[left] at (-1,3.8) {(4)};

        \draw[dashed] (-1,4.4) -- (3,4.4);
        \node[left] at (-1,4.4) {(5)};
        \draw[rect,line width=1.5pt] (0,0) -- (0,0.85);
        \draw[rect,line width=1.5pt] (2,0) -- (2,0.85);
        \draw[rect,line width=1.5pt] (0,0.85) -- (-0.5,0.85) -- (-0.5,3.65)--(0,3.65);
        \draw[rect,line width=1.5pt] (0,0.85) -- (0.5,0.85) -- (0.5,3.65)--(0,3.65);

        \draw[rect, line width=1.5pt] (2,0.85) -- (1.5,0.85) -- (1.5,3.65)--(2,3.65);
        \draw[rect,line width=1.5pt] (2,0.85) -- (2.5,0.85) -- (2.5,3.65)--(2,3.65);

        \draw[rect,line width=1.5pt] (0,3.65) -- (0,4.7);
        \draw[rect,line width=1.5pt] (2,3.65) -- (2,4.7);
        
        \node[rect,below] at (0,0) {A};
        \node[rect,below] at (2,0) {B};
        
        \end{tikzpicture}
    \caption{Diagrammatic representation of the FME protocol. The spatial separation of the two sources in the regions $A$ and $B$ is represented by the distance along the horizontal axis, while time increases along the vertical direction.}
    \label{fig:FME_model}
\end{figure}
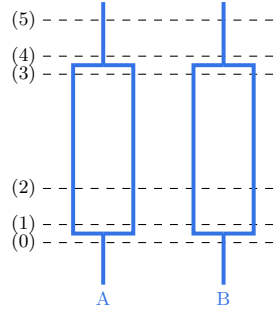

We now turn to the analysis of the FME protocol introduced in Section~\ref{sec:FME_recap}. A simplified diagram of the process is shown in Fig.~\ref{fig:FME_model}. In this section, we go through the five steps outlined in the diagram and present the main results obtained using the Split Decomposition framework of Eq.\,\eqref{split_dec_FME}, which enables an explicit treatment of the entanglement generation. The complete technical details can be found in Appendices~\ref{app:sec:creating_sup} and~\ref{app:sec:entanglement_generation}.

\paragraph*{The initial state.} 
At step $(0)$, the total system is in a product state of all degrees of freedom, as in Ref.~\cite{PhysRevLett.119.240401}. 
Each spin is independently prepared in the state $\ket{+} = \frac{1}{\sqrt{2}}(\ket{\uparrow} + \ket{\downarrow})$, 
and the matter source is in the configuration $s^0$, with one particle in each region, so that
\begin{equation}
      \ket{s^0}_M = \ket{s^0}_M^A \ket{s^0}_M^B \ket{s^0}_M^{\overline{AB}},
\end{equation}
where thus $\ket{s^0}_M^A$ and $\ket{s^0}_M^B$ are one-particle states and $\ket{s^0}_M^{\overline{AB}}$ is just the vacuum.
The field is assumed to be in its ground state, as in Eq.\,\eqref{g_state_source}, which depends on the specific matter configuration. In the Split Decomposition framework of Eq.\,\eqref{split_dec_FME}, the total initial state can thus be written as
\begin{equation}\label{init_state}
    \ket{\Psi}^{(0)} = \ket{+}_{\sigma}^A \ket{+}_{\sigma}^B \ket{s^0}_M \ket{\psi_{s^0}^0}_F.
\end{equation}

\paragraph*{Creating the superposition.} 
To proceed to the next step, we must create a spatial superposition of the two sources, each within its respective region. 
In a standard scenario, such as in Ref.~\cite{PhysRevLett.119.240401}, a Stern–Gerlach apparatus would implement a unitary $\hat{U}$ acting as
\begin{equation}\label{unphys_superp}
        \ket{\uparrow}_{\sigma}^{\bullet} \ket{s^0}_M^{\bullet} \rightarrow \ket{\uparrow}_{\sigma}^{\bullet} \ket{s^L}_M^{\bullet}, \qquad
        \ket{\downarrow}_{\sigma}^{\bullet} \ket{s^0}_M^{\bullet}\rightarrow \ket{\downarrow}_{\sigma}^{\bullet} \ket{s^R}_M^{\bullet},
\end{equation}
where $\bullet$ denotes either region $A$ or $B$, and the superscripts $L$ and $R$ indicate a displacement of the corresponding particle to the left or right.

Within our framework, the transformation must preserve the constraint in Eq.\,\eqref{physical_projection_matt} and acting only on spin and matter degrees of freedom would violate it, since $[\hat{U},\hat{\mathcal{C}}^{i,j}_{\rho}] \neq 0$.  
To obtain a physical operation, $\hat{U}$ must therefore be extended by a corresponding \emph{field dressing}. As shown in Appendix~\ref{app:sec:creating_sup}, this dressing implements the required shift of the field variable adjacent to each particle, ensuring that the full constraint $\hat{\mathcal{C}}^{i,j}_{\rho}$ is satisfied.  
The resulting physical state after step~(1) is
\begin{equation}\label{state_step1}
    \ket{\Psi}^{(1)} 
    = \hat{U} \ket{\Psi}^{(0)}
    = \sum_s 
    \ket{\sigma(s)}_{\sigma}^{AB} 
    \ket{s}_{M} 
    \ket{\psi'_s}_{F},
\end{equation}
where $\sigma(s)\in\{\uparrow\uparrow,\uparrow\downarrow,\downarrow\uparrow,\downarrow\downarrow\}$.  
A detailed derivation is given in Appendix~\ref{app:sec:entanglement_generation}.

Due to the dressing action dependent on the specific matter configuration, each field state $\ket{\psi'_s}_F$ is, in general, an excited state. We thus now wait until step~$(2)$, allowing the field to relax, approximately reaching its ground state, so that
\begin{equation}\label{state_step2}
    \ket{\Psi}^{(2)} = \sum_s e^{i\gamma(s)} \ket{\sigma(s)}^{AB}_{\sigma} \ket{s}_M \ket{\psi^0_s}_F.
\end{equation}
This relaxation may introduce relative phases $\gamma(s)$ between the configurations of matter and field, originating from the non-adiabatic character of the superposition creation process. In our discrete model, such phases are unavoidable due to the effectively instantaneous nature of the transformation. Although their precise values are generally difficult to compute, in the continuum limit a quasi-static and adiabatic implementation would render them negligible (see Ref.\,\cite{Giacomini2023quantumstatesof}). In any case, as we show below, their presence does not affect the essential conclusions of our analysis.

\paragraph*{Time evolution and field mediation.}
After the relaxation process, the full state evolves under the quantized Hamiltonian between steps $(2)$ and $(3)$. We let the system evolve according to Eq.\,\eqref{quantum_ham} for a duration $\tau = t^{(3)} - t^{(2)}$:
\begin{equation}\label{state_3}
    \ket{\Psi}^{(3)} 
    = \exp\!\left(-\text{i}\frac{\hat{H}}{\hbar}\tau\right)\!\ket{\Psi}^{(2)} 
    = \sum_s \text{e}^{i[\gamma(s) + \phi(s)]} 
    \ket{\sigma(s)}^{AB}_{\sigma} \ket{s}_M \ket{\psi^0_s}_F.
\end{equation}
Since $\hat{H}$ acts only on the field degrees of freedom, the evolution introduces an additional relative phase $\phi(s)$, determined by the energy difference between the ground states associated with the various matter configurations, computed in Section~\ref{ground_state} and given by Eq.\,\eqref{energy_differe},
\begin{equation}
    \phi(s) := -\frac{\mathcal{E}_{\rho(s)} - \mathcal{E}_0}{\hbar}\tau.
\end{equation}
This result is fully analogous to that obtained in Ref.\,\cite{Giacomini2023quantumstatesof} for both the electromagnetic and linearized gravity cases.
\paragraph*{Detecting entanglement.}
To finally detect eventual entanglement between the sources in regions $A$ and $B$ through measurements on the spins, we must recombine the spatial superposition and restore the matter to its definite initial configuration. This is achieved, analogously to the creation of the superposition, by applying a similarly dressed operator. The resulting state at step (4) is
\begin{equation}\label{state_step4}
    \ket{\Psi}^{(4)} = \hat{U}' \ket{\Psi}^{(3)} 
    = \ket{s^0}_M \sum_s \text{e}^{i[\gamma(s) + \phi(s)]} 
    \ket{\sigma(s)}^{AB}_{\sigma} \ket{\psi'_{s^0}(s)}_F,
\end{equation}
where $\ket{\psi'_{s^0}(s)}_F$ are, in general, excited field states whose form depends on the dressing associated with each matter configuration (see Appendix \ref{app:sec:entanglement_generation} for more details).  

After allowing the field to relax back to its ground state, we obtain the final state at step~(5):
\begin{equation}
    \ket{\Psi}^{(5)} 
    = \ket{s^0}_M \ket{\psi^0_{s^0}}_F 
    \sum_s \text{e}^{i[\gamma(s) + \gamma'(s) + \phi(s)]} 
    \ket{\sigma(s)}^{AB}_{\sigma},
\end{equation}
where $\gamma'(s)$ denotes the additional phase possibly introduced during the relaxation.  
At this stage, the entanglement has been effectively \emph{transferred} from the matter-field systems to the spins. Defining the generally entangled spin state as
\begin{equation}\label{final_spin}
    \ket{\chi}_{\sigma}^{AB} := 
    \sum_s \text{e}^{i[\gamma(s) + \gamma'(s) + \phi(s)]} 
    \ket{\sigma(s)}^{AB}_{\sigma},
\end{equation}
the total state can be written compactly as
\begin{equation}\label{final_state}
    \ket{\Psi}^{(5)} 
    = \ket{\chi}_{\sigma}^{AB} 
      \ket{s^0}_M 
      \ket{\psi^0_{s^0}}_F.
\end{equation}
With this form, the entanglement generated during the protocol can be directly revealed through spin measurements alone, using standard techniques from quantum information theory.

\subsection{Applying the generalized LOCC theorem}
As outlined in Section~\ref{sec:FME_recap}, gauge theories with non-factorizable Hilbert spaces, such as QED and linearized quantum gravity, raise two main objections to the FME protocol: (i) the absence of a mechanism describing the preparation of superposed matter states and entanglement generation, and (ii) the questionable applicability of the LOCC theorem in the absence of an explicitly local tensor-product structure.

Through the Split Decomposition, we have already addressed the first issue by explicitly describing how the superposed source state is produced and how entanglement arises between regions $A$ and $B$. This was enabled by the interplay between the gauge constraint, which tightly couples fields and sources, and the field’s independent quantum dynamics, which mediates the interaction. The resulting state closely resembles that found in Ref.\,\cite{Giacomini2023quantumstatesof} for QED, suggesting that a Hilbert-space decomposition analogous to the one in our discrete model could apply to electromagnetic and linearized gravitational field theories.

We now turn to the second objection, concerning the validity of the LOCC theorem. The argument used in the FME proposals~\cite{PhysRevLett.119.240401, PhysRevLett.119.240402} fundamentally relies on the principle that entanglement cannot increase under local operations and classical communication. If such an increase occurs through a field-mediated interaction, the mediator cannot be classical. The concern in gauge theories is that the lack of a factorized Hilbert space may invalidate this reasoning. However, as we now show, this is not the case: given the Operational Decomposition of Eq.\,\eqref{oper_dec_FME} and the associated local algebraic structure in Eq.\,\eqref{local_algebras_FME}, the generalized form of the LOCC theorem formulated in Box~\ref{box:general_LOCC} still applies within each $K$-sector.

Using this result, we demonstrate that the entanglement generated in the final state of the FME protocol (Eq.\,\eqref{final_state}) falls precisely within the paradigm of non-classical communication, fully consistent with the LOCC-based argument of Ref.\,\cite{PhysRevLett.119.240401}, as adapted to our toy model.

\paragraph*{Local Operations and Field Mediation.}
To begin with, we reinterpret the entire protocol in Fig.\,\ref{fig:FME_model} within the algebraic framework defined by Eq.\,\eqref{local_algebras_FME}, which acts on the Operational Decomposition and encompasses both local operations and field-mediated dynamics. 

Within this framework, as expressed in Eq.\,\eqref{oper_dec_FME}, the separation between matter and field degrees of freedom is no longer explicit. The associated algebras describe local operations in each $K$-sector independently, without mixing them. As previously discussed, this structure naturally supports a sector-wise extension of the LOCC theorem, which applies separately to the tensor product of subsystems within each $K$-sector. 

\begin{figure}[ht]
\centering
    \centering
    \begin{tikzpicture}[scale=0.85, every node/.style={draw, minimum width=0.6cm, minimum height=0.4cm, anchor=center, scale=0.85} ]

    \draw[dashed] (-1,0.7) -- (5,0.7);
    \node[draw=none, left] at (-1,0.7) {(0)};

    \draw[dashed] (-1,1.8) -- (5,1.8);
    \node[draw=none,left] at (-1,1.8) {(1)};

    \draw[dashed] (-1,2.4) -- (-0.4,2.4);
    \draw[dashed] (4.4,2.4) -- (5,2.4);
    \node[draw=none,left] at (-1,2.4) {(2)};
        
    \draw[dashed] (-1,3) -- (5,3);
    \node[draw=none,left] at (-1,3) {(3)};

    \draw[dashed] (-1,4.2) -- (5,4.2);
    \node[draw=none,left] at (-1,4.2) {(4)};

    \draw[dashed] (-1,5.35) -- (5,5.35);
    \node[draw=none,left] at (-1,5.35) {(5)};    
    \foreach \x in {0,4} {
        \draw[ thick] (\x,0.4) -- (\x,0.9);
        \draw[thick](\x,1.5)--(\x,2.1);
        \draw[thick](\x,2.7)--(\x,3.3);
        \draw[thick](\x,3.9)--(\x,4.5);  
        \draw[thick](\x,5.1)--(\x,5.6);

    }
    \draw[thick] (2,0.4) -- (2,2.1);
    \draw[thick] (2,2.7) -- (2,4.5);  \draw[thick](2,5.1)--(2,5.6);  
    
    \node[draw=none] at (0,0) {$A$};
    \node[draw=none] at (2,0) {$\overline{AB}$};
    \node[draw=none] at (4,0) {$B$};
    
    \node[ minimum width=158pt, minimum height=20pt, scale=0.8] at (2,4.8) {field mediation};
    
    \node[ minimum width=20pt, minimum height=20pt, scale=0.8] at (0,3.6) {$\hat{U}'_A$};
    \node[ minimum width=20pt, minimum height=20pt, scale=0.8] at (4,3.6) {$\hat{U}'_B$};
    
    \node[ minimum width=158pt, minimum height=20pt, scale=0.8] at (2,2.4) {field mediation};
    \node[ minimum width=20pt, minimum height=20pt, scale=0.8] at (0,1.2) {$\hat{U}_A$};
    \node[ minimum width=20pt, minimum height=20pt, scale=0.8] at (4,1.2) {$\hat{U}_B$};

    \draw [decorate, decoration={brace, mirror, amplitude=6pt}, thick]
    (4.5,5.8) -- ( -0.5,5.8);

    \draw [decorate, decoration={brace, mirror, amplitude=6pt}, thick]
    ( -0.5,-0.3) -- (4.5,-0.3);

    \draw [decorate, mypurple, decoration={brace,mirror, amplitude=4pt}, thick]
    (5.2,0.7) -- (5.2,1.75);
    \node[draw=none] at (5.7,1.25) {\textbf{\textcolor{mypurple}{LO}}};

    \draw [decorate, mypurple, decoration={brace,mirror, amplitude=4pt}, thick]
    (5.2,3.05) -- (5.2,4.15);
    \node[draw=none] at (5.7,3.6) {\textbf{\textcolor{mypurple}{LO}}};

    \draw [decorate,myorange, decoration={brace,mirror, amplitude=4pt}, thick]
    (5.2,1.85) -- (5.2,2.95);
    \node[draw=none] at (5.6,2.4) {\textbf{\textcolor{myorange}{C}}};
    
    \draw [decorate,myorange, decoration={brace,mirror, amplitude=4pt}, thick]
    (5.2,4.25) -- (5.2,5.35);
    \node[draw=none] at (5.6,4.8) {\textbf{\textcolor{myorange}{C}}};

    \node[draw=none] at (2.2,-0.8) {$ \ket{\Psi}^{(0)}  $};
    \node[draw=none] at (2.2,6.4) {$ \ket{\Psi}^{(5)}  $};
    \end{tikzpicture}

    \caption{Diagram depicting the FME protocol from the perspective of the Operational Decomposition. The horizontal lines correspond to the same time instants shown in Fig.~\ref{fig:FME_model}, while the vertical lines represent the three subsystems in tensor product with each other within each $K$-sector.}
    \label{fig:fme_locc}
\end{figure}
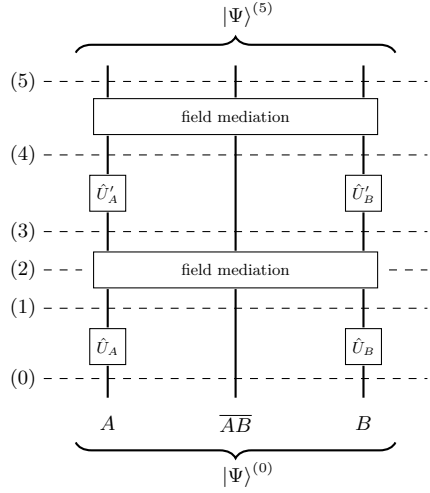

Fig.\,\ref{fig:fme_locc} illustrates the protocol as a transformation between the following two quantum states, corresponding to Eqs.\,\eqref{init_state} and~\eqref{final_state},
\begin{equation}\label{start_to_final}
 \ket{\Psi}^{(0)} = \ket{+}_{\sigma}^A \ket{+}_{\sigma}^B \ket{\psi^0}_{F,M} 
 \xrightarrow{\text{FME}} 
 \ket{\Psi}^{(5)} = \ket{\chi}_{\sigma}^{AB} \ket{\psi^0}_{F,M},
\end{equation}
and emphasizes the spatially local character of the operations involved.
The joint field–matter state $\ket{\psi^0}_{F,M}$ is identical in the initial and final steps. While it factorizes under the Split Decomposition, in the basis of the Operational Decomposition it may take a fully general form. The restoration of the field–matter system to its initial configuration is a key feature for the argument that follows.

It is important to note that the dressed operators $\hat{U}$ and $\hat{U}'$ used in Eqs.\,\eqref{state_step1} and~\eqref{state_step4} to create and merge the spatial superposition can be factorized as tensor products acting independently on each region $A$ and $B$, i.e.
\begin{equation}
    \hat{U} = \hat{U}_A \otimes \hat{U}_B, 
    \qquad 
    \hat{U}' = \hat{U}'_A \otimes \hat{U}'_B.
\end{equation}
All these operators belong to the respective local algebras defined in the Operational Decomposition of Eq.\,\eqref{local_algebras_FME} (see Appendix~\ref{app:sec:creating_sup} for a detailed derivation). This justifies representing them as local boxes acting on individual subsystems in Fig.\,\ref{fig:fme_locc}. In contrast, the field relaxation and time evolution occurring between steps~$(1)$–$(3)$ and~$(4)$–$(5)$ constitute genuinely nonlocal operations, acting solely on the field degrees of freedom.

Since our toy model is intrinsically quantum, the nonlocal field evolution illustrated in Fig.\,\ref{fig:fme_locc} is inherently of quantum nature. To formulate an argument analogous to the FME proposals~\cite{PhysRevLett.119.240401, PhysRevLett.119.240402}, we momentarily step back from the specific details of this interaction and regard it as a generic, potentially unknown, form of field mediation acting solely on the field degrees of freedom. Assuming that the Operational Decomposition of Eq.\,\eqref{oper_dec_FME} remains valid, as in gauge theories such as our model, the spin-controlled transformations can still be interpreted as local operations, in the standard quantum information sense, within each $K$-sector. Within this framework, the generalized LOCC theorem applies, allowing us to identify the field-mediated evolution as the non-classical communication channel responsible for the entanglement generation.
\paragraph*{LOCC theorem for bipartite subsystems.} 
We now apply the generalized LOCC theorem introduced in Box~\ref{box:general_LOCC}. As already mentioned, the key idea is the following: within the Operational Decomposition, local operations belong to algebras that act independently across the direct-sum sectors, allowing the LOCC theorem to be applied independently within each $K$-sector. As shown in Eq.\,\eqref{oper_dec_FME} and Fig.\,\ref{fig:fme_locc}, each sector involves three subsystems, $A$, $B$, and the rest, while the standard LOCC framework is bipartite. To adapt it, we define the bipartition
\begin{equation}\label{bip_def}
        L : \sigma_A + M_A + F, \qquad
        R : \sigma_B + M_B + F_B + M_{\overline{AB}} + F_{\overline{AB}},
\end{equation}
where $M$ and $F$ respectively refer to matter and field degrees of freedom. Note that here the choice to include the matter and field subsystems corresponding to the ``outside'' of $AB$ in the $R$ part, $M_{\overline{AB}}$ and $F_{\overline{AB}}$, is arbitrary and does not affect the argument. This choice of bipartition translates naturally to the Operational Decomposition and corresponding local algebras, ensuring that the operations at the beginning and end of the protocol can still be regarded as standard local operations on the two subsystems within each sector. Furthermore, The generalized LOCC theorem in Box~\ref{box:general_LOCC} applies directly: standard LOCC operations between the subsystems $L$ and $R$ remain valid independently within each $K$-sector. Hence, any entanglement increase observed at the end of the protocol in Fig.\,\ref{fig:fme_locc} must result from non-classical communication. Since such an increase can be detected through standard local spin measurements, it necessarily corresponds to accessible entanglement within the local tensor structure of each sector, implying that non-classical correlations were generated during the protocol.

This can now be seen explicitly by examining the initial and final states in Eq.\,\eqref{start_to_final}. Since both are pure bipartite states, acting on $\mathcal{H}_{A}^{K,\text{tot}} \otimes \mathcal{H}_{B}^{K,\text{tot}}$ in sector of the Operational Decomposition in Eq.\,\eqref{oper_dec_FME}, their entanglement can be quantified through the von Neumann entropy of either reduced subsystem~\cite{Nielsen_Chuang_2010}. For a subsystem $S$ in a state $\hat{\rho}$, entanglement is thus given by the von Neumann entropy defined as~\cite{Wilde_2013}
\begin{equation}
    \text{ent}_{\rho} = H(S)_{\rho} = -\text{Tr}(\hat{\rho} \ln \hat{\rho}),
\end{equation}
where $\text{Tr}(\bullet)$ denotes the trace over the corresponding Hilbert space.  
By computing the density matrices and using the fact that the matter and field states coincide in the initial and final configurations (see Appendix~\ref{app:sec:gen_BMV} for details), we obtain
\begin{equation}\label{ent_diff}
   \text{ent}_{LR,K}^{(5)} - \text{ent}_{LR,K}^{(0)} = H(\sigma_A)_{\chi}.
\end{equation}
If the final spin state $\ket{\chi}$ is entangled, as in Eq.\,\eqref{final_spin}, its reduced state $\hat{\chi}_A$ is mixed, implying
\begin{equation}\label{ent_inc}
    H(\sigma_A)_{\chi} > 0, 
\end{equation}
and thus 
\begin{equation}\label{ent_increase}
     \text{ent}_{LR,K}^{(5)} > \text{ent}_{LR,K}^{(0)}.
\end{equation}
Hence, the entanglement within each $K$-sector necessarily increases during the protocol, indicating the occurrence of non-classical communication, as established by the generalized LOCC theorem (Box~\ref{box:general_LOCC}). This conclusion holds independently of the chosen sector, since the entanglement increase in Eq.\,\eqref{ent_inc} is sector-invariant. In the context of the protocol in Fig.\,\ref{fig:fme_locc}, the only possible mechanism mediating such an effect is the field interaction itself\footnote{%
One may wonder whether the observed entanglement increase could be attributed to \emph{entanglement embezzlement}~\cite{PhysRevA.67.060302,vanluijk2024embezzlingentanglementquantumfields}, i.e., the apparent generation of entanglement from a highly entangled catalyst state via local operations. However, as discussed in Appendix~\ref{app:sec:emb}, embezzlement cannot be responsible for the entanglement. The protocol in Fig.\,\ref{fig:fme_locc} crucially involves an intermediate stage of autonomous field evolution, absent in genuine embezzling processes; the latter relies solely on local operations. Indeed, applying only the local operators $\hat{U}$ and $\hat{U}'$ sequentially leaves the global state unchanged, showing that the entanglement observed arises solely from the field’s non-local quantum dynamics rather than from embezzlement.}.

We have shown that the LOCC theorem can be consistently extended to gauge theories with a locally unfactorizable Hilbert space, such as the toy model considered here. This extension allows the protocol to be interpreted within the framework of local operations and classical communication, despite the absence of a local tensor product structure due to gauge constraints. While field mediation might have been expected in a quantum field theoretic setting, our analysis explicitly confirms that the entanglement generated in the discrete version of the FME protocol~\cite{PhysRevLett.119.240401,PhysRevLett.119.240402} necessarily arises from non-classical, non-local field-mediated interactions. The entanglement generation computed via the Split Decomposition in Section~\ref{sec:FME_protocol} can thus be fully understood as a genuine manifestation of non-classical communication~\cite{PhysRevLett.119.240401}.

We conclude this section illustrating how the arugment used in the FME proposals~\cite{PhysRevLett.119.240401,PhysRevLett.119.240402} can be consistently applied within our toy model. Consider two experimenters, Alice and Bob, performing an FME experiment as in Fig.\,\ref{fig:FME_model}, without any knowledge of the underlying field-theoretic structure. At the end of the protocol, they obtain a total spin state $\ket{\chi}_{\sigma}^{AB}$, reconstructed through repeated measurements to estimate, for instance, the entropy $H(A)_{\chi}$. Finding $H(A)_{\chi} > 0$, they would infer that entanglement was generated during the process. Since only local operations were performed, they would naturally invoke the LOCC theorem to conclude that the interaction field must have mediated non-classical communication.

The usual objection to this reasoning is that, in gauge theories with non-factorizable Hilbert spaces, local operations are ill-defined and the LOCC theorem ceases to apply. However, we have shown that in our toy model this concern does not arise: their inference remains valid even though the Hilbert space lacks a local tensor product structure. The operations Alice and Bob regard as local correspond precisely to the accessible local algebras of our model (Eq.\,\eqref{oper_algebras}). Hence, the observed spin entanglement ($H(\sigma_A) > 0$) directly reflects an increase in the total system’s entanglement in each $K$-sector (Eq.\,\eqref{ent_increase}), implying non-classical communication according to the generalized LOCC theorem in Box~\ref{box:general_LOCC}. This effectively resolves the objection raised against the arguments used in the FME proposals.

\section{Outlook}

The framework developed here provides a concrete operational setting in which locality, dressing, and entanglement generation can be consistently formulated in an example of gauge theory with locally non-factorizing Hilbert spaces. By introducing a tractable lattice gauge model that preserves the structural features of theories such as QED and linearized quantum gravity, we identified accessible local algebras and the associated superselection sectors, enabling a sector-wise local tensor product structure. Thanks to such a decomposition of the local algebra, the LOCC theorem can be applied independently to each sector, which leads to a generalization of the theorem to Hilbert spaces without a local factorization. 
Applied to field-mediated entanglement (FME), this structure yields an explicit operational mechanism for the creation of spatial superpositions and for entanglement generation through matter–field coupling. These results resolve the two main objections against the application of the LOCC theorem to FME protocols, namely the missing mechanism for generating matter superpositions and entanglement, and the unclear validity of LOCC-based reasoning when no standard local tensor-product structure is available.

Several directions naturally follow from this work. A first avenue of investigation concerns the extension of the toy model to progressively more general situations. For instance, the present analysis focused on effectively point-like semiclassical sources, for which quantum coherence appears only through superpositions of classical configurations. Describing sources with non-negligible position and momentum uncertainties, relevant for gravitational fields sourced by delocalized quantum states \cite{chen2024quantumeffectsgravitynewton}, will require refined dressing prescriptions and a more general operational formulation.

Secondly, the continuum limit of the model raises additional conceptual challenges. While the limit presented in Appendix \ref{app:cont_limit} reproduces two-dimensional QED, the underlying operator-algebraic structure changes: the type-I von Neumann algebras supporting the discrete-sector decompositions typically become type-III in the continuum \cite{takesaki2002theory,Yngvason_2005,Sorce_2023}. Such algebras do not admit a Hilbert-space tensor factorization, and therefore the discrete decomposition cannot be directly promoted to the strict continuum theory. An operationally motivated approach is to interpret the ``continuum limit relevant to experiments'' as one in which the lattice spacing is reduced only to a finite, resolution-dependent scale. At such finite resolution the effective description remains type-I, allowing the discrete-sector structure and the associated operational LOCC framework to persist. Making this argument precise and characterizing when type-I operational structures emerge from fundamentally type-III theories is an important direction for future work.

A complementary direction is to clarify the physical interpretation of the edge terms that label the superselection sectors. Their structural similarity to edge modes in gauge theories with boundaries \cite{PhysRevB.25.2185,Donnelly_2016,Donnelly_2015,donnelly2016local} and to recent proposals relating such modes to quantum reference frames \cite{Carrozza_2022, Kabel2024} suggests a possible unifying perspective. Establishing a concrete correspondence between these notions could deepen the operational meaning of the sectors emerging from the local algebraic decomposition.

Finally, extending the present framework to linearized quantum gravity remains a central open goal. The structural analogy between the lattice model and electromagnetism, together with the role of matter–field dressing in FME, suggests that an operational account of locality and entanglement should also exist in the gravitational case, but an explicit construction is still lacking and might present difficulties \cite{howl2025classicalgravityinducesentanglement, boulle2025subsystemsindependencegieproposals}. Such an extension would test the robustness of the generalized LOCC framework and could provide further clarity regarding gravitationally mediated entanglement and the operational signatures of the quantum nature of gravity.

In summary, the results presented here establish a minimal yet fully operational setting in which entanglement generation and locality can be consistently analyzed in gauge theories with non-factorizing Hilbert spaces. Developing continuum formulations, understanding the role of edge degrees of freedom, extending the treatment to delocalized quantum matter, and generalizing the framework to gravity represent promising next steps toward a comprehensive operational formulation of quantum gauge fields in an information-theoretic language.
\acknowledgments{The authors would like to thank Luca Ciambelli, Laurent Freidel, Apostolos Giovanakis, Marina Marinkovic, Miguel Navascu{\'e}s, Lukas Schmitt for useful discussions.
F.G. and S.G. acknowledge support from the Swiss National Science Foundation via the Ambizione Grant PZ00P2-208885, the Grant~20QU-1\_225171, and the SwissMap NCCR; furthermore, we acknowledge support from the ETH Zurich Quantum Center.
This work was made possible through the support of the WOST, WithOut SpaceTime project (https://withoutspacetime.org), supported by Grant ID\#~63683 from the John Templeton Foundation (JTF) and of the \href{https://www.templeton.org/grant/the-quantuminformation-structure-ofspacetime-qiss-second-phase}{QISS 2} ‘The Quantum Information Structure of Spacetime, Second Phase’ Project,  Grant ID\#~62312 from the John Templeton Foundation. The opinions expressed in this work are those of the author(s) and do not necessarily reflect the views of the John Templeton Foundation. 
This work was carried out with the support of the Italian Ministry of University and Research (MUR) under the Funding for Research Projects (FIS 2), pursuant to Ministerial Decree No. 23314 of 11/12/2024 (Project Q-GraSp, FIS-2023-02629). This research was funded in whole or in part by the Austrian Science Fund (FWF) [10.55776/F71] and [10.55776/COE1].
F.G. is grateful for the hospitality of Perimeter Institute where part of this work was carried out. Research at Perimeter Institute is supported in part by the Government of Canada through the Department of Innovation, Science and Economic Development Canada and by the province of Ontario through the Ministry of Economic Development, Job Creation and Trade. This research was also supported in part by the Simons Foundation through the Simons Foundation Emmy Noether Fellows Program at Perimeter Institute. This work contributes to the European Union COST Action CA23130 \textit{Bridging high and low energies in search of quantum gravity}.}

\bibliographystyle{bibstyle.bst}
\bibliography{references.bib}

\appendix

\renewcommand{\thesubsection}{\!.\arabic{subsection}}
\titleformat{\subsection}[hang]
  {\normalfont\bfseries}
  {\thesection.\arabic{subsection}} 
  {0.5em}{}

\section{\label{app:discrete_deriv}Discrete derivatives}
As also discussed in a similar way in Ref.\,\cite{ele_lattice}, there are two straightforward definitions of discrete derivatives, namely the forward and the backward derivatives. Given a generic function on the lattice, $ f^{i,j}$, we define them as
\begin{equation}
    \begin{aligned}
        \partial^+_xf^{i,j}&=\frac{1}{a}(f^{i,j+1}-f^{i,j}),\qquad \partial^-_xf^{i,j}=\frac{1}{a}(f^{i,j}-f^{i,j-1}),\\
        \partial^+_yf^{i,j}&=\frac{1}{a}(f^{i+1,j}-f^{i,j}),\qquad \partial^-_yf^{i,j}=\frac{1}{a}(f^{i,j}-f^{i-1,j}),
    \end{aligned}
\end{equation}
where we adopt the site labeling shown in Figure~\ref{fig:lattice}, and $a$ denotes the lattice spacing. It appears evident that a choice between one of these two definitions would inevitably create an asymmetry. To prevent this, we adopt a more symmetric definition by taking the average of the two. Specifically,
\begin{equation}
    \begin{aligned}
        \bar{\partial}_x f^{i,j}& = \frac{1}{2} \left( \partial^+_x f^{i,j} + \partial^-_x f^{i,j} \right) = \frac{1}{2a} \left( f^{i,j+1} - f^{i,j-1} \right), \\
        \bar{\partial}_y f^{i,j}& = \frac{1}{2} \left( \partial^+_y f^{i,j} + \partial^-_y f^{i,j} \right) = \frac{1}{2a} \left( f^{i+1,j} - f^{i-1,j} \right),
    \end{aligned}
\end{equation}
which is the definition we choose in Eq.\,\eqref{disc_der_def} for our toy model. Such a choice of discrete derivative is compelling, as it satisfies several useful properties analogous to those of the standard continuous derivative.

First, due to the symmetric construction, discrete partial derivatives commute (Schwarz theorem):
\begin{equation}
    \bar{\partial}_x \bar{\partial}_y f^{i,j} = \frac{1}{4a^2} \left( f^{i+1,j+1} - f^{i+1,j-1} - f^{i-1,j+1} + f^{i-1,j-1} \right) = \bar{\partial}_y \bar{\partial}_x f^{i,j}.
\end{equation}

Secondly, it is straightforward to verify the discrete analogue of the product rule. We illustrate this for the $x$-direction as an example. Let $f^{i,j}$ and $g^{i,j}$ be two generic functions on the lattice. The discrete derivative of their product reads
\begin{equation}
    \bar{\partial}_x(f^{i,j} g^{i,j}) = \frac{f^{i,j+1} g^{i,j+1} - f^{i,j-1} g^{i,j-1}}{2a}.
\end{equation}
By adding and subtracting suitable terms, we can rewrite this expression in two different ways
\begin{equation}
    \bar{\partial}_x(f^{i,j} g^{i,j}) 
    = g^{i,j+1} \bar{\partial}_x f^{i,j} + f^{i,j-1} \bar{\partial}_x g^{i,j} 
    = g^{i,j-1} \bar{\partial}_x f^{i,j} + f^{i,j+1} \bar{\partial}_x g^{i,j},
\end{equation}
where the first identity is obtained by adding and subtracting $f^{i,j-1} g^{i,j+1}$, and the second by using $f^{i,j+1} g^{i,j-1}$. Taking the average of these two expressions yields a symmetric form
\begin{equation}
    \bar{\partial}_x(f^{i,j} g^{i,j}) 
    = \frac{g^{i,j+1} + g^{i,j-1}}{2} \bar{\partial}_x f^{i,j} 
    + \frac{f^{i,j+1} + f^{i,j-1}}{2} \bar{\partial}_x g^{i,j},
\end{equation}
where the coefficients in front of the derivatives correspond to averages at the midpoints between lattice sites, thus reproducing the familiar structure of the product rule in the continuum. A completely analogous result holds in the $y$-direction
\begin{equation}
    \bar{\partial}_y(f^{i,j} g^{i,j}) 
    = \frac{g^{i+1,j} + g^{i-1,j}}{2} \bar{\partial}_y f^{i,j} 
    + \frac{f^{i+1,j} + f^{i-1,j}}{2} \bar{\partial}_y g^{i,j}.
\end{equation}

Finally, it is immediate from the definition that this discrete derivative operator is linear, as it is defined purely through differences of lattice-valued quantities. Linearity allows us to derive an additional property that is particularly useful for the present work, namely how to differentiate the discrete derivative of a function with respect to the function itself. We demonstrate this explicitly in the $x$-direction by directly applying the definition:
\begin{equation}
    \frac{\text{d}(\bar{\partial}_x f^{n,m})}{\text{d} f^{i,j}} = \frac{1}{2a} \left( \frac{\text{d} f^{n,m+1}}{\text{d} f^{i,j}} - \frac{\text{d} f^{n,m-1}}{\text{d} f^{i,j}} \right) = \frac{1}{2a} \left( \delta^{i,n} \delta^{j,m+1} - \delta^{i,n} \delta^{j,m-1} \right).
\end{equation}

We define this object as 
\begin{equation}
 \bar{\partial}_x^{\{n,m\}} \left( \delta^{i,n} \delta^{j,m} \right) := \frac{\text{d}(\bar{\partial}_x f^{n,m})}{\text{d} f^{i,j}}.   
\end{equation}
Using this definition and the linearity of summation, we obtain
\begin{equation}
    \sum_{n,m} \left[ \bar{\partial}_x^{\{n,m\}}(\delta^{i,n} \delta^{j,m}) \cdot f^{n,m} \right] 
    = \frac{1}{2a} \sum_{n,m} f^{n,m} \left( \delta^{i,n} \delta^{j,m+1} - \delta^{i,n} \delta^{j,m-1} \right)
    = \frac{f^{i,j-1} - f^{i,j+1}}{2a} = -\bar{\partial}_x f^{i,j},
\end{equation}
where we have also used the definition from Eq.\,\eqref{disc_der_def}.
A similar computation can be performed for the expression
$\sum_{i,j} \left[ \bar{\partial}_x^{\{n,m\}}(\delta^{i,n} \delta^{j,m}) \cdot f^{i,j} \right]$, yielding an analogous result but with an opposite sign. We conclude that this object behaves exactly like the discrete counterpart of derivatives of Dirac delta distributions in the continuum. In compact form, for either direction $s = x, y$, we have
\begin{equation}\label{diff_dev2}
    \sum_{n,m} \left[ \bar{\partial}_s^{\{n,m\}}(\delta^{i,n} \delta^{j,m}) \cdot f^{n,m} \right] = - \bar{\partial}_s f^{i,j}, \qquad
    \sum_{i,j} \left[ \bar{\partial}_s^{\{n,m\}}(\delta^{i,n} \delta^{j,m}) \cdot f^{i,j} \right] = \bar{\partial}_s f^{n,m}.
\end{equation}

\section{\label{app:class_model}Constructing the toy model}
Here, we present a detailed construction and analysis of the classical toy model introduced in Section~\ref{sec:toymodel}, and explicitly derive the results summarized in Tables~\ref{tab:e-m} and~\ref{tab:eom}.

\paragraph*{The Hamiltonian.}
We begin by reconstructing the desired Hamiltonian for the toy model in the sourceless case. This is achieved by selecting a suitable Lagrangian and performing a canonical Legendre transformation. Taking into account dimensional consistency, we consider the following Lagrangian
\begin{equation}\label{lagrangian_dime}
    L = \frac{1}{2} \sum_{i,j} \big[ m (\dot{q}_x^{i,j})^2 + m (\dot{q}_y^{i,j})^2 
    - \kappa a^2 (\bar{\partial}_x q_y^{i,j} - \bar{\partial}_y q_x^{i,j})^2 \big] 
    - \kappa a^2 \sum_{i,j} (\bar{\partial}_x \dot{q}_x^{i,j} + \bar{\partial}_y \dot{q}_y^{i,j}) q_0^{i,j} 
    + F(q_0),
\end{equation}
where $F(q_0)$ is a generic function of the $q_0^{i,j}$ fields, which can be freely chosen to match the desired Hamiltonian structure.

In order to maintain a precise analogy with electromagnetism, we fix the dimensional constants $m$ and $\kappa$ so as to connect the dimension of $q$ (taken to be length) with that of the electromagnetic vector potential $A$. To this end, we choose
\begin{equation}\label{fixed_const}
    m = 1, \qquad \kappa a^2 = 1,
\end{equation}
so that the resulting Lagrangian assumes a form directly analogous to that of classical electromagnetism:
\begin{equation}
    L = \frac{1}{2} \sum_{i,j} \left[ (\dot{q}_x^{i,j})^2 + (\dot{q}_y^{i,j})^2 
    - (\bar{\partial}_x q_y^{i,j} - \bar{\partial}_y q_x^{i,j})^2 \right] 
    - \sum_{i,j} (\bar{\partial}_x \dot{q}_x^{i,j} + \bar{\partial}_y \dot{q}_y^{i,j}) q_0^{i,j} 
    + F(q_0).
\end{equation}
This choice is not too restrictive, since it amounts to selecting either time ($T$) or length ($L$) as the fundamental unit of dimension in the model. Since $q$ has the dimension of length, our choice implies
\begin{equation}
    [\kappa a^2] = E = 1,
\end{equation}
and
\begin{equation}
    [m] = E \, T^2 \, L^{-2} = T^2 \, L^{-2} = 1.
\end{equation}
The specific values of $m$ and $\kappa$ do not influence any fundamental physical result in our analysis and their choice become relevant only when discussing the continuum limit in Appendix \ref{app:cont_limit}.

We now proceed with the canonical procedure and compute the conjugate momenta. First, note that
\begin{equation}
    p_0^{i,j} = \frac{\delta L}{\delta \dot{q}_0^{i,j}} = 0 \qquad \forall\, i,j,
\end{equation}
which defines the primary constraint.
The remaining conjugate momenta are
\begin{equation}\label{momenta}
    p_s^{i,j} = \frac{\delta L}{\delta \dot{q}_s^{i,j}} = \dot{q}_s^{i,j} + \bar{\partial}_s q_0^{i,j} \qquad \forall\, i,j,\qquad s \in \{x,y\},
\end{equation}
where we used Eqs.\,\eqref{diff_dev2}.

The Hamiltonian is obtained via the Legendre transform:
\begin{equation}
    H = \sum_{s \in \{0,x,y\}} \sum_{i,j} p_s^{i,j} \dot{q}_s^{i,j} - L(q, p).
\end{equation}
Substituting the velocities (inverted from Eq.\,\eqref{momenta}), simplifying, and using the definition of the discrete magnetic field $b^{i,j}$ from Eq.\,\eqref{b_def}, we find
\begin{equation}
    H =\sum_{i,j} p_0^{i,j} \dot{q}_0^{i,j}
    + \frac{1}{2} \sum_{i,j} \left[ (p_x^{i,j})^2 + (p_y^{i,j})^2 + (b^{i,j})^2 \right] 
    + \sum_{i,j} (\bar{\partial}_x p_x^{i,j} + \bar{\partial}_y p_y^{i,j}) q_0^{i,j}
    + G(q_0) - F(q_0),
\end{equation}
where $G(q_0)$ collects additional terms depending on $q_0^{i,j}$. By choosing $F(q_0) = G(q_0)$, we recover the desired Hamiltonian form in a generic gauge, i.e.
\begin{equation}\label{hamiltonian}
        H=\sum_{i,j}(p_0^{i,j}\dot q_0^{i,j})+\frac{1}{2}\sum_{i,j}[(p_x^{i,j})^2+(p_y^{i,j})^2+(b^{i,j})^2]+\sum_{i,j}[(\bar{\partial}_xp_x^{i,j}+\bar{\partial}_yp_y^{i,j})q_0^{i,j}].
\end{equation}

\paragraph*{Equations of motion and secondary constraint.}
 Having obtained the Hamiltonian, we now turn to Hamilton's equations. First, consider the equation for $\dot{p}_0^{i,j}$. Since, as we see in Eq.\,\eqref{prim_constr}, this component is not dynamical, its time derivative must vanish, which leads to the conservation of the primary constraint
\begin{equation}\label{ham1}
    \dot{p}_0^{i,j} = -\frac{\delta H}{\delta q_0^{i,j}} = 0 
    \quad \Longrightarrow \quad 
    \mathcal{C}^{i,j} := \bar{\partial}_x p_x^{i,j} + \bar{\partial}_y p_y^{i,j} = 0.
\end{equation}
This gives a secondary constraint of the theory which corresponds, as expected, to the discrete analogue of Gauss’s law.

The remaining Hamilton's equations yield the equations of motion for the system. By differentiating the Hamiltonian with respect to $q_x^{i,j}$ and $q_y^{i,j}$, and using the property of discrete derivatives in Eq.\,\eqref{diff_dev2}, we obtain
\begin{equation}\label{p_dot}
\begin{cases}
\displaystyle
\dot{p}_x^{i,j} = -\frac{\delta H}{\delta q_x^{i,j}} 
= -\sum_{n,m} b^{n,m} \frac{\text{d}b^{n,m}}{\text{d}q_x^{i,j}} 
= \sum_{n,m} b^{n,m} \bar{\partial}^{\{n,m\}}_y(\delta^{i,n}\delta^{j,m}) 
= -\bar{\partial}_y b^{i,j}, \\[1em]

\displaystyle
\dot{p}_y^{i,j} = -\frac{\delta H}{\delta q_y^{i,j}} 
= -\sum_{n,m} b^{n,m} \frac{\text{d}b^{n,m}}{\text{d}q_y^{i,j}} 
= -\sum_{n,m} b^{n,m} \bar{\partial}^{\{n,m\}}_x(\delta^{i,n}\delta^{j,m}) 
= \bar{\partial}_x b^{i,j}.
\end{cases}
\end{equation}
These equations are the discrete analogues of the source-free Maxwell equation $\partial_t \mathbf{E} = \nabla \times \mathbf{B}$. 

To further reinforce the analogy with Maxwell's equations, we can express the dynamics in terms of $p$ and $b$ only. Using Eq.\,\eqref{momenta} and Schwarz’ theorem for the discrete derivative, the time evolution of $b^{i,j}$ reads
\begin{equation}\label{b_ind_q0}
    \dot{b}^{i,j}=\bar{\partial}_x\dot{q}_y^{i,j}-\bar{\partial}_y\dot{q}_x^{i,j}=\bar{\partial}_xp_y^{i,j}-\bar{\partial}_x\bar{\partial}_y\dot{q}_0^{i,j}-\bar{\partial}_yp_x^{i,j}+\bar{\partial}_y\bar{\partial}_x\dot{q}_0^{i,j}=\bar{\partial}_xp_y^{i,j}-\bar{\partial}_yp_x^{i,j},
\end{equation}
which corresponds to the $z$-component of $\partial_t \mathbf{B} = -\nabla \times \mathbf{E}$.

We are now left to verify that the Gauss constraint is preserved under time evolution:
\begin{equation}
\dot{\mathcal{C}}^{i,j} = \bar{\partial}_x \dot{p}_x^{i,j} + \bar{\partial}_y \dot{p}_y^{i,j} = -\bar{\partial}_x \bar{\partial}_y b^{i,j} + \bar{\partial}_y \bar{\partial}_x b^{i,j} = 0,
\end{equation}
where we have again used Schwarz theorem for the discrete derivative. This confirms that the secondary constraint is automatically conserved in the vacuum case, reflecting the identity $\nabla \cdot (\nabla \times \bullet) = 0$ in electromagnetism. This consistency is a key geometric feature of the model and motivates our choice of a two-dimensional lattice, as such structure would not be possible in lower dimensions.
\paragraph*{Adding the sources.}
We conclude the classical analysis of the toy model by introducing the external sources. To do so, we extend the Hamiltonian in Eq.\,\eqref{hamiltonian} by adding two source terms, maintaining the analogy with electromagnetism:
\begin{equation}
\begin{split}
    H =& \sum_{i,j} p_0^{i,j} \dot{q}_0^{i,j} 
    + \frac{1}{2} \sum_{i,j} \left[ (p_x^{i,j})^2 + (p_y^{i,j})^2 + (b^{i,j})^2 \right] 
    + \sum_{i,j} (\bar{\partial}_x p_x^{i,j} + \bar{\partial}_y p_y^{i,j}) q_0^{i,j} \\
    & - \sum_{i,j} \rho^{i,j} q_0^{i,j} 
    - \sum_{i,j} \left( J_x^{i,j} q_x^{i,j} + J_y^{i,j} q_y^{i,j} \right).
\end{split}
\end{equation}

These modifications lead to corrected constraints and equations of motion. In particular, the Gauss-like constraint acquires an additional term from the charge density,
\begin{equation}
\mathcal{C}_{\rho}^{i,j}:=\mathcal{C}^{i,j}+\rho^{i,j}=0 \qquad \forall i,j,
\end{equation}
which clearly corresponds to Gauss law in the presence of a source. We must now check its conservation under time evolution by imposing
\begin{equation}
\dot{\mathcal{C}}_{\rho}^{i,j}=0\quad \Rightarrow\quad \bar{\partial}_x\dot{p}_x^{i,j}+\bar{\partial}_y\dot{p}_y^{i,j}+\dot{\rho}^{i,j}=\bar{\partial}_xJ_x^{i,j}+\bar{\partial}_yJ_y^{i,j}+\dot{\rho}^{i,j}=0.
\end{equation}
This condition represents a new secondary constraint, which is exactly the discrete analogue of the continuity equation in electromagnetism.

\section{\label{app:gauge_inv}Gauge transformations and gauge-invariant quantities}
In this appendix, we explicitly derive the results summarized in Tab.\,\ref{tab:g_trans}, which describe the variation of the fundamental variables of our toy model under gauge transformations generated by the primary and secondary constraints in the absence of sources.

It is straightforward to verify that all primary and secondary constraints Poisson-commute with one another, classifying them as first-class. As established in the standard treatment of gauge theories (see, e.g. Ref.\,\cite{quant_gauge_sys}), first-class constraints typically generate infinitesimal gauge transformations. These transformations leave the physical content of the system unchanged, implying that multiple configurations of the variables $q$ and $p$ may correspond to the same physical state.

To identify the true physical observables, meaning those that are free of gauge redundancy, it is therefore essential to determine which combinations of the dynamical variables remain invariant under these gauge transformations.

We begin with the gauge transformations generated by the primary constraints, i.e., the $p_0^{i,j}$. Our goal is to compute how the canonical variables transform under these symmetries.

Following Dirac's formalism \cite{quant_gauge_sys}, the infinitesimal variation of a function $f$ under a transformation generated by a constraint $\phi$ is given by the Poisson bracket
\begin{equation}\label{classical_gauge_transf_app}
\delta f = \epsilon \{f, \phi\} = \epsilon \sum_{i,j,s} \left( \frac{\partial f}{\partial q_s^{i,j}} \frac{\partial \phi}{\partial p_s^{i,j}} - \frac{\partial f}{\partial p_s^{i,j}} \frac{\partial \phi}{\partial q_s^{i,j}} \right),
\end{equation}
where $\epsilon$ is the infinitesimal gauge parameter.

In our case, the only non-zero Poisson bracket involving $p_0^{i,j}$ is
\begin{equation}
\delta q_0^{i,j} = \epsilon^{i,j} \{q_0^{i,j}, p_0^{i,j}\} = \epsilon^{i,j},
\end{equation}
which implies the transformation
\begin{equation}
q_0^{i,j} \rightarrow q_0^{i,j} + \epsilon^{i,j}.
\end{equation}
Since the variation is linear in the arbitrary function $\epsilon^{i,j}$, one can generate finite transformations by summing infinitesimal ones. This implies that $\epsilon^{i,j}$ can be taken as an arbitrary function on the lattice. In particular, it can be chosen such that $q_0^{i,j} = 0$ for all $i,j$, which constitutes a partial gauge fixing equivalent to adopting the temporal gauge ($A_0 = 0$) in electromagnetism. As already shown in Eq.\,\eqref{b_ind_q0}, the $b^{i,j}$ fields are independent of $q_0^{i,j}$, confirming that this gauge choice does not affect the equations of motion expressed in terms of $p$ and $b$.

We now turn to the gauge transformations generated by the secondary constraints $\mathcal{C}^{i,j}$ (defined in Eq.\,\eqref{sourceless_constr}). As before, we begin by examining their action on the canonical variables. First, we note that
\begin{equation}
\{p_s^{n,m}, \mathcal{C}^{i,j}\} = 0 \qquad \forall, n,m,i,j \qquad \text{and} \qquad s \in {x,y}.
\end{equation}
This shows that the momenta $p_s^{i,j}$, along with all functions constructed from them, are automatically gauge-invariant and thus correspond to physical observables.

Now let us proceed with the $q$'s. Let us start from computing
\begin{equation}
\{q_s^{i,j}, \mathcal{C}^{n,m} \} = \frac{\partial \mathcal{C}^{n,m}}{\partial p_s^{i,j}} = \frac{\partial \left[ \sum_r \bar{\partial}_r p_r^{n,m} \right]}{\partial p_s^{i,j}} = \sum_r \bar{\partial}^{\{n,m\}}_r \left( \delta_{r,s} \delta^{i,n} \delta^{j,m} \right) = \bar{\partial}^{\{n,m\}}_s (\delta^{i,n} \delta^{j,m}).
\end{equation}
Since these constraints are acting non-locally on quantities associated to different close-by sites, we must compute the full variation of $q_s^{i,j}$ as
\begin{equation}
\delta q_s^{i,j} = \sum_{n,m} \left[\epsilon^{n,m} \{ q_s^{i,j}, \mathcal{C}^{n,m} \}\right]=\sum_{n,m} \left[\epsilon^{n,m}\cdot\bar{\partial}^{\{n,m\}}_s (\delta^{i,n} \delta^{j,m})\right],
\end{equation}
where we can consider $\epsilon^{n,m}$ as an arbitrary function again, due to the linearity of the transformation. Using the previous result and the discrete derivative property in Eq.\,\eqref{diff_dev2}, we obtain
\begin{equation}\label{q_gauge_tranf}
\delta q_s^{i,j} = -\bar{\partial}_s \epsilon^{i,j}.
\end{equation}

This result also allows us to explicitly verify that the $b$'s are gauge-invariant. Computing the variation of the definition in Eq.\,\eqref{b_def}, we find
\begin{equation}
\delta b^{i,j} = \bar{\partial}_x (\delta q_y^{i,j}) - \bar{\partial}_y (\delta q_x^{i,j}) = -\bar{\partial}_x \bar{\partial}_y \epsilon^{i,j} + \bar{\partial}_y \bar{\partial}_x \epsilon^{i,j} = 0,
\end{equation}
where we again employed the discrete Schwartz theorem as usual. This confirms that the $b$'s are gauge-invariant quantities and therefore valid physical observables, just as in electromagnetism.

\section{\label{app:graph_def}Graphical notation}
In this appendix we introduce a graphical representation of quantities in the quantized version of the toy model, referred to as \emph{Graphs}. This notation offers an intuitive visualization of operator actions on the lattice and proves useful in simplifying several derivations. The following definitions are adopted
\begin{equation}
\vcenter{\hbox{
\begin{tikzpicture}[scale=0.8]
  \draw[mycyan, line width=0.7mm] (0,0) circle (3pt);
  \fill[black] (0,0) circle (2pt);
  \node[right] at (0.5,0) {$\hat{q}_x$};

  \draw[red, line width=0.7mm] (0,-1) circle (3pt);
  \fill[black] (0,-1) circle (2pt);
  \node[right] at (0.5,-1) {$\hat{q}_y$};

  \fill[mycyan] (2,0) circle (2pt);
  \node[right] at (2.5,0) {$\hat{p}_x$};

  \fill[red] (2,-1) circle (2pt);
  \node[right] at (2.5,-1) {$\hat{p}_y$};
\end{tikzpicture}
}}
\label{graph_def}
\end{equation}

With these conventions, quantities on the grid are visualized by placing operator symbols at the relevant sites. For example, $\hat{b}^{i,j}=\frac{1}{2a}(\hat{q}_y^{i,j+1}-\hat{q}_y^{i,j-1}-\hat{q}_x^{i+1,j}+\hat{q}_x^{i-1,j})$ and $\hat{\mathcal{C}}^{i,j}=\frac{1}{2a}(\hat{p}_x^{i,j+1}-\hat{p}_x^{i,j-1}+\hat{p}_y^{i+1,j}-\hat{p}_y^{i-1,j})$ at site $(i,j)$ are combinations of, respectively, $\hat{q}$'s and $\hat{p}$'s at neighboring sites. Hence, we represent them as
\begin{equation}
\vspace{-10pt}
\vcenter{\hbox{
\begin{tikzpicture}[scale=0.8]

\def\rows{2}
\def\cols{0}
\definecolor{pointcolor}{rgb}{0,0,0}

\foreach \x in {-3,-2,-1} {
    \foreach \y in {0,...,\rows} {
        \fill[pointcolor] (\x,\y) circle (1.5pt);
    }
}
\foreach \x in {2,3,4} {
    \foreach \y in {0,...,\rows} {
        \fill[pointcolor] (\x,\y) circle (1.7pt);
    }
}

\node[above right] at (3,0.8) {\scriptsize $i,j$};
\node[above right] at (-2,0.81) {\scriptsize $i,j$};

\fill[mycyan] (4,1) circle (2pt);
\node[text=mycyan, right] at (4,1) {$+$};
\fill[mycyan] (2,1) circle (2pt);
\node[text=mycyan, left] at (2,1) {$-$};  
\fill[red] (3,2) circle (2pt);
\node[text=red, above] at (3,2) {$+$};
\fill[red] (3,0) circle (2pt);
\node[text=red, below] at (3,0) {$-$};  

\draw[red, line width=0.4mm] (-1,1) circle (2.5pt);
\node[text=red, right] at (-1,1) {$+$};
\draw[red, line width=0.4mm] (-3,1) circle (2.5pt);
\node[text=red, left] at (-3,1) {$-$};  
\draw[mycyan, line width=0.4mm] (-2,2) circle (2.5pt);
\node[text=mycyan, above] at (-2,2) {$-$};
\draw[mycyan, line width=0.4mm] (-2,0) circle (2.5pt);
\node[text=mycyan, below] at (-2,0) {$+$};  

\node[below] at (-2,-0.75) {$\hat{b}^{i,j}$};
\node[below] at (3,-0.75) {$\hat{\mathcal{C}}^{i,j}$};   

\draw[lightgreen,rounded corners=6pt, line width=2pt]
  (-2, 2.6) -- (-2.3, 2.6) -- (-2.3, 1.3) -- (-3.6, 1.3) -- (-3.6, 0.7)
  -- (-2.3, 0.7) -- (-2.3, -0.6) -- (-1.7, -0.6) -- (-1.7, 0.7)
  -- (-0.4,0.7) -- (-0.4,1.3) -- (-1.7,1.3) -- (-1.7,2.6) -- (-2,2.6);

\draw[mypurple,rounded corners=6pt, line width=2pt]
  (3, 2.6) -- (3.3, 2.6) -- (3.3, 1.3) -- (4.6, 1.3) -- (4.6, 0.7)
  -- (3.3, 0.7) -- (3.3, -0.6) -- (2.7, -0.6) -- (2.7, 0.7)
  -- (1.4,0.7) -- (1.4,1.3) -- (2.7,1.3) -- (2.7,2.6) -- (3,2.6);
  
\draw[->, thick] (-4.5, -1.6) -- (-3.5, -1.6)  node[right] {\small  $j$};
\draw[->, thick] (-4.5, -1.6) -- (-4.5, -0.6) node[above] {\small $i$};
\end{tikzpicture}
}}
\label{graph_b_C}
\end{equation}
Finally, we introduce the following compound quantities,
\begin{equation}
  \vcenter{\hbox{  
     \begin{tikzpicture}[scale=1]
    \draw[lightgreen, line width=0.7mm] (0,0) circle (3pt); 
    \fill[black] (0,0) circle (2pt); 
    \node[right] at (0.5,0) {$\hat{b}$ ,};

    \fill[myyellow] (3,0) circle (2pt); 
   \node[right] at (3.5,0) {$\hat{p}_x \wedge\hat{p}_y$ ,};
    \end{tikzpicture}   
    }}\label{graph_def_2}
\end{equation}
where $\hat{b}$ is as in Graph\,\eqref{graph_b_C}, and $\hat{p}_x \wedge \hat{p}_y$ denotes the simultaneous inclusion of both $x$ and $y$ components ($\wedge$ indicates logical conjunction). These symbols further simplify the notation used in the following analysis.

\section[Minimal gauge-invariant observables: the b operator]{\label{app:b_inv}Minimal gauge-invariant observables: the $\boldsymbol b$ operator}

We aim to demonstrate that, by requiring a generic function of $q$ on nearby sites $i,j$ to be gauge-invariant, it must take the form of $b^{i,j}$. To do this, we will use a graphical argument, which will be more easily understood after quantization. Let us, therefore, map all canonical quantities to operators and Poisson brackets to commutators (for details, see Section \ref{sec:quant}). We adopt the following canonical commutation relations:
\begin{equation}
    [\hat{q}_s^{i,j},\hat{p}_r^{n,m}] = \text{i}\hbar \cdot \delta_{s,r} \delta^{i,n} \delta^{j,m}\,\,\,\, \forall s,r \in \{x,y\}\,\,\,\, \forall i,j.
\end{equation}
In this context the gauge-invariance condition 
\begin{equation}
    \{f(q),\hat{\mathcal{C}}^{i,j}\}=0\,\,\,\, \forall i,j
\end{equation}
simply becomes
\begin{equation}
    [f(\hat{q}),\hat{\mathcal{C}}^{i,j}]=0\,\,\,\, \forall i,j.
\end{equation}
We want now to construct a linear combination of $q$'s around a certain point of the grid, such that it fulfills the above condition. 

To address this we make use of a graphical proof based on the Graph notation introduced in the previous Appendix \ref{app:graph_def}, but before it is useful to discuss some additional features of such notation. Using the canonical commutation relations, we can associate to Graphs the commutators of functions involving only $\hat{q}$’s or only $\hat{p}$’s. At each lattice site, we consider the components of $\hat{q}$ and $\hat{p}$ and compute the corresponding commutators. For instance, the Graphs below represent quantities with mutually commuting operators, 
\begin{equation}
\vcenter{\hbox{
    \begin{tikzpicture}[scale=0.8]

    \node at (-2.5,2.2) {1)};
    \fill[red] (-2,2.2) circle (1.5pt);
    \draw[mycyan, line width=0.4mm] (-2,2.2) circle (2.5pt);
    \node[right] at (-1.75,2.2) {$[\hat{q}_x^{i,j},\hat{p}_y^{i,j}]=0$};

    \node at (1.5,2.2) {2)};
    \fill[mycyan] (2,2.2) circle (1.5pt);
    \draw[red, line width=0.4mm] (2,2.2) circle (2.5pt);
    \node[right] at (2.25,2.2) {$[\hat{q}_y^{i,j},\hat{p}_x^{i,j}]=0$};
    
    \fill[red] (-2,1) circle (1.5pt);
    \node[text=red, above] at (-2,1.05) {$+$};
    \draw[red, line width=0.4mm] (-2,1) circle (2.5pt);
    \fill[mycyan] (-1,0) circle (1.5pt);
    
    \node at (-2.5,0.5) {3)};
    \node[text=mycyan, right] at (-1,0) {$-$};
    \draw[mycyan, line width=0.4mm] (-1,0) circle (2.5pt);
    \fill[black] (-1,1) circle (1.5pt);
    \fill[black] (-2,0) circle (1.5pt);
    \node[right] at (-0.5,0.5) {$[\hat{q}_y^{i+1,j},\hat{p}_y^{i+1,j}]-[\hat{q}_x^{i,j+1},\hat{p}_x^{i,j+1}]=0$};
    \end{tikzpicture}
    }}
    \label{commutation_Ex}
\end{equation}
Here we have explicitly computed the commutators of operators at the same site (including their relative signs). Operators supported on different sites commute trivially, and thus no additional contributions appear.

\begin{figure}[ht]
    \centering
    \begin{tikzpicture}[scale=0.8]
    \definecolor{mycyan}{rgb}{0.00,0.50,0.50}
    \definecolor{rect}{rgb}{0.9,0.45,0}
    \def\rows{4} 
    \def\cols{4} 

    \foreach \x in {0,...,\cols} {
        \foreach \y in {0,...,\rows} {

            \fill[black] (\x,\y) circle (1.5pt);
        }
    }
    \node[above right] at (2,2) {$i,j$};
    \draw[red, line width=0.4mm] (3,2) circle (2.5pt);
    \node[text=red, right] at (3,2) {$+$};
    \draw[rect,dashed, line width=0.75mm] (3.5,3.5) rectangle (0.5,0.5); %
    
    \fill[mycyan] (3,2) circle (1.5pt);
    \node[text=mycyan, above] at (3,2) {$+$};
    \fill[mycyan] (1,2) circle (1.5pt);
    \node[text=mycyan, left] at (1,2) {$-$};  
    \fill[red] (2,3) circle (1.5pt);
    \node[text=red, above] at (2,3) {$+$};
    \fill[red] (2,1) circle (1.5pt);
    \node[text=red, below] at (2,1) {$-$};  
    \end{tikzpicture}
    \caption{A portion of the lattice is shown, with the desired area containing $\hat{f}$ explicitly indicated by the dotted line. The starting choice of $q_y^{i,j+1}$ is represented by the red circle, and the constraint $\hat{\mathcal{C}}^{i,j}$ is represented by filled points, as given by the definitions in Graphs\/\eqref{graph_def} and \eqref{graph_b_C}.}
    \label{starting_grid}
\end{figure}
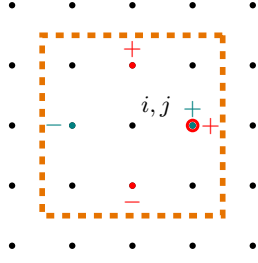

Our goal is now to find an expression for the linear combination $f(\hat{q})$ that involves the least number of $\hat{q}$'s over the smallest area possible around a site on the grid. To start, we consider a portion of the grid and select the first term of $\hat{f}(\hat{q})$. For each site, we have four possibilities corresponding to the combination of the two possible signs ($+$ and $-$) and the two components of $\hat{q}$ ($x$ and $y$). Let us begin by choosing $+$ and $y$ for the site at $i,j+1$ (see Fig.\,\ref{starting_grid}). We observe that, with this choice, the constraint centered at $i,j$ automatically commutes with this first term (see the second example in Graph\,\eqref{commutation_Ex}).

At this point, our linear combination takes the form
\begin{equation}
\hat{f} = \hat{q}_y^{i,j+1} + \dots .   
\end{equation}
To proceed, we require that all the other constraints around the site $(i,j+1)$ also commute with $\hat{f}$. As with $\hat{\mathcal{C}}^{i,j}$, the constraint centered at $(i,j+2)$ automatically commutes with the first term. However, this is not the case for the constraints $\hat{\mathcal{C}}^{i+1,j+1}$ and $\hat{\mathcal{C}}^{i-1,j+1}$.

Let us start by considering $\hat{\mathcal{C}}^{i+1,j+1}$ and draw at the Graph below:
\begin{center}
    \begin{tikzpicture}[scale=0.8]

    \definecolor{mycyan}{rgb}{0.00,0.50,0.50}
    \definecolor{rect}{rgb}{0.9,0.45,0}
    \def\rows{4} 
    \def\cols{4} 

    \foreach \x in {1,...,\cols} {
        \foreach \y in {1,...,\rows} {

            \fill[black] (\x,\y) circle (1.5pt);
        }
    }
    \node[above right] at (2,2) {$i,j$};
    \draw[red, line width=0.4mm] (3,2) circle (2.5pt);
    \draw[rect,dashed, line width=0.75mm] (3.5,3.5) rectangle (0.5,0.5); %
    
    \fill[mycyan] (4,3) circle (1.5pt);
    \node[text=mycyan, above] at (4,3) {$+$};
    \fill[mycyan] (2,3) circle (1.5pt);
    \node[text=mycyan, left] at (2,3) {$-$};  
    \fill[red] (3,4) circle (1.5pt);
    \node[text=red, above] at (3,4) {$+$};
    \fill[red] (3,2) circle (1.5pt);
    \node[text=red, below] at (3,2) {$-$};  
    \end{tikzpicture}    
\end{center}
Here, we simply represented the constraint at $(i+1,j+1)$ and multiplied the signs at the overlapping site $(i,j+1)$, yielding a minus sign. It now becomes evident that the only way to ensure that $\hat{f}$ commutes with $\hat{\mathcal{C}}^{i+1,j+1}$, while keeping it confined within the orange box, is to add a specific term at $(i+1,j)$. Based on the examples given in Graph\,\eqref{commutation_Ex}, the unique and natural choice is the one that leads to the following Graph:
\begin{center}
    \begin{tikzpicture}[scale=0.8]
    \definecolor{mycyan}{rgb}{0.00,0.50,0.50}
    \definecolor{rect}{rgb}{0.9,0.45,0}
    \def\rows{4} 
    \def\cols{4} 

    \foreach \x in {1,...,\cols} {
        \foreach \y in {1,...,\rows} {

            \fill[black] (\x,\y) circle (1.5pt);
        }
    }
    \node[above right] at (2,2) {$i,j$};
    \draw[mycyan, line width=0.4mm] (2,3) circle (2.5pt);
    \draw[rect,dashed, line width=0.75mm] (3.5,3.5) rectangle (0.5,0.5);  
    \draw[red, line width=0.4mm] (3,2) circle (2.5pt);    
    \fill[mycyan] (4,3) circle (1.5pt);
    \node[text=mycyan, above] at (4,3) {$+$};
    \fill[mycyan] (2,3) circle (1.5pt);
    \node[text=mycyan, left] at (2,3) {$+$};  
    \fill[red] (3,4) circle (1.5pt);
    \node[text=red, above] at (3,4) {$+$};
    \fill[red] (3,2) circle (1.5pt);
    \node[text=red, below] at (3,2) {$-$};  
    \end{tikzpicture}    
\end{center}
This means that we need to add precisely $-\hat{q}_x^{i+1,j}$, which updates the function to
\begin{equation}
    \hat{f} = \hat{q}_y^{i,j+1} - \hat{q}_x^{i+1,j} + \dots
\end{equation}
It is again clear from inspecting the Graph in Fig.\,\ref{starting_grid} that this updated $\hat{f}$ still commutes with $\hat{\mathcal{C}}^{i,j}$.

We can now proceed analogously around the point $(i,j)$, by first choosing the term at $(i,j-1)$ and then at $(i-1,j-1)$. Recalling that for overlapping sites we multiply the signs, we obtain the following Graphs:
\begin{center}
    \begin{tikzpicture}[scale=0.8]
    \definecolor{mycyan}{rgb}{0.00,0.50,0.50}
    \definecolor{rect}{rgb}{0.9,0.45,0}
    \def\rows{4}
    \def\cols{3}

    \foreach \x in {0,...,\cols} {
        \foreach \y in {1,...,\rows} {

            \fill[black] (\x,\y) circle (1.5pt);
        }
    }
    \node[left] at (-1,2.5) {$\rightarrow$};
    \node[above right] at (2,2) {$i,j$};
    \node[text=mycyan, above] at (8,4) {$-$};
    \draw[mycyan, line width=0.4mm] (2,3) circle (2.5pt);
    \draw[rect,dashed, line width=0.75mm] (3.5,3.5) rectangle (0.5,0.5);
    \draw[red, line width=0.4mm] (3,2) circle (2.5pt);
    \node[text=red, right] at (3,2) {$+$};  
    \draw[red, line width=0.4mm] (1,2) circle (2.5pt);   
    \fill[mycyan] (2,3) circle (1.5pt);
    \node[text=mycyan, right] at (2,3) {$-$};
    \fill[mycyan] (0,3) circle (1.5pt);
    \node[text=mycyan, left] at (0,3) {$-$};  
    \fill[red] (1,4) circle (1.5pt);
    \node[text=red, above] at (1,4) {$+$};
    \fill[red] (1,2) circle (1.5pt);
    \node[text=red, below] at (1,2) {$+$};  
    \node[below] at (1.5,0) {$-\hat{q}_y^{i,j-1}$};
    \node[right] at (4.5,2.5) {$\rightarrow$};

    \def\rows{4} 
    \def\cols{9} 

    \foreach \x in {6,...,\cols} {
        \foreach \y in {1,...,\rows} {

            \fill[black] (\x,\y) circle (1.5pt);
        }
    }
    \node[above right] at (8,3) {$i,j$};
    \draw[mycyan, line width=0.4mm] (8,4) circle (2.5pt);
    \draw[rect,dashed, line width=0.75mm] (9.5,4.5) rectangle (6.5,1.5);
    \draw[mycyan, line width=0.4mm] (8,2) circle (2.5pt);
    \draw[red, line width=0.4mm] (9,3) circle (2.5pt);
    \node[text=red, right] at (9,3) {$+$};  
    \draw[red, line width=0.4mm] (7,3) circle (2.5pt);   
    \fill[mycyan] (8,2) circle (1.5pt);
    \node[text=mycyan, right] at (8,2) {$+$};
    \fill[mycyan] (6,2) circle (1.5pt);
    \node[text=mycyan, left] at (6,2) {$-$};  
    \fill[red] (7,1) circle (1.5pt);
    \node[text=red, below] at (7,1) {$-$};
    \fill[red] (7,3) circle (1.5pt);
    \node[text=red, above] at (7,3) {$-$};  
    \node[below] at (7.5,0) {$+\hat{q}_x^{i-1,j}$};
    \end{tikzpicture}    
\end{center}
Thus we found that the final function is precisely
\begin{equation}
    f=\hat{q}_y^{i,j+1}-\hat{q}_y^{i,j-1}-\hat{q}_x^{i+1,j}+\hat{q}_y^{i-1,j}=\hat{b}^{i,j}.
\end{equation}
This concludes our proof. In fact, we can now observe that, under the requirements employed in this procedure, any alternative initial choice of sign and $\hat{q}$ component at a given site will uniquely determine the construction of one of the four possible $\hat{b}$ operators that include that site, as illustrated in Fig.\,\ref{b_constr}. 
\begin{figure}[ht]
    \centering
    \begin{tikzpicture}[scale=0.8]
    \definecolor{mycyan}{rgb}{0.00,0.50,0.50}
    \definecolor{rect}{rgb}{0.9,0.45,0}
    \def\rows{4} 
    \def\cols{4} 

    \foreach \x in {0,...,\cols} {
        \foreach \y in {0,...,\rows} {

            \fill[black] (\x,\y) circle (1.5pt);
        }
    }
    \draw[red, line width=0.4mm] (2,2) circle (2.5pt);
    \node[text=red, right] at (2,2) {$+$};
    \draw[red, line width=0.4mm] (0,2) circle (2.5pt);
    \node[text=red, left] at (0,2) {$-$};  
    \draw[mycyan, line width=0.4mm] (1,3) circle (2.5pt);
    \node[text=mycyan, above] at (1,3) {$-$};
    \draw[mycyan, line width=0.4mm] (1,1) circle (2.5pt);
    \node[text=mycyan, below] at (1,1) {$+$};  
    \draw[rect,dashed, line width=0.75mm] (2.5,3.5) rectangle (-0.5,0.5); %

    \draw[red, line width=0.4mm] (4,2) circle (2.5pt);
    \node[text=red, right] at (4,2) {$+$};
    \node[text=red, left] at (2,2) {$-$};  
    \draw[mycyan, line width=0.4mm] (3,3) circle (2.5pt);
    \node[text=mycyan, above] at (3,3) {$-$};
    \draw[mycyan, line width=0.4mm] (3,1) circle (2.5pt);
    \node[text=mycyan, below] at (3,1) {$+$};  
    \draw[rect,dashed, line width=0.75mm] (4.5,3.5) rectangle (1.5,0.5); %

    \def\rows{4}
    \def\cols{12} 
 
    \foreach \x in {8,...,\cols} {
        \foreach \y in {0,...,\rows} {

            \fill[black] (\x,\y) circle (1.5pt);
        }
    }

    \draw[red, line width=0.4mm] (11,3) circle (2.5pt);
    \node[text=red, right] at (11,3) {$+$};
    \draw[red, line width=0.4mm] (9,3) circle (2.5pt);
    \node[text=red, left] at (9,3) {$-$};  
    \draw[mycyan, line width=0.4mm] (10,4) circle (2.5pt);
    \node[text=mycyan, above] at (10,4) {$-$};
    \draw[mycyan, line width=0.4mm] (10,2) circle (2.5pt);
    \node[text=mycyan, below] at (10,2) {$+$};  
    \draw[rect,dashed, line width=0.75mm] (11.5,4.5) rectangle (8.5,1.5); %
    
    \draw[red, line width=0.4mm] (11,1) circle (2.5pt);
    \node[text=red, right] at (11,1) {$+$};
    \draw[red, line width=0.4mm] (9,1) circle (2.5pt);
    \node[text=red, left] at (9,1) {$-$};  
    \draw[mycyan, line width=0.4mm] (10,2) circle (2.5pt);
    \node[text=mycyan, above] at (10,2) {$-$};
    \draw[mycyan, line width=0.4mm] (10,0) circle (2.5pt);
    \node[text=mycyan, below] at (10,0) {$+$};  
    \draw[rect,dashed, line width=0.75mm] (11.5,2.5) rectangle (8.5,-0.5); %

    \end{tikzpicture}
    \caption{The four possible $\hat{b}$'s that are constructed around a center point. On the left, combinations coming from the two choices of signs for $\hat{q}_y$ component, while on the right the two for $\hat{q}_x$.}
    \label{b_constr}
\end{figure}
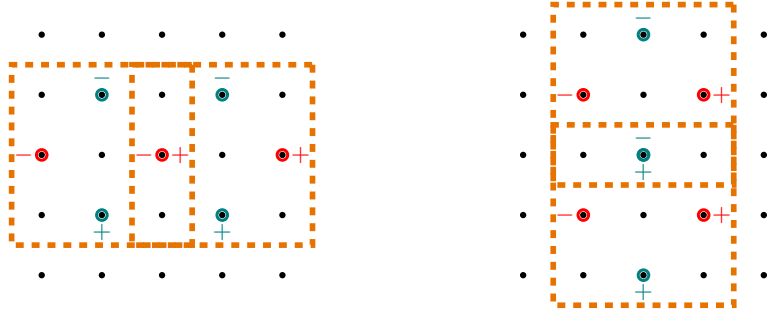

\section{\label{app:qmatter_model} Quantum model for matter}
As we already briefly introduced in Section \ref{sec:q_matter}, we choose to model the quantum matter content of our theory in the simplest possible way. Our choice is to associate to each site of the lattice a matter degree of freedom represented by a single qubit. Considering the computational basis at each site $\{\ket{0}^{i,j}_M, \ket{1}^{i,j}_M\}$, we interpret the state $\ket{1}_M^{i,j}$ as ``a particle of charge $1$ is present at site $(i,j)$'', and correspondingly $\ket{0}_M^{i,j}$ as ``no particle is present at site $(i,j)$''. 

We promote the matter densities $\rho^{i,j}$ to operators acting on the matter Hilbert space $\mathcal{H}^M$, 
\begin{equation}
    \hat{\rho}^{i,j} = \ket{1}\!\bra{1}^{i,j}_M.
\end{equation}
We can also define the total matter operator as
\begin{equation}
    \hat{\rho} = \sum_{i,j} \hat{\rho}^{i,j},
\end{equation}
which, with our choice of unit charge, is equivalent to the total particle number operator. 

Similarly to the case of the pure field (see Eq.\,\eqref{kin}), the kinematical matter Hilbert space can be written as a tensor product of the qubit Hilbert spaces at each lattice site,
\begin{equation}\label{matter_kin_hilb}
    \mathcal{H}^M = \bigotimes_{i,j} \mathcal{H}_{i,j}^q,
\end{equation}
where each local qubit Hilbert space is $\mathcal{H}_{i,j}^q = \mathbb{C}^2$.

We now choose to impose a constraint on the matter Hilbert space: conservation of total charge, i.e., the total number of particles. This condition follows from the continuity equation in Tab.\,\ref{tab:eom} for a static source. Consequently, we restrict physically accessible matter operations to those that commute with the total matter operator $\hat{\rho}$, i.e., operators conserving the total particle number across the lattice.

The matter Hilbert space can then be decomposed as a direct sum of subspaces with fixed particle number,
\begin{equation}
    \mathcal{H}^M = \bigoplus_{n=0}^{N^2} \mathcal{H}^M_n,
\end{equation}
where $\mathcal{H}^M_n$ contains all states with exactly $n$ particles. These subspaces define so called superselection sectors labeled by $n$, since accessible operators cannot induce transitions between them. This structure is analogous to a Fock space, as expected for this matter model (see, e.g., Ref.\,\cite{RevModPhys.79.555} for a similar discussion in the context of quantum reference frames).

We can further investigate $\mathcal{H}^M_n$, which consists of all coherent superpositions of matter configurations with total particle number $n$. Consider a matter state $ \ket{s_n}_M$ written in the computational basis as
\begin{equation}\label{seq_state_def}
     \ket{s_n}_M = \bigotimes_{i,j} \ket{c_{i,j}}^{i,j}_M, \qquad \text{such that} \qquad
    \begin{cases}
        c_{i,j} \in \{0,1\} \\
        \sum_{i,j} c_{i,j} = n
    \end{cases}.
\end{equation}
Here, $s_n$ labels a configuration consisting of a sequence of $0$'s and $1$'s representing one of the possible arrangements of $n$ particles over the $N \times N$ lattice. The Hilbert space $\mathcal{H}^M_n$, expressed in the computational basis, is the linear span of all $\binom{N^2}{n}$ possible configurations of $n$ particles on the lattice,
\begin{equation}
    \mathcal{H}^M_n =  \underset{s_n}{\text{span}} \left\{  \ket{s_n}_M \right\} = \bigoplus_{s_n}  \ket{s_n}_M,
\end{equation}
where the last equality holds because the states $ \ket{s_n}_M$ form an orthonormal basis. 
Note, however, that in this context the terms in the direct sum do not represent proper superselection sectors; 
rather, the notation is meant to indicate that all superpositions of the basis vectors $ \ket{s_n}_M$ are allowed.
As a result, the Hilbert space of matter fully decomposes as 
\begin{equation}\label{matt_dec_app}
    \mathcal{H} = \bigoplus_{n=0}^{N^2} \bigoplus_{s_n}  \ket{s_n}_M.
\end{equation}

\section{\label{app:g_state}Computing the ground state}
In this appendix we present the full derivation of the ground state of the toy model, first in the sourceless case and then in the presence of a static classical source configuration. 

\subsection{\label{app:sec:DFT}The discrete Fourier transform} Before getting into the actual computation of the ground state, it is useful to review the concept of discrete Fourier transform, which will be fundamental in the derivation. Usually, in the standard treatment of quantum field theory, there is a frequent use of the Fourier transform (see e.g. Ref.\,\cite{QFT-fradkin}). So, in order to conduct a similar procedure, we would need its discrete equivalent to apply on our lattice gauge model. Let us start by considering a one dimensional discrete function $f^j$ on an $N$-sites chain, and define the direct and inverse discrete Fourier transform (DFT) respectively as follows:
\begin{equation}
    \text{Direct}:\qquad \tilde{f}^{\beta}=\text{DFT}(f^j)=C\sum_{j=0}^{N-1}f^j\text{e}^{-\text{i}\frac{2\pi}{N}j\,\beta},
\end{equation}
\begin{equation}
    \text{Inverse}:\qquad f^{j}=\text{DFT}^{-1}(\tilde{f}^\beta)=\tilde{C}\sum_{\beta=0}^{N-1}\tilde{f}^{\beta}\text{e}^{\text{i}\frac{2\pi}{N}j\,\beta},
\end{equation}
where the two normalization constants can be chosen freely given that $\tilde{C}\cdot C=1/N$. The canonical choice for this constants is to fix $C=1$ and $\tilde{C}=1/N$ or vice versa (like it is done in Ref.\,\cite{dft_guide}). Another possibility is to fix them both to $1/\sqrt{N}$ and have a symmetric convention. In this case we make the choice $C=1$ and $\tilde{C}=1/N$, which, as will be shown in Section~\ref{app:cont_limit}, is convenient to perform the continuum limit.

An extremely useful property of the Fourier transform involving the derivative $f'$ is
\begin{equation}
    \text{FT}(f')=\text{i}k\tilde{f}(k).
\end{equation}
We can find the analogous property for the discrete Fourier transform and discrete derivative:
\begin{equation}
    \text{DFT}(\bar{\partial}f^j)=\frac{\text{i}}{a}\sin\bigg(\frac{2\pi}{N}\beta\bigg)\tilde{f}^{\beta}.
\end{equation}
The convention is then that Greek letters, like $\beta$, are indices labeling sites in the Fourier space, while Latin ones continue to label sites in position space.

Let us prove this by simply applying the discrete Fourier transform and recalling the definition of discrete derivative:
\begin{equation}
    \text{DFT}(\bar{\partial}f^{i,j})=\sum_{j=0}^{N-1}\bar{\partial}f^j\text{e}^{-\text{i}\frac{2\pi}{N}j\alpha}=\frac{1}{2a}\bigg(\sum_{j=0}^{N-1}f^{j+1}\text{e}^{-\text{i}\frac{2\pi}{N}j\alpha}-\sum_{j=0}^{N-1}f^{j-1}\text{e}^{-\text{i}\frac{2\pi}{N}j\alpha}\bigg).
\end{equation}
Now, by enforcing the periodic boundary conditions ($j=N\rightarrow J=0$), we can freely shift the sums and get
\begin{equation}
    \text{DFT}(\bar{\partial}f^{i,j})=\frac{1}{2a}\bigg(\sum_{j=-1}^{N-2}f^{j+1}\text{e}^{-\text{i}\frac{2\pi}{N}j\alpha}-\sum_{j=1}^{N}f^{j-1}\text{e}^{-\text{i}\frac{2\pi}{N}j\alpha}\bigg).
\end{equation}
We can also redefine the summing indices as $j_1=j+1$ and $j_2=j-1$, obtaining
\begin{equation}\label{dev_on_exp}
    \text{DFT}(\bar{\partial}f^{i,j})=\frac{1}{2a}\bigg(\sum_{j_1=0}^{N-1}f^{j_1}\text{e}^{-\text{i}\frac{2\pi}{N}(j_1-1)\alpha}-\sum_{j_2=0}^{N-1}f^{j_2}\text{e}^{-\text{i}\frac{2\pi}{N}(j_2+1)\alpha}\bigg) 
    =-\sum_{j}f^j\bar{\partial}_s(\text{e}^{-\text{i}\frac{2\pi}{N}j\alpha}),
\end{equation}
where we simply redefined the indices again.
Finally we notice that we have precisely retained the discrete Fourier transform of $f^j$ in both terms, i.e.
\begin{equation}
       \text{DFT}(\bar{\partial}f^{i,j})=\frac{1}{2a}\bigg(\text{e}^{\text{i}\frac{2\pi}{N}\alpha}\sum_{j_1=0}^{N-1}f^{j_1}\text{e}^{-\text{i}\frac{2\pi}{N}j_1\alpha}-\text{e}^{-\text{i}\frac{2\pi}{N}\alpha}\sum_{j_2=0}^{N-1}f^{j_2}\text{e}^{-\text{i}\frac{2\pi}{N}j_2\alpha}\bigg)=\frac{\text{e}^{\text{i}\frac{2\pi}{N}\alpha}-\text{e}^{-\text{i}\frac{2\pi}{N}\alpha}}{2a}\tilde{f}^{\alpha}. 
\end{equation}
We complete the proof by using the expression for the sine in terms of complex exponential, getting
\begin{equation}
     \text{DFT}(\bar{\partial}f^{i,j})=\frac{\text{i}}{a}\sin\bigg(\frac{2\pi}{N}\alpha\bigg)\tilde{f}^{\alpha}.
\end{equation}

We can now apply this results to our toy model. As already mentioned, we will consider the lattice to have $N$ sites per side and periodic boundary conditions. 
For a generic function $f^{i,j}$ on the lattice we have that
\begin{equation}\label{DFT_lattice}
        \tilde{f}^{\alpha,\beta}=\text{DFT}(f^{i,j})=\sum_{j=0}^{N-1}\sum_{i=0}^{N-1}f^{i,j}\text{e}^{-\text{i}\frac{2\pi}{N}(i\alpha+j\beta)},\qquad  
        f^{i,j}=\text{DFT}^{-1}(\tilde{f}^{\alpha,\beta})=\frac{1}{N^2}\sum_{\alpha=0}^{N-1}\sum_{\beta=0}^{N-1}\tilde{f}^{\alpha,\beta}\text{e}^{\text{i}\frac{2\pi}{N}(i\alpha+j\beta)}.
\end{equation}
We will often still just write $\sum_{i,j}$ and $\sum_{\alpha,\beta}$, for simplicity when specification of the limits is not needed.

We have an analogue for the derivative property:
\begin{equation}
        \text{DFT}(\bar{\partial}_xf^{i,j})=\frac{\text{i}}{a}\sin\bigg(\frac{2\pi}{N}\beta\bigg)\tilde{f}^{\alpha,\beta}, \qquad
        \text{DFT}(\bar{\partial}_yf^{i,j})=\frac{\text{i}}{a}\sin\bigg(\frac{2\pi}{N}\alpha\bigg)\tilde{f}^{\alpha,\beta}.
\end{equation}
  By defining
\begin{equation}\label{k_def}
    \mathbf{\bar{k}}:={\mathbf{\bar{k}}}(
    \alpha,\beta)=\frac{1}{a}\begin{pmatrix} \sin\big(\frac{2\pi}{N}\beta\big)\\ \sin\big(\frac{2\pi}{N}\alpha\big) \end{pmatrix},
\end{equation}
we can recast this property in a more compact and familiar way as
\begin{equation}\label{DFT_diff}
    \text{DFT}(\bar{\partial}_s f^{i,j})=\text{i}\bar{k}_s\tilde{f}^{\alpha,\beta}\,\,\,\, \forall s\in\{x,y\},
\end{equation}
where $\bar{k}_s$ represents one of the two components of $\mathbf{\bar{k}}$ in Eq.\,\eqref{k_def}. The choice for this name foreshadows its meaning once the continuum limit is performed (see Section~\ref{app:cont_limit}). 

Let us finally make use of the relation for the discrete Fourier transform of the Kronecker delta \cite{dft_guide} to write the following useful property
\begin{equation}\label{discrete_delta}
    \delta_{\alpha,\gamma}=\frac{1}{N}\sum_{j}\text{e}^{\text{i}\frac{2\pi}{N}(\gamma-\alpha)j}.
\end{equation}
\subsection{The Gauss-like constraint in Fourier space.} Now that we have defined our discrete equivalent of the Fourier transform, we can turn back to the problem of finding the ground state of the sourceless case. Thanks to the analogy with the electromagnetic case, we can largely follow the procedure outlined in Ref.\,\cite{QFT-fradkin} for the temporal gauge. We start by studying the Gauss-like constraint of the case with no source in Fourier space. 

Let us begin with rewriting the constraint in the $q$ field basis using Eq.\,\eqref{p_q_rep}, namely 
\begin{equation}\label{constr_pos_q}
  \hat{\mathcal{C}}^{i,j}\ket{\Psi}=\bar{\partial}_s\hat{p}_s^{i,j}\ket{\Psi}=0\rightarrow -\text{i}\,\hbar\,\bar{\partial}_s \frac{\partial}{\partial q_s^{i,j}} \Psi[q]=0,
\end{equation}
where we always sum the repeated index $s$ and we necessarily have that $\ket{\Psi}\in\mathcal{H}^{\text{phy}}$. Here, $\Psi[q]$ is a shorthand for the wave functional $\Psi(\{q_s^{i,j}\})$ coming from the field basis representation of the state
\begin{equation}
    \ket{\Psi}=\int\mathcal{D}q\,\,\Psi[q]\bigg(\prod_{i,j}\ket{q_x^{i,j}}\ket{q_y^{i,j}}\bigg).
\end{equation}
Let us now transition to the discrete Fourier space. Applying the direct transformation in Eq.\,\eqref{DFT_lattice} to $q$ gives
\begin{equation}
    \tilde{q}_s^{\alpha,\beta}=\text{DFT}(q_s^{i,j})=\sum_{i,j}q_s^{i,j}\text{e}^{-\text{i}\frac{2\pi}{N}(i\alpha+j\beta)}.
\end{equation}
We also compute the momenta in the $q$ basis by applying the same transformation to Eq.\,\eqref{p_q_rep}, i.e.
\begin{equation}
    \tilde{p}_s^{\alpha,\beta}=\text{DFT}(p_s^{i,j})=-\text{i}\,\hbar\, \text{DFT}\bigg(\frac{\partial}{\partial q_s^{i,j}}\bigg)=-\text{i}\hbar\sum_{i,j}\text{e}^{-\text{i}\frac{2\pi}{N}(i\alpha+j\beta)}\frac{\partial}{\partial q_s^{i,j}}.
\end{equation}
Now we notice that 
\begin{equation}
    \frac{\partial q^{i,j}_s}{\partial \tilde{q}_s^{\alpha,\beta}}=\frac{\partial}{\partial \tilde{q}_s^{\alpha,\beta}}\bigg(\frac{1}{N^2}\sum_{\gamma,\theta}\text{e}^{\text{i}\frac{2\pi}{N}(i\gamma+j\theta)}\tilde{q}_s^{\gamma,\theta}\bigg)=\frac{1}{N^2}\text{e}^{\text{i}\frac{2\pi}{N}(i\alpha+j\beta)},
\end{equation}
where we expressed $q^{i,j}_s$ using the inverse transform in Eq.\,\eqref{DFT_lattice}. Given this result we can rewrite the exponential in the previous equation, obtaining
\begin{equation}
   \tilde{p}_s^{\alpha,\beta}=-\text{i}\hbar\sum_{i,j}\text{e}^{-\text{i}\frac{2\pi}{N}(i\alpha+j\beta)}\frac{\partial}{\partial q_s^{i,j}}=-\text{i}\hbar N^2\sum_{i,j}\frac{\partial q^{i,j}_s}{\partial \tilde{q}_s^{-\alpha,-\beta}}\frac{\partial}{\partial q_s^{i,j}}.
\end{equation}
Recognizing the usual chain rule relation for partial derivatives, we reduce this to
\begin{equation}\label{p_four_q_bas}
     \tilde{p}_s^{\alpha,\beta}=-\text{i}\,\hbar\, \text{DFT}\bigg(\frac{\partial}{\partial q_s^{i,j}}\bigg)=-\text{i}\hbar N^2\frac{\partial}{\partial \tilde{q}_s^{-\alpha,-\beta}}.
\end{equation}
Let us now recast the constraint in Eq.\,\eqref{constr_pos_q} in Fourier space.
The differentiation property in Eq.\,\eqref{DFT_diff} and the previous result in Eq.\,\eqref{p_four_q_bas} lead to the expression
\begin{equation}
    \text{DFT}(\bar{\partial}_sp_s^{i,j})=\text{i}\bar{k}_s\tilde{p}_s^{\alpha,\beta}=\hbar N^2\bar{k}_s\frac{\partial}{\partial \tilde{q}_s^{-\alpha,-\beta}}.
\end{equation}
Therefore, by discarding all multiplicative constants, the constraint becomes
\begin{equation}
    \bar{k}_s \frac{\partial}{\partial \tilde{q}^{-\alpha,-\beta}_s}\Psi[q]=0.
\end{equation}
Since this has to be valid for every choice of $\alpha$ and $\beta$ we can safely get rid of the minus signs and write
\begin{equation}
    \bar{k}_s \frac{\partial}{\partial \tilde{q}^{\alpha,\beta}_s}\Psi[q]=0.
\end{equation}
We can further re-express it by decomposing the transformed field in the longitudinal and transverse parts as
\begin{equation}\label{T_L_dec}
    \tilde{q}_s^{\alpha,\beta}=\frac{\bar{k}_s}{|\mathbf{\bar{k}}|}\tilde{q}^{\alpha,\beta}_L+\frac{t_s}{|\mathbf{t}|}\tilde{q}^{\alpha,\beta}_T\,\,\,\text{with}\,\,\,\mathbf{\bar{k}}\cdot\mathbf{t}=0.
\end{equation}
We can perform a similar decomposition for the derivatives too. Using that $\bar{k}_st_s=0$ we find that
\begin{equation}\label{inv_T_L_dec}
   \tilde{q}^{\alpha,\beta}_L=\frac{\bar{k}_s}{|\mathbf{\bar{k}}|} \tilde{q}_s^{\alpha,\beta}\,\,\,\text{and}\,\,\,\tilde{q}^{\alpha,\beta}_T=\frac{t_s}{|\mathbf{t}|} \tilde{q}_s^{\alpha,\beta},
\end{equation}
and applying the chain rule we find
\begin{equation}\label{T_L_dev_dec}
    \frac{\partial}{\partial\tilde{q}_s^{\alpha,\beta}}=\frac{\partial\tilde{q}^{\alpha,\beta}_L}{\partial\tilde{q}_s^{\alpha,\beta}}\frac{\partial}{\partial\tilde{q}^{\alpha,\beta}_L}+\frac{\partial\tilde{q}^{\alpha,\beta}_T}{\partial\tilde{q}_s^{\alpha,\beta}}\frac{\partial}{\partial\tilde{q}^{\alpha,\beta}_T}=\frac{\bar{k}_s}{|\mathbf{\bar{k}}|}\frac{\partial}{\partial\tilde{q}^{\alpha,\beta}_L}+\frac{t_s}{|\mathbf{t}|}\frac{\partial}{\partial\tilde{q}^{\alpha,\beta}_T}.
\end{equation}
So the constraint becomes
\begin{equation}\label{constr_four}
    |\mathbf{\bar{k}}|\frac{\partial}{\partial\tilde{q}^{\alpha,\beta}_L}\Psi[q]=0.
\end{equation}
Therefore, we ultimately found that the Gauss-like constraint forces the wave function to be independent of the longitudinal component of the field, i.e.
\begin{equation}
    \Psi[q]=\Psi[q_{T}],
\end{equation}
precisely as is the case in electromagnetism.
\subsection{The Hamiltonian in Fourier space}
To find the ground state we need to solve the Schr{\"o}dinger equation
\begin{equation}\label{shc_eq}
    \hat{H}\ket{\smash{\Psi_0}}=\mathcal{E}_0\ket{\smash{\Psi_0}}.
\end{equation}
To write it in the Schr{\"o}dinger picture we need to rewrite the sourceless Hamiltonian in the field basis. Recalling again the expression for the momenta in the field basis (Eq.\,\eqref{p_q_rep}) and the Hamiltonian of the model (Eq.\,\eqref{ham_temp_gauge}), setting $J_s^{i,j} = 0$ since we are in the vacuum, we write 
\begin{equation}\label{ham_q_basis}
    H=\frac{1}{2}\sum_{i,j}\bigg[-\hbar ^2\frac{\partial}{\partial q^{i,j}_s}\frac{\partial}{\partial q^{i,j}_s}+(b^{i,j})^2\bigg].
\end{equation}
Now we follow the standard procedure and rewrite this expression in Fourier space. 

We want to rewrite both terms in the sum by using the relation for the inverse DFT in Eq.\,\eqref{DFT_lattice}. For the first one we need to start from the expression of the transformed momentum in $q$ basis in Eq.\,\eqref{p_four_q_bas}. Inverting the DFT we get that
\begin{equation}
    \frac{\partial}{\partial q^{i,j}_s}=N^2\text{DFT}^{-1}\bigg(\frac{\partial}{\partial \tilde{q}_s^{-\alpha,-\beta}}\bigg)=\frac{N^2}{N^2}\sum_{\alpha,\beta}\text{e}^{\text{i}\frac{2\pi}{N}(i\alpha+j\beta)}\frac{\partial}{\partial \tilde{q}_s^{-\alpha,-\beta}}=\sum_{\alpha,\beta}\text{e}^{-\text{i}\frac{2\pi}{N}(i\alpha+j\beta)}\frac{\partial}{\partial \tilde{q}_s^{\alpha,\beta}},  
\end{equation}
where at the end we simply renamed the indices.
So the first terms can be written as
\begin{equation}
    \frac{\partial}{\partial q^{i,j}_s}\frac{\partial}{\partial q^{i,j}_s}=\sum_{\alpha,\beta}\sum_{\gamma,\theta}\text{e}^{-\text{i}\frac{2\pi}{N}(i\alpha+j\beta)}\text{e}^{-\text{i}\frac{2\pi}{N}(i\gamma+j\theta)}\frac{\partial}{\partial \tilde{q}^{\alpha,\beta}_s}\frac{\partial}{\partial \tilde{q}^{\gamma,\theta}_s}.
\end{equation}
Rearranging the exponential and using the decomposition in Eq.\,\eqref{T_L_dev_dec} we get
\begin{equation}
\begin{split}
       \frac{\partial}{\partial q^{i,j}_s}\frac{\partial}{\partial q^{i,j}_s}=&\sum_{\alpha,\beta}\sum_{\gamma,\theta}\text{e}^{-\text{i}\frac{2\pi}{N}(\alpha+\gamma)i}\text{e}^{-\text{i}\frac{2\pi}{N}(\beta+\theta)j}\bigg(\frac{\bar{k}_s\bar{l}_s}{|\mathbf{\bar{k}}||\mathbf{\bar{l}}|}\frac{\partial}{\partial \tilde{q}^{\alpha,\beta}_L}\frac{\partial}{\partial \tilde{q}^{\gamma,\theta}_L}+\frac{\bar{k}_su_s}{|\mathbf{\bar{k}}||\mathbf{u}|}\frac{\partial}{\partial \tilde{q}^{\alpha,\beta}_L}\frac{\partial}{\partial \tilde{q}^{\gamma,\theta}_T}\\
       &+\frac{\bar{l}_st_s}{|\mathbf{\bar{l}}||\mathbf{t}|}\frac{\partial}{\partial \tilde{q}^{\alpha,\beta}_T}\frac{\partial}{\partial \tilde{q}^{\gamma,\theta}_L}+\frac{t_su_s}{|\mathbf{t}||\mathbf{u}|}\frac{\partial}{\partial \tilde{q}^{\alpha,\beta}_T}\frac{\partial}{\partial \tilde{q}^{\gamma,\theta}_T}\bigg).
\end{split}
\end{equation}
Here, $\mathbf{\bar{l}}$ is the wave vector in Eq.\,\eqref{k_def}, associated to the Fourier variables $\gamma$ and $\theta$:
\begin{equation}\label{h_def}
    {\mathbf{\bar{l}}}:={\mathbf{\bar{l}}}(
    \gamma,\theta)=\frac{1}{a}\begin{pmatrix} \sin\big(\frac{2\pi}{N}\theta\big)\\ \sin\big(\frac{2\pi}{N}\gamma\big) \end{pmatrix}.
\end{equation}
Furthermore, $\mathbf{u}$ is the corresponding orthogonal vector, such that
$\mathbf{\bar{l}}\cdot\mathbf{u}=0$.

Let us do the same for the ``magnetic field'' $b^{i,j}$, which means
\begin{equation}
    b^{i,j}=\bar{\partial}_xq^{i,j}_y-\bar{\partial}_yq^{i,j}_x=\frac{1}{N^2}\sum_{\alpha,\beta}\text{i}(\bar{k}_x\tilde{q}^{\alpha,\beta}_y-\bar{k}_y\tilde{q}^{\alpha,\beta}_x)\text{e}^{\text{i}\frac{2\pi}{N}(i\alpha+j\beta)},
\end{equation}
where we applied the inverse transform to the differentiation property of DFT (Eq.\,\eqref{DFT_diff}) and used that $\text{DFT}^{-1}$ is the inverse of $\text{DF}$, i.e.
\begin{equation}
    \bar{\partial}_sf^{i,j}=\text{DFT}^{-1}(\text{DFT}(\bar{\partial}_sf^{i,j}))=\frac{1}{N^2}\sum_{\alpha,\beta}(\text{i}\bar{k}_s\tilde{f}^{\alpha,\beta}\text{e}^{\text{i}\frac{2\pi}{N}(i\alpha+j\beta)}).
\end{equation}
Again considering the decomposition in Eq.\,\eqref{T_L_dec}, we find that
\begin{equation}
    b^{i,j}=\frac{1}{N^2}\sum_{\alpha,\beta}\text{i}\bigg(\frac{\bar{k}_x\bar{k}_y}{|\mathbf{\bar{k}}|}\tilde{q}^{\alpha,\beta}_L+\frac{\bar{k}_xt_y}{|\mathbf{t}|}\tilde{q}^{\alpha,\beta}_T-\frac{\bar{k}_y\bar{k}_x}{|\mathbf{\bar{k}}|}\tilde{q}^{\alpha,\beta}_L-\frac{\bar{k}_yt_x}{|\mathbf{t}|}\tilde{q}^{\alpha,\beta}_T\bigg)\text{e}^{\text{i}\frac{2\pi}{N}(i\alpha+j\beta)}.
\end{equation}
Evidently, the first and third terms cancel, ultimately giving
\begin{equation}
  b^{i,j}=  \frac{1}{N^2}\sum_{\alpha,\beta}\frac{\text{i}}{|\mathbf{t}|}(\bar{k}_xt_y-\bar{k}_yt_x)\tilde{q}^{\alpha,\beta}_T\text{e}^{\text{i}\frac{2\pi}{N}(i\alpha+j\beta)}.
\end{equation}
So, the second term in the Hamiltonian becomes
\begin{equation}
    (b^{i,j})^2=-\frac{1}{N^4}\sum_{\alpha,\beta}\sum_{\gamma,\theta}\frac{1}{|\mathbf{t}||\mathbf{u}|}(\bar{k}_xt_y-\bar{k}_yt_x)\cdot(\bar{l}_xu_y-\bar{l}_yu_x)
    \cdot\tilde{q}^{\alpha,\beta}_T\tilde{q}^{\gamma,\theta}_T\text{e}^{\text{i}\frac{2\pi}{N}(\alpha+\gamma)i}\text{e}^{\text{i}\frac{2\pi}{N}(\beta+\theta)j},
\end{equation}
where we rearranged the exponential terms as before. 

We can finally insert these two results in the expression for the Hamiltonian:
\begin{equation}
\begin{split}
    H=&\frac{1}{2}\sum_{i,j}\sum_{\alpha,\beta}\sum_{\gamma,\theta}\bigg[-\hbar ^2\,\text{e}^{-\text{i}\frac{2\pi}{N}(\alpha+\gamma)i}\,\text{e}^{-\text{i}\frac{2\pi}{N}(\beta+\theta)j}\bigg(\frac{\bar{k}_s\bar{l}_s}{|\mathbf{\bar{k}}||\mathbf{\bar{l}}|}\frac{\partial}{\partial \tilde{q}^{\alpha,\beta}_L}\frac{\partial}{\partial \tilde{q}^{\gamma,\theta}_L}
       +\frac{\bar{k}_su_s}{|\mathbf{\bar{k}}||\mathbf{u}|}\frac{\partial}{\partial \tilde{q}^{\alpha,\beta}_L}\frac{\partial}{\partial \tilde{q}^{\gamma,\theta}_T}+\frac{\bar{l}_st_s}{|\mathbf{\bar{l}}||\mathbf{t}|}\frac{\partial}{\partial \tilde{q}^{\alpha,\beta}_T}\frac{\partial}{\partial \tilde{q}^{\gamma,\theta}_L}\\
       &+\frac{t_su_s}{|\mathbf{t}||\mathbf{u}|}\frac{\partial}{\partial \tilde{q}^{\alpha,\beta}_T}\frac{\partial}{\partial \tilde{q}^{\gamma,\theta}_T}\bigg)
    -\frac{\text{e}^{\text{i}\frac{2\pi}{N}(\alpha+\gamma)i}\text{e}^{\text{i}\frac{2\pi}{N}(\beta+\theta)j}}{N^4|\mathbf{t}||\mathbf{u}|}(\bar{k}_xt_y-\bar{k}_yt_x)\cdot(\bar{l}_xu_y-\bar{l}_yu_x)\tilde{q}^{\alpha,\beta}_T\tilde{q}^{\gamma,\theta}_T\bigg].
\end{split}
\end{equation}
Let us now make use of the property of the Kronecker delta in Eq.\,\eqref{discrete_delta} to rewrite the Hamiltonian as
\begin{equation}
\begin{split}
    H=&\frac{1}{2}N^2\hbar ^2\sum_{\alpha,\beta}\sum_{\gamma,\theta}\delta_{\alpha,-\gamma}\delta_{\beta,-\theta}\bigg[-\frac{\bar{k}_s\bar{l}_s}{|\mathbf{\bar{k}}||\mathbf{\bar{l}}|}\frac{\partial}{\partial \tilde{q}^{\alpha,\beta}_L}\frac{\partial}{\partial \tilde{q}^{\gamma,\theta}_L}-\frac{\bar{k}_su_s}{|\mathbf{\bar{k}}||\mathbf{u}|}\frac{\partial}{\partial \tilde{q}^{\alpha,\beta}_L}\frac{\partial}{\partial \tilde{q}^{\gamma,\theta}_T}+\frac{\bar{l}_st_s}{|\mathbf{\bar{l}}||\mathbf{t}|}\frac{\partial}{\partial \tilde{q}^{\alpha,\beta}_T}\frac{\partial}{\partial \tilde{q}^{\gamma,\theta}_L}\\
       &-\frac{t_su_s}{|\mathbf{t}||\mathbf{u}|}\frac{\partial}{\partial \tilde{q}^{\alpha,\beta}_T}\frac{\partial}{\partial \tilde{q}^{\gamma,\theta}_T}
    -\frac{1}{N^4\hbar^2}\frac{1}{|\mathbf{t}||\mathbf{u}|}(\bar{k}_xt_y-\bar{k}_yt_x)\cdot(\bar{l}_xu_y-\bar{l}_yu_x)\tilde{q}^{\alpha,\beta}_T\tilde{q}^{\gamma,\theta}_T\bigg].
\end{split}
\end{equation}
By applying the deltas we can get rid of two of the four sums:
\begin{equation}\label{ugly_H}
\begin{split}
    H=&\frac{1}{2}N^2\hbar^2\sum_{\alpha,\beta}\bigg[-\frac{\bar{k}_s\bar{l}_s}{|\mathbf{\bar{k}}||\mathbf{\bar{l}}|}\frac{\partial}{\partial \tilde{q}^{\alpha,\beta}_L}\frac{\partial}{\partial \tilde{q}^{-\alpha,-\beta}_L}-\frac{\bar{k}_su_s}{|\mathbf{\bar{k}}||\mathbf{u}|}\frac{\partial}{\partial \tilde{q}^{\alpha,\beta}_L}\frac{\partial}{\partial \tilde{q}^{-\alpha,-\beta}_T}-\frac{\bar{l}_st_s}{|\mathbf{\bar{l}}||\mathbf{t}|}\frac{\partial}{\partial \tilde{q}^{\alpha,\beta}_T}\frac{\partial}{\partial \tilde{q}^{-\alpha,-\beta}_L}\\
       &-\frac{t_su_s}{|\mathbf{t}||\mathbf{u}|}\frac{\partial}{\partial \tilde{q}^{\alpha,\beta}_T}\frac{\partial}{\partial \tilde{q}^{-\alpha,-\beta}_T}
    -\frac{1}{N^4\hbar^2}\frac{1}{|\mathbf{t}||\mathbf{u}|}(\bar{k}_xt_y-\bar{k}_yt_x)\cdot(\bar{l}_xu_y-\bar{l}_yu_x)\tilde{q}^{\alpha,\beta}_T\tilde{q}^{-\alpha,-\beta}_T\bigg].
\end{split}
\end{equation}
Since the sine is an odd function, we notice that for $\gamma=-\alpha$ and $\theta=-\beta$, $\mathbf{\bar{l}}$ has the same direction of $\mathbf{\bar{k}}$, but opposite sign. In fact, the expression in Eq.\,\eqref{h_def} yields
\begin{equation}\label{k_opposite}
    {\mathbf{\bar{l}}}:={\mathbf{\bar{l}}}(
    -\alpha,-\beta)=-\frac{1}{a}\begin{pmatrix} \sin\big(\frac{2\pi}{N}\beta\big)\\ \sin\big(\frac{2\pi}{N}\alpha\big) \end{pmatrix}=-{\mathbf{\bar{k}}}.
\end{equation}
So, that for the corresponding orthogonal vector,
\begin{equation}
    \bar{k}_su_s=-\bar{l}_su_s=0.
\end{equation}
Therefore, without loss of generality, we can take $\mathbf{t}=\mathbf{u}$.
Using this, the second and third term in the Hamiltonian in Eq.\,\eqref{ugly_H} evidently cancel, reducing the first four terms in the square brackets to just
\begin{equation}
    \circ:=\frac{\partial}{\partial \tilde{q}^{\alpha,\beta}_L}\frac{\partial}{\partial \tilde{q}^{-\alpha,-\beta}_L}-\frac{\partial}{\partial \tilde{q}^{\alpha,\beta}_T}\frac{\partial}{\partial \tilde{q}^{-\alpha,-\beta}_T}.
\end{equation}
For the last term in Eq.\,\eqref{ugly_H}, we perform all multiplications and obtain
\begin{equation}
    \bullet:=-\frac{1}{N^4\hbar^2}\frac{1}{|\mathbf{t}||\mathbf{u}|}(-\bar{k}^2_xt_yu_y-\bar{k}^2_yt_xu_x+\bar{k}_x\bar{k}_yt_yu_x+\bar{k}_y\bar{k}_xt_xu_y)\tilde{q}^{\alpha,\beta}_T\tilde{q}^{-\alpha,-\beta}_T.
\end{equation}
From the relation between $\mathbf{\bar{k}}$ and $\mathbf{t}$ we find that
\begin{equation}
    \bar{k}_xt_x+\bar{k}_yt_y=0\rightarrow \bar{k}_xt_x=-\bar{k}_yt_y.
\end{equation}
Plugging this in the two final terms we obtain
\begin{equation}
    \bullet=\frac{1}{N^4\hbar^2}\frac{1}{|\mathbf{t}||\mathbf{u}|}(\bar{k}^2_xt_yu_y+\bar{k}^2_yt_xu_x+\bar{k}_x^2t_xu_x+\bar{k}^2_yt_yu_y)\tilde{q}^{\alpha,\beta}_T\tilde{q}^{-\alpha,-\beta}_T=\frac{1}{N^4\hbar^2}\frac{|\mathbf{\bar{k}}|^2}{|\mathbf{t}||\mathbf{u}|}t_su_s\tilde{q}^{\alpha,\beta}_T\tilde{q}^{-\alpha,-\beta}_T.
\end{equation}
If we also here consider the simplification $\mathbf{t}=\mathbf{u}$,
\begin{equation}
\bullet=\frac{1}{N^4\hbar^2}|\mathbf{\bar{k}}|^2\tilde{q}^{\alpha,\beta}_T\tilde{q}^{-\alpha,-\beta}_T
\end{equation}
The Hamiltonian is then simply given by
\begin{equation}
    H=\frac{N^2\hbar^2}{2}\sum_{\alpha,\beta}\bigg(\frac{\partial}{\partial \tilde{q}^{\alpha,\beta}_L}\frac{\partial}{\partial \tilde{q}^{-\alpha,-\beta}_L}-\frac{\partial}{\partial \tilde{q}^{\alpha,\beta}_T}\frac{\partial}{\partial \tilde{q}^{-\alpha,-\beta}_T}+\frac{1}{N^4\hbar^2}|\mathbf{\bar{k}}|^2\tilde{q}^{\alpha,\beta}_T\tilde{q}^{-\alpha,-\beta}_T\bigg).
\end{equation}

\subsection{Solving the Schr{\"o}dinger equation in Fourier space}
Having obtained the Hamiltonian, we can now write the Schr{\"o}dinger equation explicitly as 
\begin{equation}\label{penul_sch}
    \frac{N^2}{2}\sum_{\alpha,\beta}\bigg(\hbar ^2\frac{\partial}{\partial \tilde{q}^{\alpha,\beta}_L}\frac{\partial}{\partial \tilde{q}^{-\alpha,-\beta}_L}-\hbar ^2\frac{\partial}{\partial \tilde{q}^{\alpha,\beta}_T}\frac{\partial}{\partial \tilde{q}^{-\alpha,-\beta}_T}+\frac{|\mathbf{\bar{k}}|^2}{N^4}\tilde{q}^{\alpha,\beta}_T\tilde{q}^{-\alpha,-\beta}_T\bigg)\Psi_0[q]=\mathcal{E}_0\Psi_0[q].   
\end{equation}
By enforcing the constraint in Fourier space in Eq.\,\eqref{constr_four} the first term vanishes, giving
\begin{equation}\label{final_sch}
  \frac{N^2}{2}\sum_{\alpha,\beta}\bigg(-\hbar ^2\frac{\partial}{\partial \tilde{q}^{\alpha,\beta}_T}\frac{\partial}{\partial \tilde{q}^{-\alpha,-\beta}_T}+\frac{1}{N^4}|\mathbf{\bar{k}}|^2\tilde{q}^{\alpha,\beta}_T\tilde{q}^{-\alpha,-\beta}_T\bigg)\Psi_0[q]=\mathcal{E}_0\Psi_0[q]. 
\end{equation}
It is now finally time to solve the equation for the ground state. As shown in both Ref.\,\cite{QFT-fradkin} and Ref.\,\cite{Giacomini2023quantumstatesof} for the electromagnetic case, the equation obtained above closely resembles that of a quantum harmonic oscillator. More precisely, Eq.\,\eqref{final_sch} can be interpreted as the time-independent Schr{\"o}dinger equation for a collection of $N^2$ independent harmonic oscillators, one for each pair of discrete momentum modes $(\alpha, \beta)$, and with corresponding frequencies $|\mathbf{\bar{k}}|$. This effectively leaves us with one harmonic oscillator per $(\alpha, \beta)$ mode. Since the oscillators are decoupled, their total ground state wavefunction is simply the product of the ground states of each individual oscillator. Each of these ground states is a Gaussian function, and thus their product yields an overall wavefunction of the form \begin{equation}\label{q_basis_ansatz} \Psi_0[q]=A\exp\bigg(-\sum_{\alpha,\beta}f^{\alpha,\beta}\tilde{q}^{\alpha,\beta}_T\tilde{q}^{-\alpha,-\beta}_T\bigg), 
\end{equation} 
where $A$ is a normalization constant, and the function $f^{\alpha,\beta}$ is to be determined by solving the Schr{\"o}dinger equation. Notice also that this choice of $\Psi[q]$ automatically satisfies the constraint in Eq.\,\eqref{constr_four}, since it is independent of the longitudinal component of the field. 
 
Let us proceed by computing the two differentiations of $\Psi_0[q]$ coming from the first term in the Schr{\"o}dinger equation, i.e.
\begin{equation}
    -\frac{\partial}{\partial \tilde{q}^{\alpha,\beta}_T}\frac{\partial}{\partial \tilde{q}^{-\alpha,-\beta}_T}\Psi[q]=-\frac{\partial}{\partial \tilde{q}^{\alpha,\beta}_T}\frac{\partial}{\partial \tilde{q}^{-\alpha,-\beta}_T}\bigg[A\,\exp\bigg(-\sum_{\gamma,\theta}f^{\gamma,\theta}\tilde{q}^{\gamma,\theta}_T\tilde{q}^{-\gamma,-\theta}_T\bigg)\bigg].
\end{equation}
Differentiating the exponential gives two non-vanishing terms, coming from the derivative of the sum in the exponent,
\begin{equation}
  -\frac{\partial}{\partial \tilde{q}^{\alpha,\beta}_T}\frac{\partial}{\partial \tilde{q}^{-\alpha,-\beta}_T}\Psi[q]=\frac{\partial}{\partial \tilde{q}^{\alpha,\beta}_T}\bigg[\tilde{q}_T^{\alpha,\beta}(f^{\alpha,\beta}+f^{-\alpha,-\beta})A\,\exp\bigg(-\sum_{\gamma,\theta}f^{\gamma,\theta}\tilde{q}^{\gamma,\theta}_T\tilde{q}^{-\gamma,-\theta}_T\bigg)\bigg],  
\end{equation}
and by performing also the last derivative we obtain
\begin{equation}
  -\frac{\partial}{\partial \tilde{q}^{\alpha,\beta}_T}\frac{\partial}{\partial \tilde{q}^{-\alpha,-\beta}_T}\Psi[q]=(f^{\alpha,\beta}+f^{-\alpha,-\beta})\Psi[q]-\tilde{q}_T^{\alpha,\beta}\tilde{q}_T^{-\alpha,-\beta}(f^{\alpha,\beta}+f^{-\alpha,-\beta})^2\Psi[q]. 
\end{equation}
So, Eq.\,\eqref{final_sch} becomes
\begin{equation}
     \frac{N^2}{2}\sum_{\alpha,\beta}\bigg[\hbar ^2(f^{\alpha,\beta}+f^{-\alpha,-\beta})-\hbar ^2\tilde{q}_T^{\alpha,\beta}\tilde{q}_T^{-\alpha,-\beta}(f^{\alpha,\beta}+f^{-\alpha,-\beta})^2
    +\frac{1}{N^4}|\mathbf{\bar{k}}|^2\tilde{q}^{\alpha,\beta}_T\tilde{q}^{-\alpha,-\beta}_T\bigg]\Psi_0[q]=\mathcal{E}_0\Psi_0[q]. 
\end{equation}
Now we can divide by the wave function on both sides, and notice that the only way for it to be solved is to require that
\begin{equation}
   \hbar ^2(f^{\alpha,\beta}+f^{-\alpha,-\beta})^2=\frac{1}{N^4}|\mathbf{\bar{k}}|^2 .
\end{equation}
The obvious choice is to fix
\begin{equation}
    f^{\alpha,\beta}=f^{-\alpha,-\beta}=\frac{1}{ N^2\hbar }\frac{|\mathbf{\bar{k}}|}{2},
\end{equation}
which is well-defined, since it is clear from the definition in Eq.\,\eqref{k_def} that $|\mathbf{\bar{k}}|$ is symmetric with respect to $\alpha$ and $\beta$. Plugging $f^{\alpha,\beta}$ and $f^{-\alpha,-\beta}$ in the eigenvalue equation, we can also compute the energy associated to this state:
\begin{equation}
    \mathcal{E}_0=\frac{N^2}{2}\sum_{\alpha,\beta}\hbar^2(f^{\alpha,\beta}+f^{-\alpha,-\beta}) = \frac{1}{2}\hbar\sum_{\alpha,\beta}|\mathbf{\bar{k}}|.
\end{equation}
The obtained result is precisely what we would expect from the ground state energy of $N^2$ quantum harmonic oscillators, one at each site of the Fourier space grid.
Therefore, we have indeed found the ground state of our model in the sourceless case:
\begin{equation}\label{g_state_q_basis}
    \Psi_0[q]=A\,\exp\bigg(-\frac{1}{2N^2\hbar }\sum_{\alpha,\beta}|\mathbf{\bar{k}}|\,\tilde{q}^{\alpha,\beta}_T\tilde{q}^{-\alpha,-\beta}_T\bigg).
\end{equation}

\subsection{Ground state in position space.}
We have successfully computed the ground state in Fourier space. It is now time to transform back to position space. Furthermore, we will also need to express it in the $p$ field basis for later purposes. 

Let us start from the ground state in the $q$ basis in Eq.\,\eqref{g_state_q_basis}. To transform this back to position space, we have to rewrite the term $\tilde{q}^{\alpha,\beta}_T\tilde{q}^{-\alpha,-\beta}_T$ in the exponential. Let us start from the following expression:
\begin{equation}
\star =\frac{1}{|\mathbf{\bar{k}}|^2}(\bar{k}_x\tilde{q}^{\alpha,\beta}_y-\bar{k}_y\tilde{q}^{\alpha,\beta}_x)(\bar{k}_x\tilde{q}^{-\alpha,-\beta}_y-\bar{k}_y\tilde{q}^{-\alpha,-\beta}_x),
\end{equation}
which by using the decomposition in Eq.\,\eqref{T_L_dec} becomes
\begin{equation}
\star =\frac{1}{|\mathbf{\bar{k}}|^2|\mathbf{t}|^2}(\bar{k}_xt_y\tilde{q}^{\alpha,\beta}_T-\bar{k}_yt_x\tilde{q}^{\alpha,\beta}_T)(\bar{k}_xt_y\tilde{q}^{-\alpha,-\beta}_T-\bar{k}_yt_x\tilde{q}^{-\alpha,-\beta}_T),
\end{equation}
where the terms proportional to the longitudinal got simplified automatically and we again supposed that the transverse vector is the same for both $\mathbf{\bar{k}}$ and $-\mathbf{\bar{k}}$. Now we perform all multiplication and collect common terms, finding that
\begin{equation}
\star =\frac{1}{|\mathbf{\bar{k}}|^2|\mathbf{t}|^2}(\bar{k}_x^2t_y^2+ \bar{k}_y^2t_x^2 - \bar{k}_xt_x\bar{k}_yt_y -\bar{k}_xt_x\bar{k}_yt_y )\tilde{q}^{\alpha,\beta}_T\tilde{q}^{-\alpha,-\beta}_T.
\end{equation}
Recalling $\mathbf{\bar{k}}\cdot\mathbf{t}=0$ we can use that $\bar{k}_xt_x=-\bar{k}_yt_y$ and rewrite the last two terms in parenthesis, i.e.
\begin{equation}
\star =\frac{1}{|\mathbf{\bar{k}}|^2|\mathbf{t}|^2}(\bar{k}_x^2t_y^2+ \bar{k}_y^2t_x^2 + \bar{k}^2_xt^2_x+\bar{k}^2_yt^2_y )\tilde{q}^{\alpha,\beta}_T\tilde{q}^{-\alpha,-\beta}_T.
\end{equation}
By collecting the common terms we notice that the parenthesis perfectly simplifies the normalization in front, obtaining
\begin{equation}
\star =\frac{1}{|\mathbf{\bar{k}}|^2|\mathbf{t}|^2}(\bar{k}_x^2+\bar{k}_y^2)(t_x^2+t_y^2)\tilde{q}^{\alpha,\beta}_T\tilde{q}^{-\alpha,-\beta}_T=\tilde{q}^{\alpha,\beta}_T\tilde{q}^{-\alpha,-\beta}_T.
\end{equation}
So we showed that
\begin{equation}
\tilde{q}^{\alpha,\beta}_T\tilde{q}^{-\alpha,-\beta}_T = \frac{1}{|\mathbf{\bar{k}}|^2}(\bar{k}_x\tilde{q}^{\alpha,\beta}_y-\bar{k}_y\tilde{q}^{\alpha,\beta}_x)(\bar{k}_x\tilde{q}^{-\alpha,-\beta}_y-\bar{k}_y\tilde{q}^{-\alpha,-\beta}_x).
\end{equation}
We recall the differentiation property of the discrete Fourier transform in Eq.\,\eqref{DFT_diff}, which in our case reads 
\begin{equation}
    \bar{k}_s\tilde{q}_r^{\alpha\beta}=-\text{i}\sum_{i,j}(\bar{\partial}_sq_r^{i,j})\text{e}^{-\text{i}\frac{2\pi}{N}(i\alpha+j\beta)},
\end{equation}
and
\begin{equation}
    \bar{l}_s(-\alpha,-\beta)\tilde{q}_r^{-\alpha-\beta}=-\bar{k}_s\tilde{q}_r^{-\alpha-\beta}=-\text{i}\sum_{i,j}(\bar{\partial}_sq_r^{i,j})\text{e}^{\text{i}\frac{2\pi}{N}(i\alpha+j\beta)}.
\end{equation}
Using these results we obtain that
\begin{equation}
  \tilde{q}^{\alpha,\beta}_T\tilde{q}^{-\alpha,-\beta}_T =\frac{1}{|\mathbf{\bar{k}}|^2}\sum_{i,j}\sum_{n,m}(\bar{\partial}_xq^{i,j}_y-\bar{\partial}_yq^{i,j}_x)(\bar{\partial}_xq^{n,m}_y-\bar{\partial}_yq^{n,m}_x)\text{e}^{-\text{i}\frac{2\pi}{N}(i\alpha+j\beta)}\text{e}^{\text{i}\frac{2\pi}{N}(n\alpha+m\beta)} . 
\end{equation}
Finally we recall the definition of $b^{i,j}$, obtaining
\begin{equation}
 \tilde{q}^{\alpha,\beta}_T\tilde{q}^{-\alpha,-\beta}_T =\frac{1}{|\mathbf{\bar{k}}|^2}\sum_{i,j}\sum_{n,m}b^{i,j}b^{n,m}\text{e}^{-\text{i}\frac{2\pi}{N}(i\alpha+j\beta)}\text{e}^{\text{i}\frac{2\pi}{N}(n\alpha+m\beta)}.   
\end{equation}
With this result, the ground state wave function in the position space can be rewritten as
\begin{equation}\label{g_state_q_pos}
    \Psi_0[q]=A\exp\bigg[-\frac{1}{2\hbar }\sum_{i,j}\sum_{n,m}G(i-n,j-m)b^{i,j}b^{n,m}\bigg],
\end{equation}
where $G$ is defined as
\begin{equation}\label{G_func_app}
  G(i-n,j-m):=\frac{1}{N^2}\sum_{\alpha,\beta}\frac{1}{|\mathbf{\bar{k}}|} \text{e}^{-\text{i}\frac{2\pi}{N}[(i-n)\alpha+(j-m)\beta)]}.
\end{equation}
\subsection[Ground state in the p field representation]{Ground state in the $\boldsymbol{p}$ field representation}
Now we proceed by changing basis from $q$ to $p$ field basis. In this basis the Gauss-like constraint in Eq.\,\eqref{constr_four} becomes
\begin{equation}\label{constr_four_p}
    |\bar{\mathbf{k}}|\,\tilde{p}^{\alpha,\beta}_L\Psi[p]=0,
\end{equation}
and the $q$ is defined as
\begin{equation}
    \hat{q}^{i,j}_s:=\text{i}\hbar\frac{\partial}{\partial p_s^{i,j}},
\end{equation}
In Fourier space we similarly have that
\begin{equation}
    \tilde{q}_s^{\alpha,\beta}=\text{DFT}(q_s^{i,j})=\text{i}\,\hbar \,\text{DFT}\bigg(\frac{\partial}{\partial p_s^{i,j}}\bigg)=\text{i}\hbar N^2\frac{\partial}{\partial p_s^{-\alpha,-\beta}}.
\end{equation}
Using this and inverting Eq.\,\eqref{p_four_q_bas}, the Schr{\"o}dinger equation in Eq.\,\eqref{final_sch}, can be written in $p$ basis as
\begin{equation}\label{final_schr_p}
      \frac{N^2}{2}\sum_{\alpha,\beta}\bigg(\frac{1}{N^4}\tilde{p}^{\alpha,\beta}_T\tilde{p}^{-\alpha,-\beta}_T-\hbar ^2|\mathbf{\bar{k}}|^2\frac{\partial}{\partial \tilde{p}^{\alpha,\beta}_T}\frac{\partial}{\partial \tilde{p}^{-\alpha,-\beta}_T}\bigg)\Psi_0[p]=\mathcal{E}_0\Psi_0[p], 
\end{equation}
where again this holds only if the wave function $\Psi_0[p]$ also fulfills the constraint in Eq.\,\eqref{constr_four_p}, and $\tilde{p}^{\alpha,\beta}_T$ represents the transverse part of Eq.\,\eqref{T_L_dev_dec} written in the $p$ basis.

Repeating a procedure analogous to the $q$ basis case, we can pick an ansatz for the wave function of the ground state as written in $p$ basis, namely
\begin{equation}
    \Psi_0[p]=A\,\delta(\tilde{p}_L)\exp\bigg(-\sum_{\alpha,\beta}g^{\alpha,\beta}\,\tilde{p}^{\alpha,\beta}_T\tilde{p}^{-\alpha,-\beta}_T\bigg),
\end{equation}
where $g^{\alpha,\beta}$ is another function to be determined. This time we also had to multiply the gaussian exponential by the delta function, defined as the product of delta functions that fixes all $\tilde{p}^{\alpha,\beta}_L$ to zero, i.e.
\begin{equation}
    \delta(\tilde{p}_L)=\prod_{\alpha,\beta}\delta(\tilde{p}^{\alpha,\beta}_L).
\end{equation}
This is done to ensure that $\Psi[q]$, not only solves the Schr{\"o}dinger equation, but also the constraint in Eq.\,\eqref{constr_four_p}. Differently from the $q$ field basis case, where it was sufficient to pick a wave function independent of $\tilde{q}_L^{\alpha,\beta}$, here it is necessary to fix $p^{\alpha,\beta}_L$ exactly to zero. 

We now plug the ansatz in Eq.\,\eqref{final_schr_p} and repeat a similar procedure to the one done for the wave function written in the $q$ basis. This time the $|\mathbf{\bar{k}}|^2$ factor is multiplied to the derivatives, giving that
\begin{equation}
    g^{\alpha,\beta}=\frac{1}{2N^2\hbar \,|\mathbf{\bar{k}}|},
\end{equation}
and the wave function is
\begin{equation}
    \Psi_0[p]=A\,\delta(\tilde{p}_L)\exp\bigg(-\frac{1}{2N^2\hbar }\sum_{\alpha,\beta}\frac{1}{|\mathbf{\bar{k}}|}\,\tilde{p}^{\alpha,\beta}_T\tilde{p}^{-\alpha,-\beta}_T\bigg),
\end{equation}
Thanks to the delta function fixing $p_L^{\alpha,\beta}$ to zero everywhere, the state can also be simply re-cast into
\begin{equation}
    \Psi_0[p]=A\,\delta(\tilde{p}_L)\exp\bigg(-\frac{1}{2N^2\hbar }\sum_{\alpha,\beta}\frac{1}{|\mathbf{\bar{k}}|}\,\tilde{p}^{\alpha,\beta}_s\tilde{p}^{-\alpha,-\beta}_s\bigg).
\end{equation}
Finally, by using the inverse discrete Fourier transform relation in Eq.\,\eqref{DFT_lattice} for $\tilde{p}^{\alpha,\beta}$ and $\tilde{p}^{-\alpha,-\beta}$, it is straightforward to find the wave function in position space
\begin{equation}\label{g_state_p_basis_app}
    \Psi_0[p]=A\,\delta(\mathcal{C})\,\exp\bigg[-\frac{1}{2\hbar }\sum_{i,j}\sum_{n,m}G(i-n,j-m)\,p^{i,j}_sp^{n,m}_s\bigg].    
\end{equation}
Here $G$ is again defined as in Eq.\,\eqref{G_func_app} and 
\begin{equation}
    \delta(\mathcal{C})=\prod_{i,j}\delta(\bar{\partial}_xp^{i,j}_x+\bar{\partial}_yp^{i,j}_y).
\end{equation}

\subsection{Ground state with static source}
We finally conclude this appendix by using the result obtained in the previous section to derive the ground state of our model in the case of a static source $\rho^{i,j}$. This means that we consider
\begin{equation}
\begin{cases}
       J_x^{i,j}=0 \\
        J_y^{i,j}=0
\end{cases}
\,\,\,\forall i,j.
\end{equation}
Looking at the summary Tables \ref{tab:e-m} and \ref{tab:eom}, it is easy to conclude that with this choice the resulting theory is exactly equivalent to the sourceless case except for the constraint. In particular this means that the quantized Hamiltonian in the temporal gauge is equal to the sourceless one in Eq.\,\eqref{ham_q_basis}, implying that we can repeat the same procedure and yield exactly the same Schr{\"o}dinger equation in Eq.\,\eqref{penul_sch}. The only difference is found in the constraint condition, which now becomes
\begin{equation}
  \hat{\mathcal{C}}_\rho^{i,j}\ket{\Psi}=(\bar{\partial}_s\hat{p}_s^{i,j}+\rho^{i,j})\ket{\Psi}=0\rightarrow\bigg(-\text{i}\,\hbar \,\bar{\partial}_s \frac{\partial}{\partial q_s^{i,j}} +\rho^{i,j}\bigg)\Psi[q]=0.
\end{equation}
We can follow the same procedure we performed in the previous section and go to Fourier space. If we define
\begin{equation}\label{dft_rho}
    \tilde{\rho}^{\alpha,\beta}=\text{DFT}(\rho^{i,j})=\sum_{i,j}\rho^{i,j}\text{e}^{-\text{i}\frac{2\pi}{N}(i\alpha+j\beta)},
\end{equation}
the constraint is given by
\begin{equation}\label{constr_four_source}
\bigg(\hbar N^2|\mathbf{\bar{k}}|\frac{\partial}{\partial\tilde{q}_L^{-\alpha,-\beta}}-\tilde{\rho}^{\alpha,\beta}\bigg)\Psi[q],
\end{equation}
where we used the decomposition in Eq.\,\eqref{T_L_dev_dec}. Therefore, in this case, the wave function is not independent of the longitudinal component. This implies that the ground state found in the previous section is not a valid state anymore. We thus need to add a term to the exponential in Eq.\,\eqref{g_state_q_basis} such that it is a solution of Eq.\,\eqref{constr_four_source} and, at the same time, results in the state with lowest energy.

We are now in a situation completely analogous to the one exposed in Ref.\,\cite{Giacomini2023quantumstatesof} for the electromagnetic case. Their approach was to translate the sourceless ground state by an amount corresponding to the Coulomb field, which corresponds to the lowest energy solution of Gauss-law. We proceed in a similar way and we take the new ground state to be 
\begin{equation}
        \Psi_{0,\rho}[q]=A\,\exp\bigg(-\frac{1}{2N^2\hbar }\sum_{\alpha,\beta}|\mathbf{\bar{k}}|\,\tilde{q}^{\alpha,\beta}_T\tilde{q}^{-\alpha,-\beta}_T+\frac{1}{N^2\hbar}\sum_{\alpha,\beta}\frac{1}{|\mathbf{\bar{k}}|}\tilde{q}_L^{-\alpha,-\beta}\tilde{\rho}^{\alpha,\beta}\bigg).
\end{equation}
It is easy to verify that this function satisfies the constraint in Eq.\,\eqref{constr_four_source}. Furthermore, we notice that such a Gaussian state again represents a set of $N^2$ independent harmonic oscillators, each in their ground state, exactly as in the sourceless case, with the only difference being that the longitudinal components of the Gaussian are now shifted by a term that accounts for the presence of matter, leaving the harmonic oscillators in the unchanged ground state. Hence, this is effectively the ground state of the model with matter density $\rho^{i,j}$. 

Plugging this state into Eq.\,\eqref{penul_sch} yields the new ground state energy. The first term in the exponential gives the ground state energy $\mathcal{E}_0$, which we already computed in the previous section, while the source term produces the following difference:
\begin{equation}
   \mathcal{E}_\rho-\mathcal{E}_0=\frac{N^2}{2\Psi[q]}\sum_{\alpha,\beta}\hbar ^2\frac{\partial}{\partial \tilde{q}^{\alpha,\beta}_L}\frac{\partial}{\partial \tilde{q}^{-\alpha,-\beta}_L}\Psi[q]=\frac{1}{2N^2}\sum_{\alpha,\beta}\frac{\tilde{\rho}^{\alpha,\beta}\tilde{\rho}^{-\alpha,-\beta}}{|\mathbf{\bar{k}}|^2}.
\end{equation}
By using the expression in Eq.\,\eqref{dft_rho}, we rewrite it as
\begin{equation}
   \mathcal{E}_\rho-\mathcal{E}_0=\frac{1}{2}\sum_{i,j}\sum_{n,m}D(i-n,j-m)\rho^{i,j}\rho^{n,m}, 
\end{equation}
where we defined
\begin{equation}
    D(i-n,j-m)=\frac{1}{N^2}\sum_{\alpha,\beta}\frac{1}{|\mathbf{\bar{k}}|^2}\text{e}^{-\text{i}\frac{2\pi}{N}[(i-n)\alpha+(j-m)\beta]}.
\end{equation}
The energy obtained is thus analogous to the potential energy of the Coulomb field created by a classical charge density in QED.

We now can also rewrite the wave function in position space. The first term in the exponential obviously results in the same of the one in Eq.\,\eqref{g_state_q_pos}, while we use a similar procedure for the second one:
\begin{equation}
   \circ= \frac{1}{N^2\hbar}\sum_{\alpha,\beta}\frac{1}{|\mathbf{\bar{k}}|}\tilde{q}^{-\alpha,-\beta}_L\tilde{\rho}^{\alpha,\beta}.
\end{equation}
To rewrite this in position space we use the usual decomposition in Eq.\,\eqref{inv_T_L_dec} and Eq.\,\eqref{k_opposite}. We get
\begin{equation}
    \circ= -\frac{1}{N^2\hbar}\sum_{\alpha,\beta}\frac{\bar{k}_s}{|\mathbf{\bar{k}}|^2}\tilde{q}^{-\alpha,-\beta}_s\tilde{\rho}^{\alpha,\beta}.
\end{equation}
Now we can use the direct DFT and differentiation property to write
\begin{equation}
    \tilde{\rho}^{\alpha,\beta}=\sum_{i,j}\rho^{i,j}\text{e}^{-\text{i}\frac{2\pi}{N}(i\alpha+j\beta)},
\end{equation}
and
\begin{equation}
    \bar{k}_s\tilde{q}^{-\alpha,-\beta}_s=-\text{i}\sum_{i,j}(\bar{\partial}_sq_s^{i,j})\text{e}^{\text{i}\frac{2\pi}{N}(i\alpha+j\beta)}=\text{i}\sum_{i,j}q_s^{i,j}[\bar{\partial}_s\text{e}^{\text{i}\frac{2\pi}{N}(i\alpha+j\beta)}],
\end{equation}
where we also used the result in Eq.\,\eqref{dev_on_exp}.
The source term thus becomes
\begin{equation}
    \circ= \frac{\text{i}}{\hbar}\sum_{i,j}\sum_{n,m}[\bar{\partial}_sD(i-n,j-m)]q_s^{i,j}\rho^{n,m},
\end{equation}
where we have that
\begin{equation}
    D(i-n,j-m)=\frac{1}{N^2}\sum_{\alpha,\beta}\frac{1}{|\mathbf{\bar{k}}|^2}\text{e}^{-\text{i}\frac{2\pi}{N}[(i-n)\alpha+(j-m)\beta]}.
\end{equation}
Inspired by Ref.\,\cite{Giacomini2023quantumstatesof}, we define
\begin{equation}
    p^{i,j}_{s,\rho}=\sum_{n,m}\bar{\partial}_s^{\{i,j\}}D(i-n,j-m)\rho^{n,m},
\end{equation}
with $\bar{\partial}_s^{\{i,j\}}$ being the discrete derivative acting just on the function depending on $i,j$. This represents the analogous of the Coulomb potential for our toy model.
With this result the source term can be cast in its final form:
\begin{equation}
    \circ=- \frac{\text{i}}{\hbar}\sum_{i,j}p^{i,j}_{s,\rho}q_s^{i,j}.
\end{equation}

The full wave function in position space is thus given by
\begin{equation}\label{source_g_state_q}
   \Psi_{0,\rho}[q]= A\exp\bigg[-\frac{1}{2\hbar }\sum_{i,j}\sum_{n,m}G(i-n,j-m)b^{i,j}b^{n,m}- \frac{\text{i}}{\hbar}\sum_{i,j}p^{i,j}_{s,\rho}q_s^{i,j}\bigg].
\end{equation}

In conclusion, we notice that, as was found in Ref.\,\cite{Giacomini2023quantumstatesof} for electromagnetism, the additional source term simply leads to a translation of the $p$ field. Therefore, we can introduce the familiar translation operator as
\begin{equation}
    T_{\rho}=\exp\left(-\frac{\text{i}}{\hbar}\sum_{i,j}p^{i,j}_{s,\rho}q_s^{i,j}\right),
\end{equation}
which represents a translation of each $p_s^{i,j}$ field by an amount equal to the Coulomb field $p^{i,j}_{s,\rho}$. We can thus interpret the ground state wavefunction in Eq.\,\eqref{source_g_state_q} as the translated sourceless wave function in Eq.\,\eqref{g_state_q_pos}, i.e.
\begin{equation}
     \Psi_{0,\rho}[q]=  T_{\rho}\Psi_{0}[q].
\end{equation}

Consequently, if we want to obtain the ground state wave function as written in the $p$ field basis, we just have to shift the $p^{i,j}_s$ and $p^{n,m}_s$ in the exponent of the sourceless one in Eq.\,\eqref{g_state_p_basis_app} by the Coulomb field depending on $\rho$
\begin{equation}\label{source_g_state_p_app}
\Psi_{0,\rho}[p] = T_{\rho}\Psi_{0}[p] =A\,\delta(\mathcal{C_{\rho}})\, \exp\bigg[ 
- \frac{1}{2\hbar} \sum_{i,j}\sum_{n,m} G(i-n,j-m)  (p^{i,j}_s - p^{i,j}_{s,\rho})(p^{n,m}_s - p^{n,m}_{s,\rho}) 
\bigg].
\end{equation}
Here $G$ is again the same as in Eq.\,\eqref{G_func_app} and this time we defined
\begin{equation}
    \delta(\mathcal{C}_{\rho})=\prod_{i,j}\delta(\bar{\partial}_xp^{i,j}_x+\bar{\partial}_yp^{i,j}_y+\rho^{i,j}).
\end{equation}
\section{\label{app:la}Deriving the local algebras}
In this appendix we provide a pedagogical discussion and detailed derivation of all results outlined in Section \ref{sec:alg_def}.

\subsection{\label{app:sec:alg_def} Defining local algebras}
We now go through the procedure to obtain the local algebra $\mathcal{A}_A$ of operations accessible to a finite lattice region. We consider the region $A$ as defined at the beginning of Section \ref{sec:alg_def}. For clarity, we first treat the sourceless case and then include quantum sources. 

\paragraph*{No source.} In the absence of a source, only operations involving the pure field degrees of freedom are relevant. Following lattice gauge theory approaches (see, e.g., Ref.\,\cite{Casini_2014,casini2023lectures}), $\mathcal{A}_A$ is a subalgebra of the gauge-invariant operators $\mathcal{A}$, that are operators that commute with the constraints. In the sourceless case, these constraints are simply the pure field operators $\hat{\mathcal{C}}^{i,j}$, representing the discrete analogue of Gauss's law in vacuum, i.e.
\begin{equation}
    \mathcal{A}_A \subset \mathcal{A} := \left\{ \hat{X} \in \mathcal{O} \,\middle|\, [\hat{X}, \hat{\mathcal{C}}^{i,j}] = 0 \quad \forall i,j \right\}.
\end{equation}

Here, $\mathcal{O}$ is the full algebra generated by the position and momentum operators at each lattice site:
\begin{equation}
    \mathcal{O} := \left\langle \hat{q}_s^{i,j}, \hat{p}_s^{i,j} \,\middle|\, i,j \in [0,N-1],\ s \in \{x,y\} \right\rangle = \mathcal{L}(\mathcal{H}_{\text{kin}}),
\end{equation}
where $\mathcal{L}(\mathcal{H}_{\text{kin}})$ denotes the space of all linear operators on the kinematical Hilbert space.

A natural choice for $\mathcal{A}_A$ is then the set of all gauge-invariant operators generated by positions and momenta acting solely within region $A$. Defining the algebra of all operators supported in $A$ as
\begin{equation}\label{local_algebra_kin}
    \mathcal{O}_{A} := \left\langle \hat{q}_s^{i,j}, \hat{p}_s^{i,j} \,\middle|\, (i,j) \in A,\, s \in \{x,y\} \right\rangle \subset \mathcal{O},
\end{equation}
the algebra of physical operators accessible in $A$ is the gauge-invariant subset of $\mathcal{O}_A$, namely
\begin{equation}\label{la_nosource}
    \mathcal{A}_A := \left\{ \hat{X} \in \mathcal{O}_A \,\middle|\, [\hat{X}, \hat{\mathcal{C}}^{i,j}] = 0 \quad \forall\, (i,j) \in [0,N-1] \right\}.
\end{equation}
As discussed in Section \ref{sec:alg_def}, this case admits a clear visualization using the Graph notation introduced in Appendix \ref{app:graph_def}, where momentum and position generators (i.e. $\hat{p}$'s and $\hat{b}$'s) are represented on the lattice, as illustrated in Fig.\,\ref{graph_algrebra_def}.

\paragraph*{Static quantum source.} We now introduce quantum sources on the lattice following the model in Section \ref{sec:q_matter} and Appendix \ref{app:qmatter_model}. 

Given the structure of $\mathcal{H}^M$ in Eq.\,\eqref{matter_kin_hilb}, the full matter operator algebra is generated by all operators acting on individual qubits, i.e.
\begin{equation}
    \mathcal{O}^M = \bigvee_{i,j} \mathcal{O}^M_{i,j} \subset \mathcal{O},
\end{equation}
where $\mathcal{O}^M_{i,j} = \mathcal{L}(\mathbb{C}^2)$, and $\mathcal{O}$ denotes the full algebra of matter and field operators. The subalgebra of matter operators localized in region $A$ is
\begin{equation}
    \mathcal{O}^M_{A} = \bigvee_{(i,j) \in A} \mathcal{O}^M_{i,j} \subset \mathcal{O}^M.
\end{equation}

The field subalgebra $\mathcal{O}^{f}_{A}$ is defined as in the sourceless case via Eq.\,\eqref{local_algebra_kin} as
\begin{equation}
    \mathcal{O}^F_{A} := \left\langle \hat{q}_s^{i,j},\, \hat{p}_s^{i,j} \,\middle|\, (i,j) \in A,\ s \in \{x,y\} \right\rangle \subset \mathcal{O}^F.
\end{equation}
The combined algebra of all operators localized in $A$ is then
\begin{equation}
    \mathcal{O}_{A} = \mathcal{O}^M_{A} \vee \mathcal{O}^F_{A} \subset \mathcal{O}.
\end{equation}

Finally, the local algebra of physical observables is defined as the subset of $\mathcal{O}_{A}$ that commutes with all constraints $\hat{\mathcal{C}}_{\rho}^{i,j}$, accounting for the presence of sources as in Eq.\,\eqref{physical_projection_matt}, getting
\begin{equation}\label{source_algerba_def}
    \mathcal{A}_A := \left\{ \hat{X} \in \mathcal{O}_{A}\,\middle|\, [\hat{X}, \hat{\mathcal{C}}^{i,j}_{\rho}] = 0 \quad \forall i,j \in [0,N-1] \right\}.
\end{equation}

In this case, explicitly identifying the joint matter-field generators is less trivial than in the sourceless, pure field case. Consequently, a straightforward graphical representation of the algebra, such as in Fig.\,\ref{graph_algrebra_def}, is no longer directly available. 

\section{\label{app:Hss}Deriving the local Hilbert space decompositions}
In the previous appendix we defined the local algebra of observables accessible to region $A$. We now build on that result by studying in detail the corresponding Hilbert space structure. To this end, we follow the procedure outlined in Section~\ref{sec:struct_hilb}, originally developed in the context of gauge theories \cite{dec_ent_donnelly_2012,Casini_2014,casini2023lectures,Van_Acoleyen_2016}. For clarity, we first introduce the method in the sourceless case, and then extend the analysis to include a quantum source, where we explicitly derive the two possible decompositions discussed in that section.

\subsection{\label{app:sec:simple_Hss}Inducing a Hilbert space structure} We begin by introducing the concept of inducing a Hilbert space structure from the algebra of observables acting on it, following the approach of \cite{Zanardi_2004}.
Consider a local algebra $\mathcal{A}_A$, and its \textit{commutant} $\mathcal{A}'_A$, defined as
\begin{equation}
    \mathcal{A}'_A := \left\{ \hat{X} \in \mathcal{O} \,\middle|\, [\hat{X}, \hat{Y}] = 0 \quad \forall \,\hat{Y} \in \mathcal{A}_A \right\}.
\end{equation}
Note that, in general, $\mathcal{A}'_A$ includes operators that are not part of the gauge-invariant subalgebra, i.e., $\mathcal{A}'_A \not\subset \mathcal{A}$.

We now define the \textit{center} of $\mathcal{A}_A$ as the set of operators which are both in the algebra and commute with every of its elements, namely
\begin{equation}
    \mathcal{Z}_A := \left\{ \hat{X} \in \mathcal{A}_A \,\middle|\, [\hat{X}, \hat{Y}] = 0 \quad \forall\, \hat{Y} \in \mathcal{A}_A \right\}.
\end{equation}
It is easy to see that the center coincides with the intersection of the algebra and its commutant,
\begin{equation}
    \mathcal{Z}_A = \mathcal{A}_A \cap \mathcal{A}'_A.
\end{equation}
If the center is \textit{trivial}, meaning it contains only scalar multiples of the identity operator,
\begin{equation}
   \mathcal{Z}_A = \mathcal{A}_A \cap \mathcal{A}'_A = \mathbb{I}, 
\end{equation} 
then the Hilbert space on which these algebras act admits a local tensor product decomposition \cite{Casini_2014,casini2023lectures,Bianchi_2024} given by
\begin{equation}\label{tens_prod_struc}
    \mathcal{H} = \mathcal{H}_A \otimes \mathcal{H}_{A'}.
\end{equation}
In this case, the algebras $\mathcal{A}_A$ and $\mathcal{A}'_A$ are composed of operators that act as
\begin{equation}
    \hat{X}_A = \hat{X}_A \otimes \mathbb{I}_{A'}, \qquad \hat{X}_{A'} = \mathbb{I}_A \otimes \hat{X}_{A'}.
\end{equation}
This is the standard in quantum mechanics and quantum information theory. However in gauge theories, such as the toy model considered here, the center is \textit{not} trivial, and the Hilbert space exhibits a more intricate structure.

\subsection{\label{app:sec:sourceless_struct}Sourceless case}
We now illustrate the sourceless case and consider the local algebra $\mathcal{A}_A$ defined in Eq.\,\eqref{la_nosource} and depicted in Fig.\,\ref{graph_algrebra_def}.  

\paragraph*{Finding the center.} By construction, $\mathcal{A}_A$ contains only physical observables, so every element commutes with the constraint operators $\hat{\mathcal{C}}^{i,j}$ over the whole lattice. Moreover, since the generators of $\mathcal{A}_A$ include all $\hat{p}_s^{i,j}$ within region $A$, it necessarily contains some of the $\hat{\mathcal{C}}^{i,j}$ operators too. As a result, the center $\mathcal{Z}_A$ is nontrivial: it at least contains the constraint operators fully supported in $A$.  

These operators are essentially given by
\begin{equation}
\{\hat{\mathcal{C}}^{i,j}\in\mathcal{A}_A\}=    \{\hat{\mathcal{C}}^{i,j}\in\mathcal{A}\,|\,([i-1,i+1],j),(i,[j-1,j+1])\in A\}\subset\mathcal{Z}_A,
\end{equation}
which we represent simply with the symbol \,\,\,\tikz[baseline=-0.5ex,scale=0.8]{
    \draw[mypurple,line width=2pt](0,-0.4)--(0,0.4);
    \draw[mypurple,line width=2pt](-0.4,0)--(0.4,0);
    \draw[rect, dotted, line width=1.5pt,rounded corners=0.5pt](0.5,0.5)--(0.5,-0.5)--(-0.5,-0.5)--(-0.5,0.5)--(0.5,0.5);
}\,\,.

Additionally, we must include in $\mathcal{Z}_A$ those parts of constraint operators that cross the boundary of region $A$. We refer to these elements as \textit{edge terms}. Recalling the Graph notation in Section~\ref{app:graph_def}, we illustrate some examples in the diagrams below:

\begin{center}
         \begin{tikzpicture}[scale=0.8]
        \definecolor{rect}{rgb}{0.2,0.45,0.9}

        \foreach \x in {-3,...,-1} {
            \foreach \y in {0,...,2} {
                \fill[black] (\x,\y) circle (1.5pt);
            }
        }

        \foreach \x in {2,...,4} {
            \foreach \y in {0,...,2} {
                \fill[black] (\x,\y) circle (1.5pt);
            }
        }
        \fill[mycyan] (-3,1) circle (2pt);
         \node[text=mycyan, left] at (-3,1){$-$};  
        \fill[red] (-2,2) circle (2pt);
         \node[text=red, above] at (-2,2) {$+$};
         \fill[red] (-2,0) circle (2pt);
        \node[text=red, below] at (-2,0) {$-$};  

        \fill[mycyan] (2,1) circle (2pt);
         \node[text=mycyan, left] at (2,1){$-$};  
        \fill[red] (3,2) circle (2pt);
         \node[text=red, above] at (3,2) {$+$};
         \fill[mycyan] (4,1) circle (2pt);
        \node[text=mycyan, right] at (4,1) {$+$};

        \draw[mypurple,rounded corners=6pt, line width=1.5pt] (-2, 2.6) -- (-2.3, 2.6) -- (-2.3, 1.3) -- (-3.6, 1.3) -- (-3.6, 0.7) -- (-2.3, 0.7) -- (-2.3, -0.6) -- (-1.7, -0.6) -- (-1.7, 0.7);
        
        \draw[mypurple,dotted, rounded corners=6pt, line width=1.5pt]
        (-1.7, 0.7)--(-0.4,0.7)--(-0.4,1.3)--(-1.7,1.3);

        \draw[mypurple,rounded corners=6pt, line width=1.5pt]
        (-1.7,1.3)--(-1.7,2.6)--(-2,2.6);

        \draw[rect,dashed, line width=1.5pt] (-1.5, 2.5) -- (-1.5, -0.5);

        \draw[mypurple,rounded corners=6pt, line width=1.5pt] (3, 2.6) -- (3.3, 2.6) -- (3.3, 1.3) -- (4.6, 1.3) -- (4.6, 0.7) -- (3.3, 0.7);

        \draw[mypurple,dotted,rounded corners=6pt, line width=1.5pt]
        (3.3, 0.7)--(3.3, -0.6) -- (2.7, -0.6) -- (2.7, 0.7);
        
        \draw[mypurple,rounded corners=6pt, line width=1.5pt]
        (2.7, 0.7)--(1.4,0.7)--(1.4,1.3)--(2.7,1.3)--(2.7,2.6)--(3,2.6);

        \draw[rect,dashed, line width=1.5pt] (4.5, 0.5) -- (1.5, 0.5);

         \node at (-2.7,2.8) {\large $A$};
         \node at (4.2,2.8) {\large $A$};

    \end{tikzpicture}
\end{center}
It is not hard to show that these ``truncated crosses'' commute with all $\hat{b}$ operators in $\mathcal{A}_A$. We can demonstrate this using a graphical method similar to the one introduced in Appendix~\ref{app:b_inv}. Here, we briefly illustrate the case for the leftmost example above.

Since our local algebra includes only those $\hat{b}$ operators that are fully contained within the boundary (as defined in Fig.\,\ref{graph_algrebra_def}), the only potentially nontrivial commutators to consider arise from the top-left and bottom-left $\hat{b}$ operators. We mark these with green circles following the convention in Graph\,\eqref{graph_def_2}. Using again the diagram from Graph\,\eqref{graph_b_C}, we draw their relevant components. By multiplying the signs associated with the common sites, we obtain the following two Graphs:
\begin{center}
         \begin{tikzpicture}[scale=0.8]

        \foreach \x in {-3,...,-1} {
            \foreach \y in {0,...,2} {
                \fill[black] (\x,\y) circle (1.5pt);
            }
        }

        \foreach \x in {2,...,4} {
            \foreach \y in {0,...,2} {
                \fill[black] (\x,\y) circle (1.5pt);
            }
        }
        \fill[mycyan] (-3,1) circle (2pt);
         \node[text=mycyan, left] at (-3,1){$-$};  
        \fill[red] (-2,2) circle (2pt);
         \node[text=red, above] at (-2,2) {$+$};
         \fill[red] (-2,0) circle (2pt);
        \node[text=red, below] at (-2,0) {$-$};

        \draw[mycyan,line width=1pt ] (-3,1) circle (3pt);
        \draw[red, line width=1pt ] (-2,2) circle (3pt);
         \node[text=red, above] at (-2,2) {$+$};
        \draw[lightgreen,line width=1.5pt ] (-3,2) circle (3pt);

        \fill[mycyan] (2,1) circle (2pt);
         \node[text=mycyan, left] at (2,1){$+$};  
        \fill[red] (3,2) circle (2pt);
         \node[text=red, above] at (3,2) {$+$};
         \fill[red] (3,0) circle (2pt);
        \node[text=red, below] at (3,0) {$-$};

        \draw[mycyan,line width=1pt ] (2,1) circle (3pt);
        \draw[red, line width=1pt ] (3,0) circle (3pt);
         \node[text=red, above] at (3,2) {$+$};
        \draw[lightgreen,line width=1.5pt ] (2,0) circle (3pt);

        \draw[mypurple,rounded corners=6pt, line width=1.5pt] (-2, 2.6) -- (-2.3, 2.6) -- (-2.3, 1.3) -- (-3.6, 1.3) -- (-3.6, 0.7) -- (-2.3, 0.7) -- (-2.3, -0.6) -- (-1.7, -0.6) -- (-1.7, 0.7);
        
        \draw[mypurple,dotted, rounded corners=6pt, line width=1.5pt]
        (-1.7, 0.7)--(-0.4,0.7)--(-0.4,1.3)--(-1.7,1.3);

        \draw[mypurple,rounded corners=6pt, line width=1.5pt]
        (-1.7,1.3)--(-1.7,2.6)--(-2,2.6);

        \draw[rect,dashed, line width=1.5pt] (-1.5, 2.5) -- (-1.5, -0.5);

        \draw[mypurple,rounded corners=6pt, line width=1.5pt] (3, 2.6) -- (3.3, 2.6) -- (3.3, 1.3);

        \draw[mypurple,dotted,rounded corners=6pt, line width=1.5pt]
         (3.3, 1.3)--(4.6, 1.3) -- (4.6, 0.7) -- (3.3, 0.7); 
        
        \draw[mypurple,rounded corners=6pt, line width=1.5pt]
         (3.3, 0.7)--(3.3, -0.6) -- (2.7, -0.6) -- (2.7, 0.7)--(1.4,0.7)--(1.4,1.3)--(2.7,1.3)--(2.7,2.6)--(3,2.6);

        \draw[rect,dashed, line width=1.5pt] (3.5, 2.5) -- (3.5, -0.5);

         \node at (-2.7,2.8) {\large $A$};
         \node at (4.2,2.8) {\large $A$};

    \end{tikzpicture}    
\end{center}
Comparing these results with the ones shown in Graphs\,\eqref{commutation_Ex}, we find that the commutator vanishes in both cases. Therefore, we have identified additional operators that belong to the center $\mathcal{Z}_A$. We can represent these elements with the following symbols: 

\begin{center}
 \,\,\,\tikz[baseline=-0.5ex,scale=0.8]{
    \draw[mypurple,line width=2pt](0,-0.4)--(0,0.4);
    \draw[mypurple,line width=2pt](-0.4,0)--(0,0);
    \draw[rect, dotted, line width=1.5pt,rounded corners=0.5pt](0.2,0.5)--(0.2,-0.5)--(-0.5,-0.5)--(-0.5,0.5)--(0.2,0.5);
}, 
\,\,\,\tikz[baseline=-0.5ex,scale=0.8]{
    \draw[mypurple,line width=2pt](0,-0.4)--(0,0.4);
    \draw[mypurple,line width=2pt](0,0)--(0.4,0);
    \draw[rect, dotted, line width=1.5pt,rounded corners=0.5pt](0.5,0.5)--(0.5,-0.5)--(-0.2,-0.5)--(-0.2,0.5)--(0.5,0.5);
}, 
\,\,\,\tikz[baseline=0.2ex,scale=0.8]{
    \draw[mypurple,line width=2pt](0,0)--(0,0.4);
    \draw[mypurple,line width=2pt](-0.4,0)--(0.4,0);
    \draw[rect, dotted, line width=1.5pt,rounded corners=0.5pt](0.5,0.5)--(0.5,-0.2)--(-0.5,-0.2)--(-0.5,0.5)--(0.5,0.5);
}\,\,\, and \,\,\,\tikz[baseline=-1.5ex,scale=0.8]{
    \draw[mypurple,line width=2pt](0,-0.4)--(0,0);
    \draw[mypurple,line width=2pt](-0.4,0)--(0.4,0);
    \draw[rect, dotted, line width=1.5pt,rounded corners=0.5pt](0.5,0.2)--(0.5,-0.5)--(-0.5,-0.5)--(-0.5,0.2)--(0.5,0.2);
}\,\,.   
\end{center}
Recalling the constraint of the theory, these terms can be essentially interpreted as the electric field components orthogonal to the boundary, located right outside region $A$.  

Following a similar proof, we find that these are not the only possible edge terms. Indeed, we must also consider ``truncated crosses'' with only a single arm inside $A$, which correspond to the electric field components orthogonal to the boundary but lying \textit{inside} region $A$. These are represented simply by one $\hat{p}$ component, depending on whether they sit along the vertical or horizontal border. In accordance with the definition in Graph\/\eqref{graph_def}, we identify these as
\begin{center}
    \,\,\,\tikz[baseline=-0.5ex,scale=0.8]{
    
    \fill[mycyan] (0,0) circle (2pt);
    \draw[rect, dotted, line width=1.5pt,rounded corners=0.5pt](0.3,0.5)--(0.3,-0.5);
}\,, 
\,\,\,\tikz[baseline=-0.5ex,scale=0.8]{
    
    \fill[mycyan] (0,0) circle (2pt);
    \draw[rect, dotted, line width=1.5pt,rounded corners=0.5pt](-0.3,0.5)--(-0.3,-0.5);
}\,,\,\,\,\tikz[baseline=-0.5ex,scale=0.8]{
    
    \fill[red] (0,0) circle (2pt);
    \draw[rect, dotted, line width=1.5pt,rounded corners=0.5pt](0.5,0.3)--(-0.5,0.3);
}\,\,\, and \,\,\,\tikz[baseline=-0.5ex,scale=0.8]{
    
    \fill[red] (0,0) circle (2pt);
    \draw[rect, dotted, line width=1.5pt,rounded corners=0.5pt](0.5,-0.3)--(-0.5,-0.3);
}\,\,.
\end{center}

At the corners we do not include any $\hat{q}$ in the local algebra, and therefore both $\hat{p}_x$ and $\hat{p}_y$ must be part of the center. Using the definition in Graph\/\eqref{graph_def_2}, we can represent these terms with the symbols:

\begin{center}
  \,\,\,\tikz[baseline=-0.5ex,scale=0.8]{
    
    \fill[mypurple] (0,0) circle (2pt);
    \draw[rect, dotted, line width=1.5pt,rounded corners=0.5pt](0.3,0.3)--(0.3,-0.3)--(-0.3,-0.3);
}\,, 
\,\,\,\tikz[baseline=-0.5ex,scale=0.8]{
    
    \fill[mypurple] (0,0) circle (2pt);
    \draw[rect, dotted, line width=1.5pt,rounded corners=0.5pt](0.3,0.3)--(-0.3,0.3)--(-0.3,-0.3);
}\,,\,\,\,\tikz[baseline=-0.5ex,scale=0.8]{
    
    \fill[mypurple] (0,0) circle (2pt);
    \draw[rect, dotted, line width=1.5pt,rounded corners=0.5pt](-0.3,0.3)--(0.3,0.3)--(0.3,-0.3);
}\,\,\, and \,\,\,\tikz[baseline=-0.5ex,scale=0.8]{
    
    \fill[mypurple] (0,0) circle (2pt);
    \draw[rect, dotted, line width=1.5pt,rounded corners=0.5pt](0.3,-0.3)--(-0.3,-0.3)--(-0.3,0.3);
}\,\,.
   
\end{center}

Therefore, the center of our local algebra is non-trivial and it is constituted by all gauge-invariant observables generated by the complete constraints fully contained in $A$ and all the edge terms:
\begin{equation}\label{A_center}
   \mathcal{Z}_A=\left\langle\tikz[baseline=-0.5ex, scale=0.7]{
        \draw[mypurple, line width=2pt](0,-0.4)--(0,0.4);
        \draw[mypurple, line width=2pt](-0.4,0)--(0.4,0);
        \draw[rect, dotted, line width=1.5pt,rounded corners=0.5pt](0.5,0.5)--(0.5,-0.5)--(-0.5,-0.5)--(-0.5,0.5)--(0.5,0.5);
    }\,,\,\,\,\tikz[baseline=-0.5ex,scale=0.6]{
    \draw[mypurple,line width=2pt](0,-0.4)--(0,0.4);
    \draw[mypurple,line width=2pt](-0.4,0)--(0,0);
    \draw[rect, dotted, line width=1.5pt,rounded corners=0.5pt](0.2,0.5)--(0.2,-0.5)--(-0.5,-0.5)--(-0.5,0.5)--(0.2,0.5);
}\,, 
\,\,\,\tikz[baseline=-0.5ex,scale=0.6]{
    \draw[mypurple,line width=2pt](0,-0.4)--(0,0.4);
    \draw[mypurple,line width=2pt](0,0)--(0.4,0);
    \draw[rect, dotted, line width=1.5pt,rounded corners=0.5pt](0.5,0.5)--(0.5,-0.5)--(-0.2,-0.5)--(-0.2,0.5)--(0.5,0.5);
}\,, 
\,\,\,\tikz[baseline=0.2ex,scale=0.6]{
    \draw[mypurple,line width=2pt](0,0)--(0,0.4);
    \draw[mypurple,line width=2pt](-0.4,0)--(0.4,0);
    \draw[rect, dotted, line width=1.5pt,rounded corners=0.5pt](0.5,0.5)--(0.5,-0.2)--(-0.5,-0.2)--(-0.5,0.5)--(0.5,0.5);
}\,,\,\,\,\tikz[baseline=-1.5ex,scale=0.6]{
    \draw[mypurple,line width=2pt](0,-0.4)--(0,0);
    \draw[mypurple,line width=2pt](-0.4,0)--(0.4,0);
    \draw[rect, dotted, line width=1.5pt,rounded corners=0.5pt](0.5,0.2)--(0.5,-0.5)--(-0.5,-0.5)--(-0.5,0.2)--(0.5,0.2);
}\,,\,\,\,\tikz[baseline=-0.5ex,scale=0.6]{
    
    \fill[mycyan] (0,0) circle (2pt);
    \draw[rect, dotted, line width=1.5pt,rounded corners=0.5pt](0.3,0.5)--(0.3,-0.5);
}\,, 
\,\,\,\tikz[baseline=-0.5ex,scale=0.6]{
    
    \fill[mycyan] (0,0) circle (2pt);
    \draw[rect, dotted, line width=1.5pt,rounded corners=0.5pt](-0.3,0.5)--(-0.3,-0.5);
}\,,\,\,\,\tikz[baseline=-0.5ex,scale=0.6]{
    
    \fill[red] (0,0) circle (2pt);
    \draw[rect, dotted, line width=1.5pt,rounded corners=0.5pt](0.5,0.3)--(-0.5,0.3);
}\,,\,\,\,\tikz[baseline=-0.5ex,scale=0.6]{
    
    \fill[red] (0,0) circle (2pt);
    \draw[rect, dotted, line width=1.5pt,rounded corners=0.5pt](0.5,-0.3)--(-0.5,-0.3);
}\,,  \,\,\,\tikz[baseline=-0.5ex,scale=0.6]{
    
    \fill[mypurple] (0,0) circle (2pt);
    \draw[rect, dotted, line width=1.5pt,rounded corners=0.5pt](0.3,0.3)--(0.3,-0.3)--(-0.3,-0.3);
}\,, 
\,\,\,\tikz[baseline=-0.5ex,scale=0.6]{
    
    \fill[mypurple] (0,0) circle (2pt);
    \draw[rect, dotted, line width=1.5pt,rounded corners=0.5pt](0.3,0.3)--(-0.3,0.3)--(-0.3,-0.3);
}\,,\,\,\,\tikz[baseline=-0.5ex,scale=0.6]{
    
    \fill[mypurple] (0,0) circle (2pt);
    \draw[rect, dotted, line width=1.5pt,rounded corners=0.5pt](-0.3,0.3)--(0.3,0.3)--(0.3,-0.3);
}\,, \,\,\,\tikz[baseline=-0.5ex,scale=0.6]{
    
    \fill[mypurple] (0,0) circle (2pt);
    \draw[rect, dotted, line width=1.5pt,rounded corners=0.5pt](0.3,-0.3)--(-0.3,-0.3)--(-0.3,0.3);
}\right\rangle.
\end{equation}
As we already discussed, in such a case the local algebra and its commutant have a non trivial intersection, i.e.
\begin{equation}
        \mathcal{A}_A \cap \mathcal{A}'_A=\mathcal{Z}_A \neq\mathbb{I},
\end{equation}
and thus the Hilbert space cannot be simply decomposed in a tensor product like in Eq.\,\eqref{tens_prod_struc}.

\paragraph*{Superselection sectors.} 

By construction, the center we just identified is an abelian algebra of normal operators.
Therefore, it is possible to find a basis of eigenvectors that simultaneously diagonalize all its elements. In our case, the center in Eq.\,\eqref{A_center} consists exclusively of operators depending on the $\hat{p}$'s, and not on the $\hat{q}$'s. Hence, the natural choice for the diagonalizing basis of $\mathcal{Z}_A$ is the $p$-field basis introduced in Eq.\,\eqref{p_q_rep}, formed by the states $|p^{i,j}_s\rangle$ at each site. In this basis, all elements of $\mathcal{Z}_A$ act diagonally, and their eigenspaces can be labeled by an index $k$, i.e.
\begin{equation}
\mathcal{Z}_A\ni\hat{Z}=\bigoplus_k z_k\,\mathbb{I}^k.
\end{equation}

Within this decomposition, the center acts as a multiple of the identity in each eigenspace. Consequently, the analysis from Appendix~\ref{app:sec:simple_Hss} for the trivial-center case applies sector by sector. Both the local algebra and its commutant then admit a block-diagonal structure,
\begin{equation}\label{algebra_dec}
\mathcal{A}_A = \bigoplus_k \mathcal{O}_A^k \otimes \mathbb{I}^k_{A'} , \qquad
\mathcal{A}'_A = \bigoplus_k \mathbb{I}^k_A \otimes \mathcal{O}_{A'}^k. 
\end{equation}
As in Appendix~\ref{app:sec:simple_Hss}, such an algebraic structure induces a corresponding decomposition of the kinematical Hilbert space:
\begin{equation}\label{hilb_space_struc_1}
    \mathcal{H} = \bigoplus_k \mathcal{H}_{A}^k \otimes \mathcal{H}_{A'}^k,
\end{equation}
where $\mathcal{O}_A^k$ and $\mathcal{O}_{A'}^k$ are the full operator algebras acting on the Hilbert spaces $\mathcal{H}_{A}^k$ and $\mathcal{H}_{A'}^k$, respectively. This decomposition promotes the subspaces labeled by $k$ to superselection sectors: locally accessible operations (i.e., elements of $\mathcal{A}_A$) cannot generate transitions between different sectors.

So far, we have derived the Hilbert space structure induced by the local algebra, without imposing any additional conditions on the commutant algebra $\mathcal{A}'_A$. As a consequence, $\mathcal{A}'_A$ still contains operators that fail to commute with the constraints and are therefore non-physical. To make this explicit, each subalgebra $\mathcal{O}_{A'}^k$ can be further decomposed as
\begin{itemize}
    \item the subalgebra $\mathcal{A}^k_{\text{sym}}\subset \mathcal{O}_{A'}^k$ generated by the gauge symmetry transformations,
    \item the gauge-invariant subalgebra $\bar{\mathcal{A}}_{A}^k \subset \mathcal{O}_{A'}^k$, defined as the commutant of $\mathcal{A}^k_{\text{sym}}$ within~$\mathcal{O}^k_{A'}$.
\end{itemize}
Our next step is to apply the same decomposition procedure to each $\mathcal{O}_{A'}^k$.

In general, $\mathcal{A}^k_{\text{sym}}$ could be non-abelian, with a non-trivial center $\mathcal{Z}_{\text{sym}}$. In our case, however, the only generators of the gauge symmetry are the commuting constraint operators $\hat{\mathcal{C}}^{i,j}$. Hence $\mathcal{A}^k_{\text{sym}}$ is abelian and coincides with its center
\begin{equation}\label{symm_center}
  \mathcal{A}^k_{\text{sym}}= \mathcal{Z}_{\text{sym}} = \left\langle \hat{\mathcal{C}}^{i,j} \,\middle|\, i,j \in [0, N-1] \right\rangle.
\end{equation}

Diagonalizing $\mathcal{Z}_{\text{sym}}$ in the $p$-field basis therefore simultaneously diagonalizes $\mathcal{A}^k_{\text{sym}}$, leading to a further decomposition into a direct sum of sectors:
\begin{equation}
        \mathcal{A}^k_{\text{sym}}= \bigoplus_{r} \mathbb{I}^r_{\text{sym}} \otimes \mathbb{I}^{r,k}_{\bar{A}},\qquad
        \bar{\mathcal{A}}_{A}^k = \bigoplus_{r} \mathbb{I}^r_{\text{sym}} \otimes \mathcal{O}_{\bar{A}}^{r,k}.
\end{equation}
The eigenspaces of $\mathcal{Z}_{\text{sym}}$ are labeled by $r$, in analogy with the irreducible representations of a symmetry group. Importantly, $\mathcal{A}^k_{\text{sym}}$ acts trivially within each $r$-sector, independently of $k$. This is because $\mathcal{A}^k_{\text{sym}}$ commutes with $\mathcal{Z}_A$, and both admit the same diagonalizing basis. The Hilbert space on which $\mathcal{A}^k_{\text{sym}}$ acts is therefore one-dimensional, $\dim(\mathcal{H}^r_{\text{sym}})=1$, and the action reduces to multiplication by a phase.

Substituting this result into the previous decomposition gives
\begin{equation}\label{algebra_dec_2} 
\mathcal{A}_A = \bigoplus_{k,r} \mathcal{O}_A^k \otimes \mathbb{I}^{r,k}_{\bar{A}} \otimes\mathbb{I}^r_{\text{sym}}, \quad
\bar{\mathcal{A}}_A = \bigoplus_{k,r} \mathbb{I}^k_A \otimes \mathcal{O}_{\bar{A}}^{r,k}\otimes\mathbb{I}^r_{\text{sym}}. 
\end{equation}
The Hilbert space accordingly decomposes as
\begin{equation}\label{hilb_space_struc_2}
    \mathcal{H} = \bigoplus_{k,r} \mathcal{H}_{A}^k \otimes \mathcal{H}_{\bar{A}}^{r,k}\otimes\mathbb{C}_r
    =\bigoplus_{k,r} \mathcal{H}_{A}^k \otimes \mathcal{H}_{\bar{A}}^{r,k},
\end{equation}
where the one-dimensional factor $\mathbb{C}_r$ was dropped as irrelevant.

Thus, the Hilbert space does not exhibit a local tensor product structure as in Eq.\,\eqref{tens_prod_struc}, but it does admit a decomposition that preserves a local tensor product form within each direct-sum sector.

\paragraph*{Projecting onto the physical subspace.}
In the absence of sources, the constraint condition takes the form
\begin{equation}
    \hat{\mathcal{C}}^{i,j} \ket{\psi}_{\text{phy}} = 0 \qquad \forall i,j.
\end{equation}
This requires that physical states be simultaneous eigenstates of all constraint operators, with eigenvalue zero. Within the decomposition of Eq.\,\eqref{algebra_dec_2} and Eq.\,\eqref{hilb_space_struc_2}, this corresponds to projecting onto a particular eigenspace of $\mathcal{Z}_{\text{sym}}$, namely the one associated with vanishing eigenvalue for every constraint across the lattice. Selecting this eigenspace for the full set of constraints ($\mathcal{Z}_{\text{sym}}$) fixes the sector label $r$, which we denote as $r=0$.

While this projection uniquely determines $r$, it still leaves freedom in the choice of $k$. As shown in Eq.\,\eqref{A_center}, the label $k$ not only identifies the eigenspaces of the constraints fully contained within $A$ (which are automatically fixed by the choice $r=0$), but also the edge terms. Since the edge terms consist of truncated constraints with support straddling the boundary, they may take nontrivial eigenvalues even when the full constraint operator is fixed to zero.

Consequently, the physical Hilbert space is obtained by projecting onto the $r=0$ sector, while interpreting $k$ as labeling the residual degrees of freedom associated with the edge terms. Explicitly,
\begin{equation}\label{sourceless_dec}
    \mathcal{H}_{\text{phy}} = \bigoplus_{k} \mathcal{H}_{A}^k \otimes \mathcal{H}_{\bar{A}}^{0,k} 
    = \bigoplus_{k} \mathcal{H}_{A}^k \otimes \mathcal{H}_{\bar{A}}^k,
\end{equation}
where the fixed index $r$ has been suppressed for notational simplicity. 

The same projection applies to the local algebras, which take the form
\begin{equation}\label{algebra_dec_3} 
\mathcal{A}_A = \bigoplus_{k} \mathcal{O}_A^k \otimes \mathbb{I}^{k}_{\bar{A}}, \qquad
\bar{\mathcal{A}}_A = \bigoplus_{k} \mathbb{I}^k_A \otimes \mathcal{O}_{\bar{A}}^k .
\end{equation}

\subsection{\label{app:sec:source_struct}Source case}
We now extend the analysis to include quantum sources, as introduced in Section \ref{sec:q_matter} and Appendix \ref{app:qmatter_model}. Before any projection, the total kinematical Hilbert space can be written as the tensor product of the matter and field spaces,
\begin{equation}\label{starting_source_kin_space}
    \mathcal{H} = \mathcal{H}^M \otimes \mathcal{H}^F.
\end{equation}
We first treat matter and field separately, deriving the Split Decomposition in Eq.\,\eqref{split_dec}, and subsequently consider them jointly to obtain the Operational Decomposition in Eq.\,\eqref{oper_dec}.

\paragraph*{Split Decomposition.}
If we treat matter and field separately we can just use the matter Hilbert space decomposition of Eq.\,\eqref{matt_dec_app} together with the pure field decomposition of Eq.\,\eqref{hilb_space_struc_2} valid in the sourceless case. Inserting these results into Eq.\,\eqref{starting_source_kin_space}, we obtain
\begin{equation}\label{source_dec_kin_app}
    \mathcal{H} = \bigoplus_{n=0}^{N^2} \bigoplus_{s_n} \ket{s_n}_M \otimes 
    \left( \bigoplus_{k,r} \mathcal{H}_A^k \otimes \mathcal{H}_{\bar{A}}^{r,k} \right).
\end{equation}

We now project onto the physical Hilbert space by enforcing the constraint operator equation, as in the sourceless case. Here, however, the constraint depends explicitly on the matter density operator:
\begin{equation}\label{physical_projection_app}
    \hat{\mathcal{C}}^{i,j}_{\rho}\ket{\psi}_{\text{phy}} = 
    \left( \hat{\mathcal{C}}^{i,j} + \hat{\rho}^{i,j} \right) \ket{\psi}_{\text{phy}} = 0 
    \qquad \forall i,j. 
\end{equation}
This condition couples matter and field degrees of freedom, and the physical Hilbert space is obtained by projecting onto the subspace of states satisfying Eq.\,\eqref{physical_projection_app}.

In contrast to the sourceless case, the projection cannot be restricted to the zero-eigenvalue sector $r = 0$, due to the terms $\hat{\rho}^{i,j}$. Since the configuration states $\ket{s_n}_M$ in the decomposition are eigenstates of the density operators, each $\hat{\rho}^{i,j}$ acts as a number, either $0$ or $1$, on such states. The appropriate field eigenspaces are therefore determined directly by the matter configuration: $r = s_n$. Projecting accordingly onto the decomposition of Eq.\,\eqref{source_dec_kin}, we obtain the Split Decomposition of the physical Hilbert space
\begin{equation}\label{split_dec_app}
    \mathcal{H}_{\text{phy}} = \bigoplus_{n=0}^{N^2} \bigoplus_{s_n} \ket{s_n}_M \otimes 
    \left( \bigoplus_{k(s_n)} \mathcal{H}_A^{k(s_n)} \otimes \mathcal{H}_{\bar{A}}^{s_n,k(s_n)} \right).
\end{equation}
Here $k$ labels the edge terms, which in general depend on the matter configuration. One may restrict to cases where matter is localized well inside region $A$, so as not to affect the edge directly. Even then, the surviving branches of the direct sum over $k$ still depend on $s_n$.

The labels $k$ and $r$ can be regarded as tuples, whose entries correspond respectively to the eigenvalues of the independent generators of the pure field centers $\mathcal{Z}_A$ and $\mathcal{Z}_{\text{sym}}$ introduced in Eq.\,\eqref{A_center} and Eq.\,\eqref{symm_center}. In this picture, $r$ fixes the eigenvalues of all constraints across the lattice, while $k$ includes only those associated with region $A$ and its boundary. Projecting onto a physical subspace with fixed $s_n$ enforces $r = s_n$, which in turn constrains part of $k$, though residual freedom in the edge terms remains, as in the sourceless case.

As discussed in Section \ref{sec:struct_hilb}, such a decomposition provides an explicit description of the interplay between field and matter, and offers a framework for analyzing the processes underlying an FME experiment. Given Eq.\,\eqref{split_dec_app}, a general physical state in $\mathcal{H}_{\text{phy}}$ takes the form
\begin{equation}
    \ket{\Upsilon} = \eta \sum_{n=0}^{N^2} \sum_{s_n} \alpha_{s_n} \ket{s_n}_M \ket{\Psi_{s_n}}_F,
\end{equation}
where $\eta$ is a normalization constant. This expression shows that matter and field states are intrinsically entangled through the gauge constraint. As we will see in the FME example of Section \ref{sec:FME}, this coupling restricts the set of locally accessible operations and governs the generation of entanglement.

The Split Decomposition, however, does not provide a complete characterization of the local algebras of physical operators—acting jointly on matter and field—that are available to an agent localized in region $A$.

\paragraph*{Operational Decomposition.} 
We now consider matter and field degrees of freedom jointly. This yields a decomposition analogous to Eq.\,\eqref{sourceless_dec} for the sourceless case, but now constructed from the local algebras including operators acting on both matter and field in region $A$, as defined in Eq.\,\eqref{source_algerba_def}.

We begin by identifying the center $\mathcal{Z}_A$. As before, $\mathcal{Z}_A$ contains the constraints fully supported within $A$ together with the edge terms arising from the truncated constraints $\hat{\mathcal{C}}^{i,j}_{\rho}$ at the boundary. This leads to the same structural situation analyzed in Section \ref{app:sec:sourceless_struct}, and the same procedure gives the analogue of Eq.\,\eqref{algebra_dec_2}, i.e.
\begin{equation}
    \mathcal{A}_A = \bigoplus_{K,R} \mathcal{O}^{K}_{A} \otimes \mathbb{I}^{R,K}_{\bar{A}} \otimes \mathbb{I}^R_{\text{sym}}, \qquad
    \bar{\mathcal{A}}_A = \bigoplus_{K,R} \mathbb{I}^K_A \otimes \mathcal{O}^{R,K}_{\bar{A}} \otimes \mathbb{I}^R_{\text{sym}}.
\end{equation}
Here $\bar{\mathcal{A}}_A$ denotes the gauge-invariant subalgebra of the commutant of $\mathcal{A}_A$, while $K$ and $R$ label, respectively, the eigenvalues of the center $\mathcal{Z}_A$ (including both matter and field) and of the full set of constraints $\hat{\mathcal{C}}^{i,j}_{\rho}$.

This structure induces the following decomposition of the kinematical Hilbert space:
\begin{equation}
    \mathcal{H} = \bigoplus_{K,R} \mathcal{H}_{A}^K \otimes \mathcal{H}_{\bar{A}}^{R,K}.
\end{equation}
Note that $\mathcal{O}^{K}_{A}$ and $\mathcal{O}^{R,K}_{\bar{A}}$ are respectively the full algebras on $\mathcal{H}_{A}^K$ and $\mathcal{H}_{\bar{A}}^{R,K}$, which are now are both matter and field Hilbert spaces.

Finally, we project onto the physical Hilbert space by imposing the condition given in Eq.\,\eqref{physical_projection}. Since $R$ now labels the eigenvalues of the full constraints $\hat{\mathcal{C}}^{i,j}_{\rho}$, rather than just the field component as in the case of $r$, this projection is again equivalent to fixing $R = 0$. The resulting decomposition of the physical Hilbert space is
\begin{equation}
    \mathcal{H}_{\text{phy}} = \bigoplus_{K} \mathcal{H}_{A}^K \otimes \mathcal{H}_{\bar{A}}^{0,K} 
    = \bigoplus_{K} \mathcal{H}_{A}^K \otimes \mathcal{H}_{\bar{A}}^K,
\end{equation}
where $K$ labels the edge terms that remain after fixing the constraints. Correspondingly, the local algebras acting on this physical space take the following form:
\begin{equation}
    \mathcal{A}_A = \bigoplus_{K} \mathcal{O}^{K}_{A} \otimes \mathbb{I}^{K}_{\bar{A}}, \qquad
    \bar{\mathcal{A}}_A = \bigoplus_{K} \mathbb{I}^K_A \otimes \mathcal{O}^{K}_{\bar{A}}.
\end{equation}
With this decomposition, the local algebras defined above act block-diagonally and cannot mix different $K$ sectors. As we discussed in Section \ref{sec:ent_loc}, this structure makes it straightforward to extend the notions of local operations, the operational definition of entanglement, and the LOCC theorem.

\section{Detailed calculations for the example of the field mediated entanglement protocol}\label{app:FME}

In this appendix we present the full computations leading to the results employed in Section \ref{sec:FME}.

\subsection{Spatial setup}\label{app:sec:spatial_setup}
We derive here the Operational and Split decompositions specialized to the spatial configuration of the FME protocol (see Fig.~\ref{fig:bmv_regions}) presented in Section~\ref{sec:spatial_setup}.

\paragraph*{Split Decomposition.} 
We here choose to start from deriving the split decomposition in Eq.\,\eqref{split_dec_FME_nospin}. Given that we are dealing with exactly two particles, we can restrict Eq.\,\eqref{split_dec} to the sector with $n=2$. With the substitution $A\rightarrow AB\!=\!A\cup B$, the physical Hilbert space becomes
\begin{equation}
    \mathcal{H}_{\text{phy}} = \bigoplus_{s_2} \ket{s_2}_M \otimes \left( \bigoplus_{k(s_2)} \mathcal{H}_{AB}^{k(s_2)} \otimes \mathcal{H}_{\overline{AB}}^{s_2,k(s_2)} \right),
\end{equation}
where $\overline{AB}$ represents the region complementary to $A\cup B$. The label $s_2$ denotes a matter configuration with one particle in $A$ and one in $B$. Following the sequence-state definition of Eq.\,\eqref{seq_state_def}, such a configuration reads
\begin{equation}
   \ket{s}_M = \ket{0}_M^{0,0}\dots \ket{1}_M^{i,j}\ket{0}_M^{i,j+1}\dots\ket{1}_M^{n,m}\dots\ket{0}_M^{N,N},
   \qquad 
   \begin{cases}
       (i,j)\in A,\\
       (n,m)\in B,
   \end{cases}
\end{equation}
and, since particle number is fixed during the protocol, we drop the subscript ``2'' for brevity.

The local algebra of field operators on $AB$ decomposes as in Eq.\,\eqref{algebra_dec_3}:
\begin{equation}
    \mathcal{A}^F_{AB} = \bigoplus_{k(s)} \mathcal{O}^{f,k(s)}_{AB} \otimes \mathbb{I}^{k}_{\overline{AB}}.
\end{equation}
Because the local field algebras of $A$ and $B$ have been chosen to have trivial intersection, we can further decompose the algebra as
\begin{equation}
    \mathcal{A}^F_{AB} = \bigoplus_{k(s)} \mathcal{O}^{f,k(s)}_{A} \otimes \mathcal{O}^{f,k(s)}_{B} \otimes \mathbb{I}^{k}_{\overline{AB}}.
\end{equation}

Consequently, the physical Hilbert space admits the refined decomposition
\begin{equation}
    \mathcal{H}_{\text{phy}} = \bigoplus_{s} \ket{s}_M \otimes \left( \bigoplus_{k(s)} \mathcal{H}_A^{k(s)} \otimes \mathcal{H}_B^{k(s)} \otimes \mathcal{H}_{\overline{AB}}^{s,k(s)} \right).
\end{equation}
This shows Eq.\,\eqref{split_dec_FME_nospin}.

\paragraph*{Operational Decomposition.} 
We now perform a similar procedure to derive the decomposition in Eq.\,\eqref{oper_dec_FME_nospin}. In this case, we consider the full algebra—including both matter and field degrees of freedom (see Eq.\,\eqref{oper_algebras}) on the joint region $AB$, obtaining
\begin{equation}\label{AB_algebra_app}
        \mathcal{A}_{AB} = \bigoplus_{K} \mathcal{O}^{K}_{AB} \otimes \mathbb{I}^{K}_{\overline{AB}}, \qquad
        \bar{\mathcal{A}}_{AB} = \bigoplus_{K} \mathbb{I}^K_{AB} \otimes \mathcal{O}^{K}_{\overline{AB}}.
\end{equation}
Assuming $\mathcal{A}_A \cap \mathcal{A}_B = \mathbb{I}$, in accordance with the condition of spatial separation between regions $A$ and $B$, we can further decompose $\mathcal{A}_{AB}$ as
\begin{equation}\label{FME_alg_dec_app}
    \mathcal{A}_{AB} = \bigoplus_{K} \mathcal{O}^{K}_{A} \otimes \mathcal{O}^{K}_{B} \otimes \mathbb{I}^{K}_{\overline{AB}}.
\end{equation}
This induces the operational decomposition of the physical Hilbert space:
\begin{equation}
    \mathcal{H}_{\text{phy}} = \bigoplus_{K} \mathcal{H}_{AB}^K \otimes \mathcal{H}_{\overline{AB}}^K 
    = \bigoplus_{K} \mathcal{H}_{A}^K \otimes \mathcal{H}_{B}^K \otimes \mathcal{H}_{\overline{AB}}^K.
\end{equation}
If the sources are well contained inside $A$ and $B$ so as not to affect the edge terms, the label $K$ reduces to pure field edge data, analogous to $k$ in the sourceless decomposition of Eq.\,\eqref{sourceless_dec}.
\subsection{Creating the superposition}\label{app:sec:creating_sup}
In this section we provide the detailed derivation of the gauge-invariant quantum operation $\hat{U}$, which evolves the initial product state at step $(0)$ into a superposition of four spatial branches at step $(1)$ (see Fig.\,\ref{fig:FME_model}). Our goal is to obtain the physical realization of the intuitive transformation introduced in Eq.\,\eqref{unphys_superp}.

\paragraph*{Initial state.}
Before proceeding with the derivation, we analyze in more detail the initial state defined in Eq.\,\eqref{init_state}. As discussed in the main text, the total system initially consists of a product state between the spin, source, and field degrees of freedom. The two spins are prepared in the state $\ket{+} = \frac{1}{\sqrt{2}}(\ket{\uparrow} + \ket{\downarrow})$, while the source is taken to be in configuration $s^0$, where one particle occupies site $(l,a)$ in region $A$ and another occupies site $(l,b)$ in region $B$, i.e.
\begin{equation}\label{s0_app}
    \ket{s^0}_M = \prod_{i,j} \ket{\delta_{i,l}\delta_{j,a} + \delta_{i,l}\delta_{j,b}}^{i,j}_M,
    \qquad
    \text{with} \qquad
    \begin{cases}
        (l,a) \in A, \\
        (l,b) \in B.
    \end{cases}
\end{equation}
This expression can be trivially recast as
\begin{equation}
  \ket{s^0}_M = \ket{1}^{l,a}\ket{1}^{l,b}
  \prod_{\substack{i \neq l \\ j \neq a,b}} \ket{0}^{i,j}
  =: \ket{s^0}_M^A \ket{s^0}_M^B \ket{s^0}_M^{\overline{AB}}.
\end{equation}
We may thus interpret the two particles as lying along the same horizontal line ($i = l$), one in region $A$ ($j = a$) and the other in region $B$ ($j = b$).

Turning to the field, we assume that the preparation time was sufficient for it to relax to its ground state. Recalling the result of Sec.\,\ref{ground_state} for a static classical source, and since the decomposition was carried out in the $p$ basis, we write the ground-state wave functional as
\begin{equation}
    \ket{\psi_{s^0}^0}_F = \int \mathcal{D}p\,
    \Psi_{0,\rho(s^0)}[p] \prod_{i,j} \ket{p_x^{i,j}, p_y^{i,j}},
\end{equation}
which depends on the matter configuration $s^0$ through the associated matter density $\rho^{i,j}$.

Within the Split Decomposition framework, the initial state of the total system can therefore be written explicitly as
\begin{equation}\label{init_state_app}
    \ket{\Psi}^{(0)} =
    \ket{+}_{\sigma}^A \ket{+}_{\sigma}^B
    \ket{s^0}_M
    \ket{\psi_{s^0}^0}_F.
\end{equation}

\paragraph*{Gauge transformations in the quantized theory.} We now turn to the construction of operator $\hat{U}$. For such an operation to be physically meaningful, it must commute with the constraints of the theory, or equivalently, remain invariant under arbitrary gauge transformations. 

We first need to adapt the notion of gauge transformations generated by constraints, previously discussed in the classical theory (see Tab.\,\ref{tab:g_trans} and Appendix \ref{app:gauge_inv}), to the quantized model. This is straightforward, as quantization replaces Poisson brackets with commutators according to following the correspondence,
\begin{equation}
    \{\,\cdot\,,\,\cdot\,\} \rightarrow -\frac{\text{i}}{\hbar} [\,\cdot\,,\,\cdot\,],
\end{equation}
in accordance with the canonical quantization procedure.

Starting from the classical equivalent of Eq.\,\eqref{classical_gauge_transf_app}, this implies that in the quantum model, the variation of an operator $\hat{X}$ under an infinitesimal gauge transformation generated by a constraint $\hat{\phi}$ is given by
\begin{equation}
    \delta \hat{X} = -\frac{\text{i}}{\hbar} \epsilon [\hat{X}, \hat{\phi}] 
    = -\frac{\text{i}}{\hbar} \epsilon \left( \hat{X}\hat{\phi} - \hat{\phi}\hat{X} \right),
\end{equation}
where $\epsilon$ is the (infinitesimal) parameter of the transformation.

As an example, we examine the behavior of $\hat{q}_s^{i,j}$ under a gauge transformation generated by the constraint of our model, which will prove to be of central importance. Following the same reasoning as in Appendix~\ref{app:gauge_inv}, this reduces to computing the cumulative effect of the gauge transformations generated by the constraints $\hat{\mathcal{C}}^{i,j}_{\rho}$ across the entire lattice, i.e.
\begin{equation}
    \delta\hat{q}_s^{i,j} = -\frac{\text{i}}{\hbar} \sum_{n,m} \theta^{n,m} \left[ \hat{q}_s^{i,j}, \hat{\mathcal{C}}^{n,m}_{\rho} \right].
\end{equation}
To proceed, we evaluate the commutators in the expression above. Using the definition of the constraint, obtaining
\begin{equation}
    \left[ \hat{q}_s^{i,j}, \hat{\mathcal{C}}^{n,m}_{\rho} \right] 
    = \sum_r \left[ \hat{q}_s^{i,j}, \bar{\partial}^{\{n,m\}}_r \hat{p}^{n,m}_r \right] 
    + \left[ \hat{q}_s^{i,j}, \hat{\rho}^{n,m} \right] 
    = \sum_r \bar{\partial}_r \left[ \hat{q}_s^{i,j}, \hat{p}^{n,m}_r \right],
\end{equation}
where we used the fact that field and matter operators act on separate Hilbert spaces and hence commute, as well as the linearity of the commutator to pull the discrete derivative outside. Applying the canonical commutation relations, we find:
\begin{equation}
    \left[ \hat{q}_s^{i,j}, \hat{\mathcal{C}}^{n,m}_{\rho} \right] 
    = \text{i}\hbar \sum_r \bar{\partial}^{\{n,m\}}_r \left( \delta_{s,r} \delta^{i,n} \delta^{j,m} \right) 
    = \text{i}\hbar\, \bar{\partial}^{\{n,m\}}_s \left( \delta^{i,n} \delta^{j,m} \right),
\end{equation}
where in the last equality we simply applied the Kronecker deltas in the index $r$. Substituting this result into the expression for $\delta \hat{q}_s^{i,j}$ yields:
\begin{equation}
    \delta\hat{q}_s^{i,j} = \sum_{n,m} \theta^{n,m} \bar{\partial}^{\{n,m\}}_s \left( \delta^{i,n} \delta^{j,m} \right).
\end{equation}
By applying the identity for discrete derivatives (see Eq.\,\eqref{diff_dev2}), we finally arrive at:
\begin{equation}
    \delta\hat{q}_s^{i,j} = -\,\bar{\partial}_s \theta^{i,j},
\end{equation}
which, as expected, matches the classical result in Tab.\,\ref{tab:g_trans}, and obtained in Eq.\,\eqref{q_gauge_tranf}. Furthermore, since the transformation is linear in the parameter, a finite transformation can be constructed by composing infinitesimal ones. This leads to a total transformation parameter given by a sum of the $\theta^{i,j}$, which we denote as $\nu^{i,j}$, and which is truly arbitrary. Thus, the general form of a gauge transformation on $\hat{q}_s^{i,j}$ is
\begin{equation}\label{q_gauge_tranf_quantum}
    \hat{q}_s^{i,j} \rightarrow \hat{q}_s^{i,j} - \bar{\partial}_s \nu^{i,j}.
\end{equation}

\paragraph*{Acting on matter.}
Now that we have a method for determining how quantities transform under gauge transformations, we can begin investigating the operations required to act on the system. We start by examining the effect that the operator $\hat{U}$ must have on the matter degrees of freedom. Since $\hat{U}$ must produce branches in which the sources are displaced along the $x$-axis of the lattice, an essential component of this operator must involve moving a particle from one site to another.

To formalize this, we introduce the creation and annihilation operators acting on the matter Hilbert space. The annihilation operator is defined as
\begin{equation}
    \hat{a}_{i,j} := \ket{0}\!\bra{1}^{i,j}_M,
\end{equation}
which intuitively removes a particle at site $(i,j)$. The corresponding creation operator is given by the Hermitian conjugate
\begin{equation}
    \hat{a}^\dagger_{i,j} := \ket{1}\!\bra{0}^{i,j}_M.
\end{equation}
It creates a particle at site $(i,j)$.

With these definitions, it is straightforward to identify the operator responsible for generating spatial displacement between branches of $\hat{U}$ as $\hat{a}^\dagger_{n,m} \hat{a}_{i,j}$. This operator effectively moves a particle from site $(i,j)$ to site $(n,m)$. It forms a key building block of our eventual gauge-invariant operator and, when controlled by the spin, can be responsible for a transformation analogous to Eq.\,\eqref{unphys_superp}. However, as previously mentioned, this operation is not gauge-invariant on its own. To see this, we compute the variation of $\hat{a}_{i,j}$ and $\hat{a}^\dagger_{i,j}$ under general gauge transformations.

As with $\hat{q}_s^{i,j}$, we begin by evaluating the change induced by an infinitesimal gauge transformation generated by all the constraints $\hat{\mathcal{C}}^{n,m}_\rho$ on the lattice:
\begin{equation}
    \delta \hat{a}_{i,j} = -\frac{\text{i}}{\hbar} \sum_{n,m} \theta^{n,m} \left[ \hat{a}_{i,j}, \hat{\mathcal{C}}^{n,m}_\rho \right],
\end{equation}
where $\theta^{n,m}$ are the infinitesimal transformation parameters. Evaluating the commutator, we find:
\begin{equation}
    \left[ \hat{a}_{i,j}, \hat{\mathcal{C}}^{n,m}_\rho \right] = 
    \sum_r \left[ \hat{a}_{i,j}, \bar{\partial}^{\{n,m\}}_r \hat{p}_r^{n,m} \right] 
    + \left[ \hat{a}_{i,j}, \hat{\rho}^{n,m} \right] 
    = \left[ \hat{a}_{i,j}, \hat{\rho}^{n,m} \right],
\end{equation}
since, once again, matter and field operators commute. Using the definition of $\hat{a}_{i,j}$ and the expression for the matter density operator $\hat{\rho}^{n,m}$ from Eq.\,\eqref{matter_dens}, we find:
\begin{equation}
    \left[ \hat{a}_{i,j}, \hat{\mathcal{C}}^{n,m}_\rho \right] 
    = \ket{0}\!\bra{1}^{i,j}_M \cdot \ket{1}\!\bra{1}^{n,m}_M 
    - \ket{1}\!\bra{1}^{n,m}_M \cdot \ket{0}\!\bra{1}^{i,j}_M 
    = \delta_{i,n} \delta_{j,m} \hat{a}_{n,m},
\end{equation}
where we used the orthonormality of the spin basis and orthogonality across different sites. Plugging this result into the expression for the variation yields
\begin{equation}
    \delta \hat{a}_{i,j} = -\frac{\text{i}}{\hbar} \sum_{n,m} \theta^{n,m} \delta_{i,n} \delta_{j,m} \hat{a}_{n,m} 
    = -\frac{\text{i}}{\hbar} \theta^{i,j} \hat{a}_{i,j}.
\end{equation}
Hence, under an infinitesimal gauge transformation, the annihilation operator transforms as:
\begin{equation}
    \hat{a}_{i,j} \rightarrow \left( 1 - \frac{\text{i}}{\hbar} \theta^{i,j} \right) \hat{a}_{i,j}.
\end{equation}
This is a multiplicative transformation, so we can derive the finite transformation using an exponentiation trick used in standard quantum field theory. Define a small parameter $\theta^{i,j} := \frac{\nu^{i,j}}{T}$ with $T \gg \nu^{i,j}$ and $\nu^{i,j}$ completely arbitrary. Applying $T$ infinitesimal transformations and taking the limit as $T \to \infty$ gives:
\begin{equation}
    \hat{a}_{i,j} \rightarrow \left(1 - \frac{\text{i}}{\hbar} \frac{\nu^{i,j}}{T} \right)^T \hat{a}_{i,j} 
    \xrightarrow{T \to \infty} \text{e}^{-\frac{\text{i}}{\hbar} \nu^{i,j}} \hat{a}_{i,j},
\end{equation}
where we just used the well known limit expression for the exponential.
By Hermitian conjugation, we find the transformation of the creation operator:
\begin{equation}
    \hat{a}^\dagger_{i,j} \rightarrow \text{e}^{\frac{\text{i}}{\hbar} \nu^{i,j}} \hat{a}^\dagger_{i,j}.
\end{equation}
Thus, a shift operation of the form $\hat{a}^\dagger_{n,m} \hat{a}_{i,j}$ transforms under a gauge transformation as:
\begin{equation}\label{g_transf_move}
    \hat{a}^\dagger_{n,m} \hat{a}_{i,j} \rightarrow 
    \text{e}^{-\frac{\text{i}}{\hbar}(\nu^{i,j} - \nu^{n,m})} 
    \hat{a}^\dagger_{n,m} \hat{a}_{i,j}.
\end{equation}
This result explicitly shows that such an operator is not gauge-invariant, as it acquires a nontrivial phase under general gauge transformations. Therefore, it cannot represent a physical operation on its own. In order for such an operation to be acceptable it needs to be \emph{dressed} with an operator on the field that accounts for such a phase.

\paragraph*{Constructing the gauge-invariant operator.} To resolve the issue outlined above, we must construct an operator acting on the field, denoted by $\hat{W}^{i,j}_F$, that compensates for the gauge-dependent phase appearing in Eq.\,\eqref{g_transf_move}. This is straightforward if we recall the transformation behavior derived in Eq.\,\eqref{q_gauge_tranf_quantum}.

Since our objective is to generate spatial configurations in which matter sources are displaced along the $x$-axis, let us consider the following operator on the matter degrees of freedom, corresponding to moving a particle left by two sites, and analyze its gauge transformation as\footnote{The reason we choose to move the particle by two sites is due to geometric considerations related to the discrete derivative definition we adopt, which prevents the construction of an appropriate dressing for a shift of just one site.} 
\begin{equation}\label{move_particle}
    \hat{a}^\dagger_{l,a-2} \hat{a}_{l,a} \rightarrow 
    \text{e}^{-\frac{\text{i}}{\hbar}(\nu^{l,a} - \nu^{l,a-2})} 
    \hat{a}^\dagger_{l,a-2} \hat{a}_{l,a},
\end{equation}
where the indices $l$ and $a$ are chosen to maintain consistency with Eq.\,\eqref{s0_app}. It is now evident that the simplest way to cancel the arising phase is to define the dressing operator as
\begin{equation}\label{dressing}
  \hat{W}^{i,j}_F := \text{e}^{-\frac{\text{i}}{\hbar}2a\,\hat{q}^{i,j}_x}.
\end{equation}
Using the result from Eq.\,\eqref{q_gauge_tranf_quantum}, and explicitly applying the discrete derivative, we in fact find that
\begin{equation}
 \hat{q}^{l,a-1}_x \rightarrow \hat{q}^{l,a-1}_x - \bar{\partial}_x \nu^{l,a-1} 
 = \hat{q}^{l,a-1}_x - \frac{1}{2a}(\nu^{l,a} - \nu^{l,a-2}).
\end{equation}
Substituting this into the transformation of the dressing operator, we obtain
\begin{equation}
   \hat{W}^{l,a-2}_F \rightarrow \text{e}^{\frac{\text{i}}{\hbar}(\nu^{l,a} - \nu^{l,a-2})} \, \hat{W}^{l,a-2}_F.
\end{equation}
Consequently, the full dressed operator transforms as
\begin{equation}
  \hat{a}^\dagger_{l,a-2} \hat{a}_{l,a} \hat{W}^{l,a-1}_F \rightarrow 
  \text{e}^{-\frac{\text{i}}{\hbar}(\nu^{l,a} - \nu^{l,a-2})}
  \text{e}^{\frac{\text{i}}{\hbar}(\nu^{l,a} - \nu^{l,a-2})} 
  \hat{a}^\dagger_{l,a-2} \hat{a}_{l,a} \hat{W}^{l,a-1}_F 
  = \hat{a}^\dagger_{l,a-2} \hat{a}_{l,a} \hat{W}^{l,a-1}_F.
\end{equation}
Thus, the dressed operator $\hat{a}^\dagger_{l,a-2} \hat{a}_{l,a} \hat{W}^{l,a-1}_F$ is gauge-invariant and therefore corresponds to a physical operation, consistently mapping physical states to physical states.

A simple and intuitive interpretation of the physical operator we have just derived can be visualized using our graphical notation. First, observe that the dressing operator in Eq.\,\eqref{dressing} effectively implements a translation of the field variable $\hat{p}^{l,a-1}_x$ by an amount $2a$. This shift of the field degrees of freedom is precisely what is required to counterbalance the change in matter configuration, as encoded in the operator $\hat{\rho}^{i,j}$, and to ensure that the full constraint $\hat{\mathcal{C}}^{i,j}_{\rho}$ remains satisfied in the final state. 

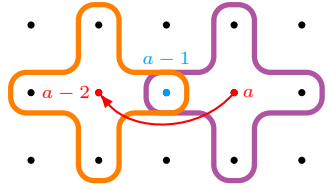
\begin{figure}[ht]
    \centering
    \begin{tikzpicture}[scale=0.9]
        \def\rows{3}
        \def\cols{5}

        \foreach \x in {1,...,\cols} {
            \foreach \y in {1,...,\rows} {
                \fill[black] (\x,\y) circle (1.5pt);
            }
        }
        \draw[mypurple,rounded corners=6pt, line width=2pt]
         (4, 3.3) -- (4.3, 3.3) -- (4.3, 2.3) -- (5.3, 2.3) -- (5.3, 1.7)--(4.3,1.7)--(4.3,0.7)--(3.7,0.7)--(3.7,1.7)--(2.7,1.7)--(2.7,2.3)--(3.7,2.3)--(3.7,3.3)--(4,3.3);
        \draw[orange,rounded corners=6pt, line width=2pt]
         (2, 3.3) -- (2.3, 3.3) -- (2.3, 2.3) -- (3.3, 2.3) -- (3.3, 1.7)--(2.3,1.7)--(2.3,0.7)--(1.7,0.7)--(1.7,1.7)--(0.7,1.7)--(0.7,2.3)--(1.7,2.3)--(1.7,3.3)--(2,3.3);
         
         \fill[red] (2,2) circle (1.5pt) ;
         \node[text=red, left] at (2,2){\scriptsize $a-2$};  

         \fill[red] (4,2) circle (1.5pt) ;
         \node[text=red, right] at (4,2){\scriptsize $a$};  

         \fill[mycyan] (3,2) circle (1.5pt) ;
         \node[text=mycyan] at (3,2.5){\scriptsize $a-1$}; 
         \draw[thinarrow, thick, red, bend left=50] (4,2) to (2,2);
    \end{tikzpicture}
    \caption{Graph showing the action of $\hat{a}^\dagger_{l,a-2} \hat{a}_{l,a} \hat{W}^{l,a-1}_F$. The red arrow represents the operation of moving a particle between the two sites, while the field variable $p_x^{i,j}$ at the blue point is shifted by $2a$. The two crosses indicate the constraints affected by the presence of the source before and after the process.}
    \label{fig:dressed_operaton}
\end{figure}

Fig.\,\ref{fig:dressed_operaton} illustrates this process. The red arrow represents the action of the operator $\hat{a}^\dagger_{l,a-2} \hat{a}_{l,a}$, which moves a particle from site $(l,a)$ to site $(l,a-2)$. The purple and orange crosses indicate the only constraints affected by the presence of the matter source before and after the operation, respectively. When the particle is annihilated at site $a$, the corresponding purple constraint must be adjusted accordingly to maintain physical consistency. Similarly, the creation of a particle at site $a-2$ necessitates a complementary adjustment of the orange constraint.

Only for intuition purposes, we can briefly examine the eigenvalues of the purple and orange constraints before and after the operation:
\begin{equation}
    \begin{cases}
        \mathcal{C}_{\rho}^{l,a} = \mathcal{C}^{l,a} + 1 = 0\\
        \mathcal{C}_{\rho}^{l,a-2} = \mathcal{C}^{l,a-2} = 0
    \end{cases}\,
    \xrightarrow{\hat{a}^\dagger_{l,a-2} \hat{a}_{l,a} \hat{W}^{l,a-1}_F}
    \,\begin{cases}
        \tilde{\mathcal{C}}_{\rho}^{l,a} = \tilde{\mathcal{C}}^{l,a} = 0\\
        \tilde{\mathcal{C}}_{\rho}^{l,a-2} = \tilde{\mathcal{C}}^{l,a-2} + 1 = 0
    \end{cases}.
\end{equation}
This implies that the field part of the constraint must transform as follows:
\begin{equation}
    \begin{cases}
        \tilde{\mathcal{C}}^{l,a} = \mathcal{C}^{l,a} + 1\\
        \tilde{\mathcal{C}}^{l,a-2} = \mathcal{C}^{l,a-2} - 1
    \end{cases}.
\end{equation}
By looking at the graphical definition of $\mathcal{C}^{i,j}$ in Graph~\ref{graph_b_C}, it easy to notice that $\hat{p}^{l,a-1}_x$, represented by the blue point in Fig.\,\ref{fig:dressed_operaton}, is the only field component that appears in both constraints, contributing with opposite signs, and does not appear in any other constraint. Taking into account the factor of $2a$ that relates the field contributions to the matter terms via the discrete derivatives, it becomes clear that the shift introduced by the dressing operator in Eq.\,\eqref{dressing} is precisely the correction required to keep all the constraint fulfilled.

Finally, it is easy to notice that if we instead consider the operator that moves the particle to the right, its gauge transformation is given by:
\begin{equation}
    \hat{a}^\dagger_{l,a} \hat{a}_{l,a-2} \rightarrow 
    \text{e}^{-\frac{\text{i}}{\hbar}(\nu^{l,a-2} - \nu^{l,a})} 
    \hat{a}^\dagger_{l,a} \hat{a}_{l,a-2},
\end{equation}
which exhibits a phase factor that is precisely the opposite of the one in Eq.\,\eqref{move_particle}. This implies that the corresponding dressing operator must be the inverse of the one defined in Eq.\,\eqref{dressing}, namely:
\begin{equation}\label{inv_dressing}
  (\hat{W}^{i,j}_F)^{-1} = \text{e}^{\frac{\text{i}}{\hbar}2a\,\hat{q}^{i,j}_x}.
\end{equation}
\paragraph*{Obtaining the superposed physical state.}
Now that we have constructed the basic building block needed to move particles between two sites while conserving the constraints, we can write down the desired form of the operator $\hat{U}$. Our goal is to generate a superposition of four branches, conditioned on the spin state. 

First, we note that the physical operations we derived, $\hat{a}^\dagger_{l,a-2} \hat{a}_{l,a} \hat{W}^{l,a-1}_F$ for example, is spatially local. This locality is evident: not only do the matter operators act locally by definition, but also the dressing operator $\hat{W}^{l,a-1}_F$, defined in Eq.\,\eqref{dressing}, acts only on the field at site $a-1$. Therefore, if $(l,a), (l,a-1), (l,a-2) \in A$, this gauge-invariant operator respects all the conditions that define the accessible local algebra on $A$, as given in Eq.\,\eqref{source_algerba_def}.

Following the transformation outlined in Eq.\,\eqref{unphys_superp}, and using the definitions of the dressing operators for left and right particle shifts in Eqs.\,\eqref{dressing} and\/\eqref{inv_dressing}, we define the unitary operator associated with region $A$ as
\begin{equation}\label{U_A}
    \hat{U}_A = \ket{\uparrow}\!\bra{\uparrow}^A_{\sigma} \, \hat{a}^\dagger_{l,a-2} \hat{a}_{l,a} \hat{W}^{l,a-1}_F + 
               \ket{\downarrow}\!\bra{\downarrow}^A_{\sigma} \, \hat{a}^\dagger_{l,a+2} \hat{a}_{l,a} (\hat{W}^{l,a+1}_F)^{-1} \in \mathcal{A}^{\text{tot}}_A,
\end{equation}
with $(l,[a-2,a+2]) \subset A$ and $\mathcal{A}^{\text{tot}}_A$ defined by the decomposition in Eq.\,\eqref{local_algebras_FME}, since the operator now acts entirely within region $A$ and includes the spin as well.

Analogously, we define the operator on region $B$ as
\begin{equation}\label{U_B}
    \hat{U}_B = \ket{\uparrow}\!\bra{\uparrow}^B_{\sigma} \, \hat{a}^\dagger_{l,b-2} \hat{a}_{l,b} \hat{W}^{l,b-1}_F + 
               \ket{\downarrow}\!\bra{\downarrow}^B_{\sigma} \, \hat{a}^\dagger_{l,b+2} \hat{a}_{l,b} (\hat{W}^{l,b+1}_F)^{-1} \in \mathcal{A}^{\text{tot}}_B,
\end{equation}
with $(l,[b-2,b+2]) \subset B$ and again $\mathcal{A}^{\text{tot}}_B$ given by Eq.\,\eqref{local_algebras_FME}.

Thus, the full operator generating the desired coherent physical superposition of four branches is simply
\begin{equation}
    \hat{U} = \hat{U}_A \otimes \hat{U}_B,
\end{equation}
which produces the evolved physical state
\begin{equation}
    \ket{\Psi}^{(1)} = \hat{U} \ket{\Psi}^{(0)},
\end{equation}
such that $ \ket{\Psi}^{(1)}$ now fulfills the constraint in Eq.\,\eqref{physical_projection} and thus it belongs to $\mathcal{H}_{\text{phy}}$.

\subsection{Generating Entanglement}\label{app:sec:entanglement_generation}
Having identified the correct gauge-invariant operation capable of generating the superposition, we can proceed with the analysis of the protocol.

\paragraph*{Superposed state.} We now perform the detailed computation leading to the state at step (1) in Eq.\,\eqref{state_step1} by directly applying $\hat{U}$. As anticipated, the action of $\hat{U}$ corresponds to the transformation outlined in Eq.\,\eqref{unphys_superp}, but now includes the additional dressing of the quantum field. We can naturally define the following matter states:
\begin{equation}
        \ket{s^L}_M^{A} = \hat{a}^\dagger_{l,a-2} \hat{a}_{l,a} \ket{s^0}_M^{A}, \qquad
        \ket{s^R}_M^{A} = \hat{a}^\dagger_{l,a+2} \hat{a}_{l,a} \ket{s^0}_M^{A}.
\end{equation}
Proceeding analogously for region $B$, we define the four global matter configurations produced by the process:
\begin{equation}
\begin{split}    
    \ket{s^{LL}}_M &= \ket{s^L}_M^{A} \ket{s^L}_M^{B} \ket{s^0}_M^{\overline{AB}}, \qquad 
    \ket{s^{RR}}_M = \ket{s^R}_M^{A} \ket{s^R}_M^{B} \ket{s^0}_M^{\overline{AB}}, \\
    \ket{s^{LR}}_M &= \ket{s^L}_M^{A} \ket{s^R}_M^{B} \ket{s^0}_M^{\overline{AB}}, \qquad 
    \ket{s^{RL}}_M = \ket{s^R}_M^{A} \ket{s^L}_M^{B} \ket{s^0}_M^{\overline{AB}}.
\end{split}
\end{equation}
Each of these matter configurations leads to a differently dressed field state, depending on the direction in which the particles move. For example, the state associated with the $LL$ configuration reads
\begin{equation}
    \ket{s^{LL}}_M \ket{\psi'_{s^{LL}}}_F = 
    \hat{a}^\dagger_{l,a-2} \hat{a}_{l,a} \, \hat{a}^\dagger_{l,b-2} \hat{a}_{l,b} \ket{s^0}_M 
    \otimes \hat{W}_F^{l,a-1} \hat{W}_F^{l,b-1} \ket{\psi_{s^0}^0}_F,
\end{equation}
and similarly for the configurations $RR$, $LR$, and $RL$.

Controlling these operations via the spin states, as specified by Eqs.\,\eqref{U_A} and \eqref{U_B}, we find that the state at time $(1)$ is
\begin{equation}
\begin{split}
    \ket{\Psi}^{(1)} = \hat{U} \ket{\Psi}^{(0)} = &\,
    \ket{\uparrow}^A_{\sigma} \ket{\uparrow}^B_{\sigma} \ket{s^{LL}}_M \ket{\psi'_{s^{LL}}}_F 
    + \ket{\uparrow}^A_{\sigma} \ket{\downarrow}^B_{\sigma} \ket{s^{LR}}_M \ket{\psi'_{s^{LR}}}_F \\
    &+ \ket{\downarrow}^A_{\sigma} \ket{\uparrow}^B_{\sigma} \ket{s^{RL}}_M \ket{\psi'_{s^{RL}}}_F 
    + \ket{\downarrow}^A_{\sigma} \ket{\downarrow}^B_{\sigma} \ket{s^{RR}}_M \ket{\psi'_{s^{RR}}}_F.
\end{split}
\end{equation}

Introducing the shorthand notation $s \in \{s^{LL}, s^{LR}, s^{RL}, s^{RR}\}$ for the four matter configurations, we can write the state compactly as
\begin{equation}
    \ket{\Psi}^{(1)} = \sum_s \ket{\sigma(s)}^{AB}_{\sigma} \ket{s}_M \ket{\psi'_s}_F,
\end{equation}
where we defined $\sigma(s)\in\{\uparrow\uparrow,\uparrow\downarrow,\downarrow\uparrow,\downarrow\downarrow\}$.

\paragraph{Relaxation and time evolution.}
After the dressing step, each conditional field state $\ket{\psi'_s}_F$ is generally excited. We let the field relax to its configuration-dependent ground state, yielding
\begin{equation}\label{state_step2_app}
\ket{\Psi}^{(2)} = \sum_s e^{i\gamma(s)} \ket{\sigma(s)}^{AB}_{\sigma} \ket{s}_M \ket{\psi^0_s}_F.
\end{equation}
where the phases $\gamma(s)$ arise from the non-adiabatic preparation of the superposition (they vanish in the adiabatic/continuum limit and do not affect our conclusions below).
Between steps $(2)$ and $(3)$ the state evolves for a time interval $\tau=t^{(3)}-t^{(2)}$ under the quantized Hamiltonian, which acts only on the field:
\begin{equation}\label{state_3_app}
    \ket{\Psi}^{(3)} 
    = \exp\!\left(-\text{i}\frac{\hat{H}}{\hbar}\tau\right)\!\ket{\Psi}^{(2)} 
    = \sum_s \text{e}^{i[\gamma(s) + \phi(s)]} 
    \ket{\sigma(s)}^{AB}_{\sigma} \ket{s}_M \ket{\psi^0_s}_F.
\end{equation}
with an additional relative phase
\begin{equation}
\phi(s):=-\frac{\mathcal{E}_{\rho(s)}-\mathcal{E}_0}{\hbar}\,\tau,
\end{equation}
determined by the difference between the field ground-state energies conditioned on the matter configuration $s$ and on the vacuum, respectively. This expression matches the structure found in related analyses, e.g., Ref.~\cite{Giacomini2023quantumstatesof}.

\paragraph*{Merging the superposition.}
As outlined in the main text, after the relaxation and time-evolution steps~(2) and~(3), the source state must be recombined to transfer the entanglement onto the spins and enable its detection. This is again done in two steps: first we apply the merging operator $\hat{U}'$. Such operation has a structure completely analogous to $\hat{U}$ and essentially undoes the operation outlined in Eq.\,\eqref{unphys_superp} on matter, i.e.
\begin{equation}
        \ket{\uparrow}_{\sigma}^{\bullet} \ket{s^L}_M^{\bullet} \rightarrow \ket{\uparrow}_{\sigma}^{\bullet} \ket{s^0}_M^{\bullet}, \qquad
        \ket{\downarrow}_{\sigma}^{\bullet} \ket{s^R}_M^{\bullet} \rightarrow \ket{\downarrow}_{\sigma}^{\bullet} \ket{s^0}_M^{\bullet}.
\end{equation}
Therefore, it essentially maps every $\ket{s}_M$ back to the original matter configuration $\ket{s^0}_M$, leaving the spin state untouched. Obviously, for it to be physically accessible it must be gauge-invariant and thus have a dressing analogous to the one of $\hat{U}$, which adjusts the field state accordingly. In particular, the role of $\hat{W}_F$ this time is to bring the field state back to the one relative to the matter configuration $\ket{s^0}_M$.

The resulting physical operators on $A$ and $B$ have the following form
\begin{equation}\label{merging_op}
    \begin{split}
        \hat{U}'_A &= \ket{\uparrow}\!\bra{\uparrow}^A_{\sigma} \, \hat{a}^\dagger_{l,a} \hat{a}_{l,a-2} (\hat{W}^{l,a-1}_F)^{-1} + 
               \ket{\downarrow}\!\bra{\downarrow}^A_{\sigma} \, \hat{a}^\dagger_{l,a} \hat{a}_{l,a+2} \hat{W}^{l,a+1}_F \in \mathcal{A}^{\text{tot}}_A, \\
        \hat{U}'_B &= \ket{\uparrow}\!\bra{\uparrow}^B_{\sigma} \, \hat{a}^\dagger_{l,b} \hat{a}_{l,b-2} (\hat{W}^{l,b-1}_F)^{-1} + 
               \ket{\downarrow}\!\bra{\downarrow}^B_{\sigma} \, \hat{a}^\dagger_{l,b} \hat{a}_{l,b+2} \hat{W}^{l,b+1}_F \in \mathcal{A}^{\text{tot}}_B,
    \end{split}
\end{equation}
which are obviously accessible local operators as well as $\hat{U}_A$ and $\hat{U}_B$. Thus the full operator is just given by $\hat{U}' = \hat{U}'_A \otimes \hat{U}'_B$. Applying this to $\ket{\Psi}^{(3)}$, we obtain
\begin{equation}
    \ket{\Psi}^{(4)} = \hat{U}' \ket{\Psi}^{(3)} = \ket{s^0}_M \sum_s \text{e}^{i\gamma(s)} \, \text{e}^{i\phi(s)} 
    \ket{\sigma(s)}^{AB}_{\sigma} \ket{\psi'_{s^0}(s)}_F,
\end{equation}
where $\ket{\psi'_{s^0}(s)}_F$ are again excited states of the field. These in general depend on the previous matter configurations through the specific dressing they undergo, i.e.
\begin{equation}
    \ket{\psi'_{s^0}(s)}_F = \hat{W}_F(s) \ket{\psi^0_s}_F,
\end{equation}
where $\hat{W}_F(s)$ stands for one of the four possible dressing operations arising from the action of $\hat{U}'$. We thus ultimately recover exactly the state given in Eq.\,\eqref{state_step4}.

\subsection{Generalizing the argument.}\label{app:sec:gen_BMV}
Here we present the calculations that lead to the generalized FME argument and the result in Eq.\,\eqref{ent_diff}. We begin by restating the bipartition introduced in the main text: system $L$ comprises the spin, matter and field degrees of freedom in region $A$, while system $R$ comprises the corresponding components in region $B$ together with the rest of the lattice,
\begin{equation}\label{bip_def_app}
        L : \sigma_A + M_A + F_A, \qquad
        R : \sigma_B + M_B + F_B + M_{\overline{AB}} + F_{\overline{AB}}.
\end{equation}
Since sources are present only in $A$ and $B$, matter degrees of freedom in the rest $M_{\overline{AB}}$ are trivial and are included only for completeness. The partition is conventional: assigning the rest to $L$ instead would not alter the argument.

Under this bipartition the Operational Decomposition becomes
\begin{equation}
    \mathcal{H}_{\text{phy}} = \bigoplus_{K} \mathcal{H}_{L}^{K,\text{tot}} \otimes \mathcal{H}_{R}^{K,\text{tot}},
\end{equation}
with $\mathcal{H}_{L}^{K,\text{tot}}:=\mathcal{H}_{A}^{K,\text{tot}}$ and $\mathcal{H}_{R}^{K,\text{tot}}:=\mathcal{H}_{B}^{K,\text{tot}}\otimes\mathcal{H}_{\overline{AB}}^K$. The associated local algebras read
\begin{equation}
        \mathcal{A}_{L}^{\text{tot}} = \bigoplus_{K} \mathcal{O}^{K,\text{tot}}_{A} \otimes \mathbb{I}^{K}_B \otimes \mathbb{I}^{K}_{\overline{AB}} = \mathcal{A}_{A}^{\text{tot}}, \qquad
        \mathcal{A}_{R}^{\text{tot}} = \bigoplus_{K} \mathbb{I}^{K}_A \otimes \mathcal{O}^{K,\text{tot}}_{B} \otimes \mathcal{O}^{K}_{\overline{AB}} = \mathcal{A}_{B}^{\text{tot}} \vee \bar{\mathcal{A}}_{AB},
\end{equation}
where $\mathcal{A}_{B}^{\text{tot}} \vee \bar{\mathcal{A}}_{AB}$ denotes the joint algebra on $B$ and $\overline{AB}$, and $\bar{\mathcal{A}}_{AB}$ is the gauge-invariant subalgebra in Eq.\,\eqref{AB_algebra_app}.

With this structure the operators we implement at the protocol endpoints remain local within each $K$-sector, i.e.
\begin{equation}
        \hat{U}_A, \hat{U}'_A \in \mathcal{A}^{\text{tot}}_A = \mathcal{A}_L^{\text{tot}}, \qquad
        \hat{U}_B, \hat{U}'_B \in \mathcal{A}^{\text{tot}}_B \subset \mathcal{A}_R^{\text{tot}}.
\end{equation}
Hence the LOCC theorem applies independently in each sector $\mathcal{H}_{L}^{K,\text{tot}}\otimes\mathcal{H}_{R}^{K,\text{tot}}$: if entanglement increases by the end of the protocol, some form of non-classical communication must have occurred. Because the increase is detectable by standard local spin measurements, it corresponds to accessible entanglement within the local tensor product structure of the corresponding sector.

To demonstrate this explicitly, recall the initial and final states of Eq.\,\eqref{start_to_final}. Under the $L$–$R$ decomposition they read
\begin{equation}
        \ket{\Psi}^{(0)} = \ket{+}_{\sigma}^L \ket{+}_{\sigma}^R \ket{\psi^0}_{F,M}^{LR}, \qquad
        \ket{\Psi}^{(5)} = \ket{\chi}_{\sigma}^{LR} \ket{\psi^0}_{F,M}^{LR}.
\end{equation}
Operationally accessible entanglement corresponds to conventional bipartite entanglement within each $K$-sector. Projecting to a specific sector yields
\begin{equation}
        \ket{\Psi}^{(0)}_K = \ket{+}_{\sigma}^L \ket{+}_{\sigma}^R \ket{\psi^0}_{F,M}^{LR,K}, \qquad
        \ket{\Psi}^{(5)}_K = \ket{\chi}_{\sigma}^{LR} \ket{\psi^0}_{F,M}^{LR,K},
\end{equation}
where the spin states are independent of $K$. Each $\ket{\Psi}^{(i)}_K$ thus defines a pure state on $\mathcal{H}_L^{K,\text{tot}} \otimes \mathcal{H}_R^{K,\text{tot}}$, and the entanglement between $L$ and $R$ can be quantified via the von Neumann entropy of either reduced state \cite{Nielsen_Chuang_2010}.

For a subsystem $S$ in state $\hat{\rho}$, the von Neumann entropy is
\begin{equation}
    H(S)_{\rho} := -\text{Tr}(\hat{\rho} \ln \hat{\rho}),
\end{equation}
where $\text{Tr}(\bullet)$ denotes the trace over the relevant Hilbert space.

The density matrices of the initial and final states are
\begin{equation}
        \hat{\rho}^{(0)}_K = \ket{+}\!\bra{+}_{\sigma}^L \otimes \ket{+}\!\bra{+}_{\sigma}^R \otimes \ket{\psi^0}\!\bra{\psi^0}_{F,M}^{LR,K}, \qquad
        \hat{\rho}^{(5)}_K = \ket{\chi}\!\bra{\chi}_{\sigma}^{LR} \otimes \ket{\psi^0}\!\bra{\psi^0}_{F,M}^{LR,K}.
\end{equation}

The entanglement at the beginning of the protocol is
\begin{equation}
    \text{ent}_{LR,K}^{(0)} = H(R)_{\rho^{(0)}_K} = H(L)_{\rho^{(0)}_K}.
\end{equation}
Tracing out $R$ gives
\begin{equation}
    \hat{\rho}^{(0)}_{L,K} = \text{Tr}_R(\hat{\rho}^{(0)}_K)
    = \ket{+}\!\bra{+}_{\sigma}^L \otimes \text{Tr}_R\!\left(\ket{\psi^0}\!\bra{\psi^0}_{F,M}^{LR,K}\right)
    =: \ket{+}\!\bra{+}_{\sigma}^A \otimes \hat{\psi}^0_{A,K},
\end{equation}
where we used the subsystem definition in Eq.\,\eqref{bip_def_app}.  
Since the entropy of a product state is additive,
\begin{equation}
   \text{ent}_{LR,K}^{(0)} = H(\sigma_A M_A F_A)_{\rho^0_{L,K}} = H(\sigma_A)_{\ket{+}\bra{+}} + H(M_A F_A)_{\psi^0_K} = H(M_A F_A)_{\psi^0_K},
\end{equation}
as $H(\sigma_A)_{\ket{+}\bra{+}}$ vanishes for the pure spin state $\ket{+}\bra{+}$.

For the final state,
\begin{equation}
    \text{ent}_{LR,K}^{(5)} = H(L)_{\rho^{(5)}_K}.
\end{equation}
The reduced state of $L$ is
\begin{equation}
 \hat{\rho}^{(5)}_{L,K} = \text{Tr}_R(\hat{\rho}^{(5)}_K)
 = \text{Tr}_R\!\left(\ket{\chi}\!\bra{\chi}_{\sigma}^{LR}\right) \otimes \text{Tr}_R\!\left(\ket{\psi^0}\!\bra{\psi^0}_{F,M}^{LR,K}\right)
 =: \hat{\chi}_{A} \otimes \hat{\psi}^0_{A,K}.
\end{equation}
Hence,
\begin{equation}
    \text{ent}_{LR,K}^{(5)} = H(\sigma_A M_A F_A)_{\rho^{(5)}_{L,K}} = H(\sigma_A)_{\chi} + H(M_A F_A)_{\psi^0_K}.
\end{equation}
Unlike the initial case, $\ket{\chi}_{\sigma}^{LR}$ is generally entangled (see Eq.\,\eqref{final_spin}), in which case its reduced state $\hat{\chi}_A$ is mixed:
\begin{equation}\label{ent_inc_app}
    H(\sigma_A)_{\chi} > 0.
\end{equation}
Since the matter and field components are unchanged, comparison yields
\begin{equation}\label{ent_increase_app}
     \text{ent}_{LR,K}^{(5)} > \text{ent}_{LR,K}^{(0)}
\end{equation}
if $\ket{\chi}_{LR}^\sigma$ is entangled, which is the desired result.

\subsection{Ruling out embezzlement.\label{app:sec:emb}}
One last possible objection to this conclusion involves the concept of \emph{entanglement embezzlement} \cite{PhysRevA.67.060302}.

Entanglement embezzlement refers to a subtle quantum phenomenon in which entanglement is effectively extracted from a highly entangled \emph{catalyst} state using only local operations, by altering the catalyst state in a minimal way. In such processes, the catalyst provides a hidden entanglement ``budget'' from which systems can borrow to become entangled themselves. While this phenomenon is highly nontrivial and typically requires a high-dimensional auxiliary system, it can in principle occur in certain regimes of quantum field theories. In fact, it has recently been argued that every state in a certain type of quantum field theories can serve as a resource for embezzlement if appropriate local operations are allowed \cite{vanluijk2024embezzlingentanglementquantumfields}.

At first glance, one might thus worry that the transformation from the initial to the final state in Eq.\,\eqref{start_to_final}, implemented via the local operators $\hat{U}$ and $\hat{U}'$, could be interpreted as an instance of such an embezzling protocol and therefore not a genuine case of entanglement generation via field-mediated interaction. However, looking at the dynamics, as visualized in Fig.\,\ref{fig:fme_locc}, shows the limitations of such argument. The protocol explicitly involves not only local operators applied at the beginning and end, but also the non-trivial period of pure field evolution
between them. This intermediate evolution cannot be part of an embezzling process, which by definition must involve only local operations and the passive use of an entangled catalyst.

A more direct argument rules out embezzlement even more conclusively. Suppose we remove the intermediate non-local field evolution and apply only the local operators $\hat{U}$ and $\hat{U}'$ in sequence, as would be prescribed to perform an embezzlement protocol. Then, as can be directly verified from their explicit forms in Eqs.\,\eqref{U_A},~\eqref{U_B}, and~\eqref{merging_op}, the combined action of $\hat{U}' \hat{U}$ simply brings the system back to its initial state:
\begin{equation}
    \hat{U}' \hat{U} \ket{\Psi}^{(0)} = \ket{\Psi}^{(0)}.
\end{equation}
This is straightforward to see: the 
matter-sector operators trivially undo the displacement of the sources, as in the original FME protocol, while the field dressing operators $\hat{W}_F(s)$ introduced during superposition creation are precisely undone by their inverses in the merging operations. As a result, if no time is allowed for the field to evolve, that is, if the system is not given a chance to acquire relative dynamical phases between branches, then no entanglement is transferred to the spin degrees of freedom. The system returns exactly to the initial pure product state, and no detectable entanglement is present.

This makes it manifest that the increase in entanglement observed in the final state cannot originate from local operations alone. The field’s autonomous evolution, uncontrolled by either party, plays a fundamental role in mediating entanglement between spatially separated regions. In other words, the process of entanglement generation in this protocol is not an artifact of embezzlement, but rather a genuine consequence of non-classical field dynamics.

\section{Continuum Limit}\label{app:cont_limit}

The discrete structure of our toy model enabled us to derive the key results present in this work and to perform otherwise involved calculations with relative ease, while the lattice provided a clear and intuitive framework for defining sources and local operations. 

In this appendix, we develop the continuum limit of our toy model, both at the classical and quantized levels, and derive the corresponding ground-state wave functions. We show that the resulting theory, together with its key features, reduces exactly to two-dimensional QED.

\subsection{Preliminary considerations}\label{sec:basic_c_limit}
Let us begin by outlining the procedure used to take the continuum limit of our discrete toy model. The natural strategy is to send the lattice spacing $a$ to zero while simultaneously letting the number of sites $N$ diverge. We first keep the physical size of the square lattice fixed at $L = Na$, and only afterwards take $L \to \infty$ to recover an unbounded continuous space. Thus
\begin{equation}\label{c_limits}
    a \to 0, \quad N \to \infty \quad \text{and then}\quad L = Na \to \infty.
\end{equation}
With this procedure in place, we can now take the continuum limit of the key quantities appearing in the toy model.

\paragraph*{Derivatives.}
We begin with the continuum limit of the discrete derivatives defined in Eq.\,\eqref{disc_der_def}. According to Eq.\,\eqref{c_limits},  a generic lattice function $f^{i,j}$ becomes a smooth field $f(\mathbf r)$ over $\mathbb R^2$, with coordinates
\begin{equation}\label{r_def} 
\mathbf{r}:=a\begin{pmatrix} j\\ i \end{pmatrix},
\end{equation}
Taking the rest of the limit, the symmetric finite differences reduce to ordinary partial derivatives, i.e.
\begin{equation}
    \begin{aligned}
        \bar{\partial}_x f(\mathbf{r}) &= \frac{f(r_x + a, r_y) - f(r_x - a, r_y)}{2a} \rightarrow \partial_x f(\mathbf{r}), \\
        \bar{\partial}_y f(\mathbf{r}) &= \frac{f(r_x, r_y + a) - f(r_x, r_y - a)}{2a} \rightarrow \partial_y f(\mathbf{r}).
    \end{aligned}
    \qquad \text{as } a \to 0.
\end{equation}

\paragraph*{Integrals.}
Lattice sums also admit a smooth continuum limit. Centering the lattice at the origin by enforcing the periodicity of boundary conditions, a generic sum over sites can be written as
\begin{equation}
    \sum_{i,j}
    = \sum_{i=-N/2}^{N/2-1} \sum_{j=-N/2}^{N/2-1}.
\end{equation}
Using Eq.\,\eqref{r_def}, we identify $r_x = a j$, $r_y = a i$, and again performing the limit it follows that
\begin{equation}\label{sum_limit}
    a^2 \sum_{i,j}
    = a^2 \sum_{i=-N/2}^{N/2-1} \sum_{j=-N/2}^{N/2-1}
    \xrightarrow[a\to 0]{L=Na}
    \int_{-L/2}^{L/2} dr_x\, dr_y
    \xrightarrow[L\to\infty]{}\!
    \int_{\mathbb R^2} d^2 r ,
\end{equation}
where we have used $dr_x = dr_y = a$.

\paragraph*{Fourier space.}
We can now proceed with sums over the discrete Fourier space. Again, a discrete sum over Fourier modes may also be written as
\begin{equation}
    \sum_{\alpha,\beta}
    = \sum_{\alpha=-N/2}^{N/2-1}\sum_{\beta=-N/2}^{N/2-1},
\end{equation}
and, defining
\begin{equation}\label{k_contin}
    \mathbf{k}
    := \frac{2\pi}{L}
    \begin{pmatrix}
        \beta \\ \alpha
    \end{pmatrix},
\end{equation}
the mode spacing is $dk_x = dk_y = 2\pi/L$. Thus,
\begin{equation}
    \frac{1}{L^2}\sum_{\alpha,\beta}
    \xrightarrow[a\to 0]{L\to\infty}
    \int\!\frac{d^2 k}{(2\pi)^2},
\end{equation}
giving the integral over continuum Fourier space.

\paragraph*{DFT to FT.}
Before turning to our toy model, we briefly use these results to show how the discrete Fourier transform defined in Eq.\,\eqref{DFT_lattice} becomes the continuous Fourier transform in the continuum limit. Using the continuum substitutions for $\mathbf{r}$, $\mathbf{k}$ and $f^{i,j}$, the direct DFT becomes
\begin{equation}
   a^2 \tilde f^{\alpha,\beta}= \sum_{i,j} f^{i,j}\,e^{- \mathrm{i}\frac{2\pi}{N}(i\alpha+j\beta)}
   = a^2\sum_{i,j} f(\mathbf{r})\,e^{-\mathrm{i}\mathbf{k} \cdot \mathbf{r}}
   \xrightarrow[a\to0]{L\to\infty}
   \!\int\! d^{2}r\, f(\mathbf{r})\,e^{-\mathrm{i}\mathbf{k}\cdot\mathbf{r}},
\end{equation}
yielding the standard Fourier transform,
\begin{equation}
   \tilde f(\mathbf{k})
   = \int d^{2}r\, f(\mathbf{r})\,e^{-\mathrm{i}\mathbf{k}\cdot\mathbf{r}} .
\end{equation}

For the inverse DFT the same substitutions give
\begin{equation}
   f^{i,j} = \frac{1}{N^{2}}
     \sum_{\alpha,\beta}
     \tilde f^{\alpha,\beta} \, e^{\mathrm{i}\frac{2\pi}{N}(i\alpha+j\beta)}
   = \frac{1}{L^{2}}
     \sum_{\alpha,\beta}
     \tilde f(\mathbf{k})\,e^{\mathrm{i}\mathbf{k}\cdot\mathbf{r}}
   \xrightarrow[a\to0]{L\to\infty}
   \int\!\frac{d^{2}k}{(2\pi)^{2}}\,
   \tilde f(\mathbf{k})\,e^{\mathrm{i}\mathbf{k}\cdot\mathbf{r}} 
   = f(\mathbf{r}) ,
\end{equation}
recovering the usual inverse Fourier transform. Finally, for the discrete wave vector, using $\mathbf{k}=\tfrac{2\pi}{L}(\beta,\alpha)$ and $L=Na$ gives
\begin{equation}
   \bar{\mathbf{k}}   = \frac{1}{a}
     \begin{pmatrix}
        \sin\!\left(\tfrac{2\pi}{N}\beta\right) \\
        \sin\!\left(\tfrac{2\pi}{N}\alpha\right)
     \end{pmatrix}
   = \frac{1}{a}
     \begin{pmatrix}
        \sin(a k_x) \\[2pt]
        \sin(a k_y)
     \end{pmatrix}
   \xrightarrow{a\to 0}
   \begin{pmatrix}
        k_x \\ k_y
     \end{pmatrix}
   = \mathbf{k}.
\end{equation}
Thus the discrete wave vector smoothly reduces to its continuous analogue.

\subsection{The classical model}
We now turn to the classical toy model and apply the continuum methods developed in the previous section. At the end we can then provide a precise and simple procedure to perform the continuum limit of any of the results found for the discrete toy model.

\paragraph*{Lagrangian.}
Starting from the sourceless Lagrangian with all constants restored (Eq.\,\eqref{lagrangian_dime}),
\begin{equation}
   L = \frac{1}{2}\sum_{i,j}\!\left[
      m(\dot q_x^{i,j})^{2}
      + m(\dot q_y^{i,j})^{2}
      - \kappa a^{2}\!
        \left(\bar\partial_x q_y^{i,j} - \bar\partial_y q_x^{i,j}\right)^{2}
   \right],
\end{equation}
and working in temporal gauge ($q_0=0$), we substitute $q^{i,j}\!\to q(\mathbf r)$ and use the continuum limits of the derivatives to get
\begin{equation}
   L = \frac{1}{2}a^{2}\!\sum_{i,j}
      \Big[
         \tfrac{m}{a^{2}}|\dot{\mathbf q}(\mathbf r)|^{2}
         - \kappa\big(\partial_x q_y - \partial_y q_x\big)^{2}
      \Big].
\end{equation}
The discrete magnetic term naturally tends to
\begin{equation}
       b^{i,j} := \bar\partial_x q_y^{i,j} - \bar\partial_y q_x^{i,j} \to \partial_x q_y - \partial_y q_x =: b(\mathbf r).
\end{equation}
Using $a^{2}\sum_{i,j}\!\to\!\int d^{2}r$ (Eq.~\eqref{sum_limit}), the continuum Lagrangian becomes
\begin{equation}
   L = \frac{1}{2}\!\int\! d^{2}r\,
       \big[\, \sigma |\dot{\mathbf q}|^{2} - \kappa b^{2} \,\big],
\end{equation}
where we introduced the \emph{field surface density} $\sigma := \frac{m}{a^{2}},$ assumed finite as $a\to0$.

\paragraph*{Hamiltonian.}
The conjugate momenta follow immediately as
\begin{equation}
   p_{s}(\mathbf r)
   = \frac{\delta L}{\delta\dot q_{s}(\mathbf r)}
   = \sigma\,\dot q_{s}(\mathbf r),
\end{equation}
implying that the discrete momenta become densities, i.e.
\begin{equation}\label{p_cont_new}
   p_{s}^{i,j} \to a^{2} p_{s}(\mathbf r).
\end{equation}
Applying the Legendre transform to the continuous Lagrangian yields the sourceless Hamiltonian in temporal gauge,
\begin{equation}
   H = \frac{1}{2}\!\int\! d^{2}r
       \left[
          \frac{|\mathbf p(\mathbf r)|^{2}}{\sigma}
          + \kappa\, b^{2}(\mathbf r)
       \right],
\end{equation}
directly analogous to the $2{+}1$-dimensional electromagnetic Hamiltonian with $J^{0}=0$ and $A_{0}=0$.

\paragraph*{Gauss constraint.}
From Eq.\,\eqref{sourceless_constr}, using the continuum limits of derivatives and momentum, the discrete constraint becomes
\begin{equation}
   \mathcal C(\mathbf r)
   := \partial_x p_x + \partial_y p_y
   = \nabla\!\cdot\!\mathbf p(\mathbf r)=0,
\end{equation}
with the following discrete–continuous relation
\begin{equation}\label{gauss_lim_new}
   \frac{1}{a^2}\mathcal C^{i,j} \to \mathcal C(\mathbf r).
\end{equation}

We now consider also the case of a static source (Eq.\,\eqref{constr_source}). Before taking the continuum limit, both terms must have the same dimensions. Since
\begin{equation}
      [\rho^{i,j}] = C,\qquad
   [\mathcal C^{i,j}] = [\bar\partial]\,[p]
   = \frac{1}{L}\,M\frac{L}{T}\!=\!\frac{M}{T}, 
\end{equation}
dimensional consistency requires introducing a constant $\nu$ with
\begin{equation}\label{nu_new}
   [\nu] = \frac{M}{T C}.
\end{equation}
The discrete constraint then becomes
\begin{equation}
     \mathcal C_{\rho}^{i,j} = \mathcal C^{i,j} + \nu \rho^{i,j}=0.  
\end{equation}
For the continuum limit, the source must scale as a density, i.e.
\begin{equation}
   \frac{1}{a^2}\rho^{i,j}\to \rho(\mathbf r),\qquad
   [\rho(\mathbf r)]=\frac{C}{L^{2}},
\end{equation}
and using Eq.~\eqref{gauss_lim_new}, the full constraint reduces to
\begin{equation}\label{cont_constr_new}
   a^{2}\mathcal C_{\rho}(\mathbf r)
   := a^{2}\big[\mathcal C(\mathbf r)+\nu\rho(\mathbf r)\big]=0\to
   \nabla\!\cdot\!\mathbf p(\mathbf r) + \nu\rho(\mathbf r) = 0,
\end{equation}
the exact $2$-dimensional analogue of the electromagnetic Gauss law.
\smallskip
\begin{highlightbox}[Recipe for the continuum limit]\label{box:cont_lim}
The discussion above yields a concise procedure to perform the continuum limit:
\begin{enumerate}
   \item Start from the discrete expression.
   \item Reinsert the fixed constants $m$, $\kappa a^{2}$ (Eq.~\eqref{fixed_const}) and $\nu$ (Eq.~\eqref{nu_new}) to ensure dimensional consistency.
   \item Apply the limit in Eq.~\eqref{c_limits} using the discrete–continuous relations summarized in Table~\ref{tab:classical_limits}.
\end{enumerate}
This algorithm systematically produces the continuous counterparts of all discrete relations used in this work.
\end{highlightbox}
\subsection{The quantized model}
We now apply the continuum recipe established to the quantized discrete toy model to compute the ground state energy, wave function, and verify their exact analogy with QED (Ref.\,\cite{Giacomini2023quantumstatesof}).

\paragraph*{Commutation relations.}
The discrete canonical commutators are
\begin{equation}
   [\hat q_s^{i,j}, \hat p_r^{n,m}] = \mathrm{i}\hbar \,\delta_{s,r} \delta^{i,n} \delta^{j,m}.
\end{equation}
Using the continuum limits $q^{i,j} \to q(\mathbf r)$, $p^{i,j} \to a^2 p(\mathbf r)$ (Table~\ref{tab:classical_limits}) and $\mathbf r = a(j,i)^T$, we obtain
\begin{equation}
   [\hat q_s(\mathbf r), \hat p_r(\mathbf r')] = \mathrm{i}\hbar \,\delta_{s,r} \frac{\delta^{i,n}}{a}\frac{\delta^{j,m}}{a} \xrightarrow{a\to 0} \mathrm{i}\hbar \,\delta_{s,r}\, \delta^2(\mathbf r - \mathbf r').
\end{equation}

\paragraph*{Sourceless case.}
The discrete ground state energy (Eq.\,\eqref{g_energy}) in the continuum limit, with dimensional constants restored, becomes
\begin{equation}
   \mathcal E_0= \frac{\hbar}{2} \sum_{\alpha,\beta} |\bar{\mathbf k}| \to \frac{\hbar}{2} \sqrt{\frac{\kappa}{\sigma}}\, L^2 \int \frac{d^2 k}{(2\pi)^2} |\mathbf k| = \frac{\hbar}{2} \sqrt{\frac{\kappa}{\sigma}}\, \delta^2(0) \int d^2 k\, |\mathbf k|,
\end{equation}
where $\delta^2(0) = L^2/(2\pi)^2$ captures the familiar volume divergence.

For the $q$-field ground state (Eq.\,\eqref{g_state_q_pos}),
\begin{equation}
   \Psi_0[q] = A \exp\Big[-\frac{1}{2\hbar} \sum_{i,j,n,m} G(i-n,j-m) b^{i,j} b^{n,m}\Big],
\end{equation}
the Green function in the continuum limit reads
\begin{equation}
  \frac{1}{a^2}\, G(i-n,j-m) \to  \int \frac{d^2 k}{(2\pi)^2} \frac{1}{|\mathbf k|} e^{-\mathrm{i} \mathbf k \cdot (\mathbf r - \mathbf r')} =:  G(\mathbf r - \mathbf r').
\end{equation}
Applying dimensional constants and the continuum prescription (Box\,\ref{box:cont_lim}) gives
\begin{equation}
   \Psi_0[q] = A \exp\Big[-\frac{\sqrt{\sigma\kappa}}{2\hbar} \int d^2 r\, d^2 r' \, G(\mathbf r - \mathbf r')\, b(\mathbf r) b(\mathbf r')\Big],
\end{equation}
the exact two-dimensional analogue of the QED ground state, with $G(\mathbf r - \mathbf r') \sim 1/|\mathbf r - \mathbf r'|$.

The $p$-field wave function follows analogously (see Table~\ref{tab:quantum_limits}), in perfect analogy with the QED electric field basis (Ref.\,\cite{Giacomini2023quantumstatesof}).

\paragraph*{Source case.}
For a static source $\rho^{i,j}$, the discrete energy difference (Eq.~\eqref{energy_differe}) reads
\begin{equation}
   \mathcal E_\rho - \mathcal E_0 = \frac{1}{2} \sum_{i,j,n,m} D(i-n,j-m) \rho^{i,j} \rho^{n,m}, \qquad
   D(i-n,j-m) = \frac{1}{N^2} \sum_{\alpha,\beta} \frac{e^{-\mathrm{i} \frac{2\pi}{N}[(i-n)\alpha + (j-m)\beta]}}{|\bar{\mathbf k}|^2}.
\end{equation}
The continuum limit gives
\begin{equation}
  \frac{1}{a^2}\, D(i-n,j-m) \to  \int \frac{d^2 k}{(2\pi)^2} \frac{e^{-\mathrm{i} \mathbf k \cdot (\mathbf r - \mathbf r')}}{|\mathbf k|^2} =: D(\mathbf r - \mathbf r'),
\end{equation}
so that, with dimensional constants,
\begin{equation}
   \mathcal E_\rho - \mathcal E_0 = \frac{\nu^2}{2\sigma} \int d^2 r\, d^2 r' \, D(\mathbf r - \mathbf r')\, \rho(\mathbf r) \rho(\mathbf r').
\end{equation}
This reproduces the two-dimensional Coulomb energy, with $D(\mathbf r - \mathbf r') \sim -\ln|\mathbf r - \mathbf r'|$.

The corresponding momenta $p_{s,\rho}^{i,j}$ (Eq.~\eqref{p_rho}) and the wave functions in the $q$- and $p$-field bases (Eqs.\,\eqref{source_g_state_q},~\eqref{source_g_state_p_app}) then follow directly, confirming the analogy with QED (again Ref.\,\cite{Giacomini2023quantumstatesof}).

\newpage
\subsection{Tables}
\vspace{-10pt}
\begin{center}

\begin{table}[ht]
    \centering
    \begin{tblr}{colspec  = {|c|c|},
                 colsep=4pt,
                 row{odd} = {bg=chillred!20},
                 row{1}   = {bg=chillred!30},      
                 rowsep=3pt,
                }    
        \hline
        \textbf{Discrete} & \textbf{Continuous} \\ 
        \hline\hline
        $\bar{\partial}_s$&$\partial_s$\\
        $a^2 \sum_{i,j}$&$\int d^2r$\\
        $\frac{1}{L^2} \sum_{\alpha,\beta}$&$\int\frac{d^2k}{(2\pi)^2}$\\
        $\tilde{f}^{\alpha,\beta}=\text{DFT}(f^{i,j})$&$\frac{1}{a^2}\tilde{f}(\mathbf{k})=\frac{1}{a^2}\text{FT}(f(\mathbf{r})) $\\
        $ \mathbf{\bar{k}} $ &$\mathbf{k}$\\
         $\mathbf{q}^{i,j}$ & $\mathbf{q}(\mathbf{r})$ \\  
         $\frac{1}{a^2}\mathbf{p}^{i,j}$&$\,\mathbf{p}(\mathbf{r})$\\ 
        $b^{i,j} = \bar{\partial}_x q_y^{i,j} - \bar{\partial}_y q_x^{i,j}$ & $b(\mathbf{r})=\partial_xq_y(\mathbf{r})-\partial_yq_x(\mathbf{r})$\\ 
        $\frac{1}{a^2}\rho^{i,j}$&$\rho(\mathbf{r})$\\
        $\mathcal{C}^{i,j}_{\rho}=\bar{\partial}_x p_x^{i,j} + \bar{\partial}_y p_y^{i,j}+\nu\rho^{i,j}$ & $\mathcal{C}(\mathbf{r})=\mathbf{\nabla}\cdot\mathbf{p}(\mathbf{r})+\nu\rho(\mathbf{r})$   \\ 
        $H = \frac{1}{2} \sum_{i,j} \left[\frac{(p_x^{i,j})^2}{m} + \frac{(p_y^{i,j})^2}{m} + \kappa a^2(b^{i,j})^2\right]$&$    H = \frac{1}{2} \int d^2r \left[ \frac{|\mathbf{p}(\mathbf{r})|^2}{\sigma} + \kappa \, b^2(\mathbf{r}) \right]$ \\[5pt]
        \hline
    \end{tblr}
    \caption{Summary of important discrete quantities in the classical model that have been taken to the continuum limit in this appendix.}
    \label{tab:classical_limits}
\end{table}
\end{center}

\begin{table}[ht]
    \centering
            \begin{center}

                \begin{tblr}{colspec  = {|c|c|},
                 colsep=4pt,
                 row{odd} = {bg=chillred!20},
                 row{1}   = {bg=chillred!30},      
                 rowsep=3pt,
                }    
                \hline
                \textbf{Discrete} & \textbf{Continuous} \\ 
                \hline\hline
                $\mathcal{E}_0 = \frac{1}{2} \hbar \sum_{\alpha,\beta} |\mathbf{\bar{k}}|$ &
                $\mathcal{E}_0 = \frac{\hbar}{2} \sqrt{\frac{\kappa }{\sigma}}\, \delta^2(0) \int \frac{d^2k}{(2\pi)^2} |\mathbf{k}|$\\
                $G(i-n, j-m) = \frac{1}{N^2} \sum_{\alpha,\beta} \frac{1}{|\mathbf{\bar{k}}|} \, \mathrm{e}^{ -\mathrm{i} \frac{2\pi}{N}[(i-n)\alpha + (j-m)\beta] }$ &
                $G(\mathbf{r} - \mathbf{r}') = \int \frac{d^2\mathbf{k}}{(2\pi)^2} \frac{1}{|\mathbf{k}|} \, \mathrm{e}^{ -\mathrm{i} \mathbf{k} \cdot (\mathbf{r} - \mathbf{r}') }\sim \frac{1}{|\mathbf{r} - \mathbf{r}'|}$\\
                $\Psi_0[q] = A \exp\left[-\frac{1}{2\hbar} \sum_{i,j} \sum_{n,m} G(i-n, j-m)\, b^{i,j} b^{n,m} \right]$ &
                $\Psi_0[q] = A \exp\left[-\frac{\sqrt{\sigma \kappa }}{2\hbar} \int d^2\mathbf{r} \, d^2\mathbf{r}' \, G(\mathbf{r} - \mathbf{r}')\, b(\mathbf{r}) b(\mathbf{r}') \right]$\\
                $\Psi_0[p]=A\,\delta(\mathcal{C})\,\exp\left[-\frac{1}{2\hbar }\sum_{i,j}\sum_{n,m}G(i-n,j-m)\,p^{i,j}_sp^{n,m}_s\right]$ &
                $\Psi_0[p]=A\,\delta(\mathcal{C})\,\exp\left[-\frac{1}{2\hbar\sqrt{\kappa \sigma} }\int d^2\mathbf{r} \, d^2\mathbf{r}' \, G(\mathbf{r} - \mathbf{r}')\,\mathbf{p}(\mathbf{r})\cdot\mathbf{p}(\mathbf{r}')\right]$\\
                $ D(i-n, j-m) = \frac{1}{N^2} \sum_{\alpha,\beta} \frac{1}{|\mathbf{\bar{k}}|^2} \, \mathrm{e}^{ -\mathrm{i} \frac{2\pi}{N}[(i-n)\alpha + (j-m)\beta] }$ &
                $D(\mathbf{r} - \mathbf{r}')=\int \frac{d^2\mathbf{k}}{(2\pi)^2} \frac{1}{|\mathbf{k}|^2} \, \mathrm{e}^{ -\mathrm{i} \mathbf{k} \cdot (\mathbf{r} - \mathbf{r}') }\sim -\ln |\mathbf{r} - \mathbf{r}'|$\\
                $\mathcal{E}_\rho - \mathcal{E}_0 = \frac{1}{2} \sum_{i,j} \sum_{n,m} D(i-n, j-m) \, \rho^{i,j} \rho^{n,m}$ &
                $\mathcal{E}_\rho - \mathcal{E}_0 = \frac{\nu^2}{2\sigma} \int d^2\mathbf{r} \, d^2\mathbf{r}' \, D(\mathbf{r} - \mathbf{r}') \, \rho(\mathbf{r}) \rho(\mathbf{r}')$\\
                $\mathbf{p}^{i,j}_{\rho}=\sum_{n,m}\bar{\partial}^{\{i,j\}}\cdot D(i-n,j-m)\rho^{n,m},$ &
                $\mathbf{p}_{\rho}(\mathbf{r})=\nu\int d^2\mathbf{r} \,\mathbf{\nabla}_{\mathbf{r}}\cdot D(\mathbf{r} - \mathbf{r}')\,\rho(\mathbf{r}')$\\
                $\Psi_{0,\rho}[q]=\Psi_0[p]\cdot\exp\left[- \frac{\text{i}}{\hbar}\sum_{i,j}p^{i,j}_{s,\rho}q_s^{i,j}\right]$ &
                $\Psi_{0,\rho}[q]=\Psi_0[p]\cdot\exp\left[- \frac{\text{i}}{\hbar}\int d^2\mathbf{r} \,\mathbf{p}_{\rho}(\mathbf{r})\cdot \mathbf{q}(\mathbf{r})\right]$\\
                \begin{tabular}[c]{@{}l@{}}
                $\Psi_{0,\rho}[p] = A\,\delta(\mathcal{C}_{\rho})\,\exp\bigg[ -\frac{1}{2\hbar } \sum\limits_{i,j}\sum\limits_{n,m} G(i-n,j-m)\,$ \\
                $\quad\quad\quad \cdot(p^{i,j}_s - p^{i,j}_{s,\rho})(p^{n,m}_s - p^{n,m}_{s,\rho})\bigg]$
                \end{tabular}
                &
                \begin{tabular}[c]{@{}l@{}}
                $\Psi_{0,\rho}[p] = A\,\delta(\mathcal{C})\,\exp\bigg[ -\frac{1}{2\hbar\sqrt{\kappa \sigma}} \int d^2\mathbf{r} \, d^2\mathbf{r}' \, G(\mathbf{r} - \mathbf{r}')$ \\
                $\quad\quad\quad \cdot(\mathbf{p} - \mathbf{p}_{\rho})(\mathbf{r}) \cdot (\mathbf{p} - \mathbf{p}_{\rho})(\mathbf{r}') \bigg]$
                \end{tabular}\\
                \hline
            \end{tblr}
            \caption{Summary of relevant discrete quantities in the quantized model and their continuum limits.}
            \label{tab:quantum_limits}

            \end{center}
\end{table}

\end{document}

%% file: preamble.tex
\usepackage[T1]{fontenc}
\usepackage[utf8]{inputenc}
\usepackage{lmodern}

\usepackage[english]{babel}
\usepackage[dvipsnames]{xcolor}
\usepackage{graphicx}
\usepackage{tikz}
\usepackage{pgfplots}
\usepackage{tikz-3dplot}
\usepackage{tkz-euclide}
\usepackage{tikz-cd}

\usetikzlibrary{arrows}
\usetikzlibrary{arrows.meta}
\usetikzlibrary{decorations.markings}
\usetikzlibrary{decorations.pathmorphing}
\usetikzlibrary{calc}
\usetikzlibrary{intersections}
\usetikzlibrary{pgfplots.fillbetween}
\usetikzlibrary{shapes.geometric}
\usetikzlibrary{quotes}

\tikzset{>={Latex[length=6pt,width=5pt]}}

\newlength\tickoffset
\pgfplotsset{compat=1.16}

\definecolor{myred}{RGB}{225, 65, 50}
\definecolor{chillred}{HTML}{da8d83}
\definecolor{darkgreen}{HTML}{006457}
\definecolor{lightgreen}{HTML}{448d3d}
\definecolor{mycyan}{HTML}{009df8} 
\definecolor{mypurple}{HTML}{ae54a2}
\definecolor{myorange}{RGB}{234, 160, 63}
\definecolor{myceleste}{HTML}{9cbfdf}
\definecolor{myblue}{HTML}{3a4856}
\definecolor{mygray}{RGB}{154, 154, 154}
\definecolor{rect}{rgb}{0.2,0.45,0.9}
\definecolor{myyellow}{HTML}{ebae34}

\usepackage{amsmath}
\usepackage{amsthm}
\usepackage{amssymb}
\usepackage{mathrsfs}
\usepackage{mathtools}
\usepackage{tensor}
\usepackage{mathrsfs}
\usepackage{array}
\usepackage{extarrows}
\usepackage{bm}
\usepackage{nicematrix}
\usepackage{slashed}
\usepackage[normalem]{ulem}
\usepackage{siunitx}
\usepackage{titlesec}
\usepackage{titlesec}
\usepackage{wrapfig}
\usepackage{enumitem}
\usepackage{needspace}
\usepackage{longtable}
\usepackage{stackengine}
\usepackage{changepage}
\usepackage{scalerel}
\usepackage{stackengine}
\usepackage{placeins}
\makeatletter
\renewenvironment{table}
{%
	\@float{table}}
{\end@float}
\makeatother

\usepackage{hyperref}
\hypersetup{
	colorlinks,
	linktoc=page,
	linkcolor={myblue},
	citecolor={myred},
	urlcolor={myred}
}
\providecommand{\doi}[1]{}
\providecommand{\eprint}[2][]{}
\renewcommand{\doi}[1]{\href{http://dx.doi.org/#1}{\nolinkurl{doi:#1}}}
\renewcommand{\eprint}[2][arXiv]{#1:#2}

\setlist{nosep}

\usepackage{xpatch}
\makeatletter
\xpatchcmd{\proof}{\topsep6\p@\@plus6\p@\relax}{}{}{}
\makeatother

\usepackage{tcolorbox}
\tcbuselibrary{theorems}
\tcbuselibrary{skins}
\tcbuselibrary{breakable}

\newtcolorbox{resultbox}{colback=myceleste!20, colframe=myceleste!55!black, breakable,before skip=7pt}
\newtcolorbox{highlightbox}[1][]{
    breakable,
    colback=chillred!15,
    colframe=chillred!55!black,
    title=#1,
    arc=4pt,
    left=6pt,
    right=6pt,
    top=4pt,
    bottom=4pt,
    before skip=0.5\baselineskip,
    after skip=0pt,
    enhanced,
    fontupper=\setlength{\emergencystretch}{2em},
}
\titlespacing*{\section}
  {0pt}{1.8ex}{0.9ex}

\titlespacing*{\subsection}
  {0pt}{1.3ex}{0.6ex}

\tikzset{
  thinarrow/.style={
    -{Latex[length=6pt, width=4pt]},
    thick
  }
}